%% file: paper.tex
\documentclass[11pt,a4paper]{article}

\pdfoutput=1

\usepackage{jheppub}
\usepackage{xcolor}
\usepackage[T1]{fontenc}
\usepackage{tocloft}
\usepackage[normalem]{ulem}
\usepackage{subcaption}
\usepackage{array}
\usepackage{enumitem}
\usepackage{extarrows}
\usepackage{mathtools}

\allowdisplaybreaks[1] 

\newcommand{\si}{SCET-1}
\newcommand{\sii}{SCET-2}
\newcommand{\softserve}{{\tt SoftSERVE}}

\newcommand{\abs}[1]{\left| #1 \right|}
\newcommand{\eps}{\epsilon}
\newcommand{\taubar}{\bar\tau}
\newcommand{\no}{\nonumber}
\def\rd{\mathrm{d}}

\title{\boldmath 
Generic dijet soft functions at two-loop order:
correlated emissions}

\author[a]{Guido Bell,}
\author[b]{Rudi Rahn,}
\author[c]{and Jim Talbert}

\note{SI-HEP-2018-37, QFET-2018-24, DESY 18-209.}

\affiliation[a]{Theoretische Physik 1, Naturwissenschaftlich-Technische Fakult\"at, 
Universit\"at Siegen,\\Walter-Flex-Strasse 3, 57068 Siegen, Germany}
\affiliation[b]{Albert Einstein Center for Fundamental Physics, Institut f\"ur Theoretische Physik, \\
Universit\"at Bern, Sidlerstrasse 5, 3012 Bern, Switzerland}
\affiliation[c]{Theory Group, Deutsches Elektronen-Synchrotron (DESY), Notkestra{\ss}e 85, \\
22607 Hamburg, Germany}

\emailAdd{bell@physik.uni-siegen.de}
\emailAdd{rahn@itp.unibe.ch}
\emailAdd{james.talbert@desy.de}

\abstract{We present a systematic algorithm for the perturbative computation of soft functions that are defined in terms of two light-like Wilson lines. Our method is based on a universal parametrisation of the phase-space integrals, which we use to isolate the singularities in Laplace space. The observable-dependent integrations can then be performed numerically, and they are implemented in the new, publicly available package \texttt{SoftSERVE}~that we use to derive all of our numerical results. Our algorithm applies to both SCET-1 and SCET-2 soft functions, and in the current version it can be used to compute two out of three NNLO colour structures associated with the so-called correlated-emission contribution. We confirm existing two-loop results for about a dozen $e^+e^-$ and hadron-collider soft functions, and we obtain new predictions for the C-parameter as well as thrust-axis and broadening-axis angularities.
}


\begin{document} 

\maketitle

\flushbottom

\input{Intro.tex}

\input{Measure.tex}

\input{BareCalculation.tex}

\input{Renorm.tex}

\input{Generalisations.tex}

\input{Numerics.tex}

\input{Results.tex}

\input{Conclude.tex}

\acknowledgments
We would like to sincerely thank the developers of {\tt{SecDec}} for their support throughout the duration
of this work, and in particular for the dedicated development of the multi-regulator functionality in
{\texttt{pySecDec}}, which facilitated crosschecks for SCET-2 observables. G.B.~is supported by the Deutsche
Forschungsgemeinschaft (DFG) within Research Unit FOR 1873, and R.R.~is supported by the Swiss National
Science Foundation (SNF) under grants CRSII2$\_$160814 and 200020$\_$182038. J.T.~and R.R.~acknowledge travel
support from the University of Oxford Department of Physics and Universit\"at Siegen. J.T.~also acknowledges
research and travel support from the Senior Scholarship Trust of Hertford College (Oxford) and DESY Hamburg. 

\appendix

\input{App_Twoloopn.tex}

\input{App_Fourier.tex}


\end{document}

%% file: Intro.tex
\newpage

\section{Introduction}
\label{sec:intro}

Soft functions are an essential ingredient of QCD factorisation theorems. They describe the low-energy
contribution to a scattering process, which is usually easier to compute than the analogous hard process 
in full QCD. Due to the eikonal form of the soft interactions, the soft functions can be represented by a 
vacuum matrix element of Wilson lines that point along the directions of the energetic, coloured particles 
in the scattering process.

As long as the underlying scale of the soft interactions is large enough, the soft functions can be calculated
order-by-order in perturbation theory. At next-to-leading order (NLO) the calculation involves one-loop
virtual and single real-emission contributions, which can be computed with standard techniques. Starting 
at NNLO and beyond, the singularity structure of the individual contributions becomes intricate and the
divergences in the phase-space integrals overlap. The calculation of NNLO soft functions is needed for 
high-precision resummations and has attracted considerable attention in the past 
years~\cite{Belitsky:1998tc,Becher:2005pd,Kelley:2011ng,Monni:2011gb,Hornig:2011iu,Li:2011zp,Kelley:2011aa,Becher:2012za,Ferroglia:2012uy,Becher:2012qc,vonManteuffel:2013vja,Ferroglia:2013awa,Czakon:2013hxa,vonManteuffel:2014mva,Boughezal:2015eha,Echevarria:2015byo,Luebbert:2016itl,Gangal:2016kuo,Li:2016tvb,Campbell:2017hsw,Wang:2018vgu,Li:2018tsq,Dulat:2018vuy,Angeles-Martinez:2018mqh}.
Very recently, first results for N$^3$LO soft functions have been presented~\cite{Li:2016ctv,Moult:2018jzp}.

Whereas most of these calculations were performed analytically on a case-by-case basis, a systematic approach
that exploits the universal structure of the soft functions is currently missing. The purpose of our work 
is to fill this gap, and as a first step we focus on soft functions that arise in processes with two 
massless, coloured, hard partons. The soft functions for these processes can be written in the form
\begin{align}
S(\tau, \mu) = \frac{1}{N_c} \; \sum_{i\in X} \;
\mathcal{M}(\tau;\lbrace k_{i} \rbrace)\;
\mathrm{Tr}\; 
|\langle X | \,T [S^{\dagger}_{n}(0) S_{\bar{n}}(0)]\, | 0 \rangle |^{2}\,,
\label{eq:softfun:definition}
\end{align}
where $S_{n}$ and $S_{\bar{n}}$ are soft Wilson lines, and $n^{\mu}$ and $\bar{n}^{\mu}$ denote the
directions of the hard partons with $n^2=\bar{n}^2=0$. For concreteness, we assume that the hard partons 
are in a back-to-back configuration ($n\cdot{\bar{n}}=2$), and that the Wilson lines are in the fundamental 
colour representation. The definition in \eqref{eq:softfun:definition} contains a trace over colour 
indices as well as a function $\mathcal{M}(\tau,\lbrace k_{i} \rbrace)$, which specifies what is measured 
on the soft radiation $X$ with parton momenta $k_{i}$ for the observable under consideration. 
We will also see later that it is irrelevant whether $n^{\mu}$ and $\bar{n}^{\mu}$ 
are incoming or outgoing directions up to the order we consider, NNLO. Our method therefore equally applies to dijet observables in 
$e^{+} e^{-}$ annihilation, single-jet observables in deep-inelastic scattering, and zero-jet observables 
at hadron colliders. For convenience, we will refer to all of these cases with two massless, coloured, hard partons 
as \emph{dijet soft functions} in the following.

The key observation of our analysis is that the soft matrix element in the definition 
\eqref{eq:softfun:definition} is universal, i.e.~independent of the considered dijet observable. It is 
therefore possible to isolate the implicit divergences in the phase-space integrals with a universal 
parametrisation, and to compute the observable-dependent coefficients in an expansion in the dimensional 
regulator \mbox{$\eps=(4-d)/2$} numerically. The dependence of the soft function on the observable is 
thus entirely confined to the measurement function $\mathcal{M}(\tau,\lbrace k_{i} \rbrace)$, which acts 
as a weight factor for the numerical integrations. We will discuss the specific form we assume for the 
measurement function in the following section, where we will also learn that it is crucial to understand 
its properties in the singular limits of the matrix element.

The goal of our analysis thus consists in devising an algorithm that allows for an auto\-mated calculation 
of dijet soft functions to NNLO in the perturbative expansion. At NNLO the double real-emission contribution 
consists of three colour structures, which are often referred to as correlated ($C_F T_F n_f$, $C_F C_A$) 
and uncorrelated ($C_F^2$) emissions. As the phase-space parametrisations that are needed to factorise 
the divergences are different in both cases, we will concentrate in this work on correlated emissions, 
leaving the uncorrelated emissions for a future study~\cite{BRT}. For observables that obey the non-Abelian 
exponentiation (NAE) theorem~\cite{Gatheral:1983cz,Frenkel:1984pz}, a dedicated calculation of the uncorrelated-emission contribution is in fact not needed, and we can therefore present complete NNLO results for a number of 
$e^+e^-$ and hadron-collider soft functions already in this work. This is, however, not true for observables 
that violate the NAE theorem, like jet-veto or grooming observables, which we will address in~\cite{BRT}
(preliminary results can be found in~\cite{Bell:2018jvf}).

As explained earlier, we aim at a numerical evaluation of bare dijet soft functions in an expansion in the 
dimensional regulator $\eps$. It is, however, well known that the phase-space integrals for certain 
soft functions suffer from rapidity divergences that are not regularised in dimensional regularisation (DR). 
This typically arises whenever the soft radiation is constrained to have small transverse momenta.
Several prescriptions for the regularisation of the rapidity divergences have been proposed in the literature (see 
e.g.~\cite{Chiu:2009yx,Becher:2011dz,Chiu:2012ir,Echevarria:2015byo,Li:2016axz}), and in this work we will use a variant 
of the analytic regulator introduced in~\cite{Becher:2011dz}. Specifically, this results in a 
modification of the generic $d$-dimensional phase-space measure of the form,
\begin{equation}
\int d^dp \; \left(\frac{\nu}{n\cdot p + \bar n \cdot p}\right)^\alpha \;  \delta(p^2) \theta(p^0) \,,
\end{equation}
where $\alpha$ is the rapidity regulator. The rapidity scale $\nu$ is introduced on dimensional 
grounds, similar to the renormalisation scale $\mu$ in conventional DR.
The rapidity divergences then show up as poles in $1/\alpha$, and the renormalised soft function $S(\tau, \mu, \nu)$  
depends on two scales $\mu$ and $\nu$. In the context of Soft-Collinear Effective Theory 
(SCET)~\cite{Bauer:2000ew,Bauer:2000yr,Bauer:2001yt,Beneke:2002ph}, these soft functions are classified as SCET-2
observables, whereas those functions $S(\tau, \mu)$ that are not sensitive to the rapidity scale $\nu$ 
(and are well-defined in DR) refer to SCET-1 observables.

With the dimensional and the rapidity regulator in place, the bare soft functions can be evaluated in a 
double expansion in $\epsilon$ and $\alpha$. The main result of our analysis is an integral representation
of a generic dijet soft function, in which all divergences are factorised. After introducing standard
plus-distributions, the expansion in the various regulators can be performed and the coefficients of this 
expansion can be evaluated numerically. For the numerical integrations, we developed a new stand-alone program 
called \softserve, which uses the Divonne integrator of the Cuba library~\cite{Hahn:2004fe}. The code
contains a number of refinements to improve the convergence of the numerical integrations, which we will 
not discuss in detail in this work, but which are explained in the user manual of \softserve.
The \softserve~package is publicly available at \url{https://softserve.hepforge.org/}.

Although the main objective of our work is the computation of bare dijet soft functions, we go one step
further and extract the ingredients that are needed in practical applications of resummation within  SCET. 
To this end, we assume that the renormalised soft function obeys a multiplicative renormalisation group 
equation (RGE) in Laplace space, which allows us to define the (non-cusp) soft anomalous dimension $\gamma^S_i$ 
for SCET-1 soft functions and the collinear anomaly exponent $d_i$ for SCET-2 soft functions. 
In a previous study~\cite{Bell:2018vaa}, we derived integral representations for these quantities at the 
two-loop level for the same class of dijet soft functions we consider in the present work. Using \softserve, which 
provides a script for the automated extraction of the resummation ingredients, it is thus possible to cross check 
the results of~\cite{Bell:2018vaa}, and to in addition obtain the finite (non-logarithmic) term $c^S_i$ 
of the renormalised soft function for SCET-1 observables and the bare soft remainder function $W^S_i$
for SCET-2 observables (both in Laplace space). According to the standard counting of logarithms for 
Sudakov problems (see Table~\ref{tab:logcounting}), the two-loop expressions of $\gamma^S_i$ and
$d_i$ are needed at next-to-next-to-leading logarithmic (NNLL) accuracy, while the two-loop
constants $c^S_i$ and $W^S_i$ enter at the NNLL$'$~level. \softserve~thus allows for increased logarithmic
accuracy of SCET resummations, as was shown for the $e^+ e^-$ event-shape angularities 
in~\cite{Bell:2018gce}, where the improvement was from NLL$'$ to NNLL$'$ (using preliminary results 
for the angularity soft function that were published in~\cite{Bell:2015lsf}).

\begin{table}[t]
\begin{center}
$
\begin{array}{ | c | c | c | c |}
\hline
\text{~~Accuracy~~}\rule[-2mm]{0mm}{6mm} & 
~~~\Gamma^{\rm cusp}_i~~~ & 
~~~\gamma^S_i, d_i~~~ &  
~~c^S_i, W^S_i~~ \\  
\hline
\hline
\text{NLL}\rule[-1.5mm]{0mm}{5.5mm} & 
~\text{2-loop}~ & 
\text{1-loop} & 
\text{tree} \\ 
\hline
\text{NLL}'\rule[-1.5mm]{0mm}{5.5mm} & 
~\text{2-loop}~ & 
\text{1-loop} & 
\text{1-loop} \\ 
\hline
\text{NNLL}\rule[-1.5mm]{0mm}{5.5mm} & 
~\text{3-loop}~ & 
\text{2-loop} & 
\text{1-loop} \\ 
\hline
\text{NNLL}'\rule[-1.5mm]{0mm}{5.5mm} & 
~\text{3-loop}~ & 
\text{2-loop} & 
\text{2-loop} \\ 
\hline
\text{N$^3$LL}\rule[-1.5mm]{0mm}{5.5mm} & 
~\text{4-loop}~ & 
\text{3-loop} & 
\text{2-loop} \\
\hline
\end{array}
$
\end{center}
\vspace{-1.0em}
\caption{Resummation ingredients that are needed at different logarithmic orders. The precise definition of the 
anomalous dimensions $\Gamma^{\rm cusp}_i, \gamma^S_i, d_i$ and the matching corrections $c^S_i, W^S_i$ 
can be found in Section~\ref{sec:renormalize}.}
\label{tab:logcounting} 
\end{table}

The outline of this paper is as follows: in Section~\ref{sec:measure} we define more precisely which 
dijet soft functions are amenable to our algorithm and we define the general properties as well as 
the specific form we assume for their measurement functions. In Section~\ref{sec:baresoftfun} 
we outline the technical aspects of the bare soft function calculation, and in Section~\ref{sec:renormalize}
we specify the form we assume for the RGEs of both SCET-1 and SCET-2 soft functions. In Section~\ref{sec:generalize} we examine several extensions of our formalism which 
are relevant, e.g., for multi-differential observables and soft functions that are defined in Fourier space. In Section~\ref{sec:numerics} we briefly discuss the numerical implementation of our algorithm
in \softserve, and in Section~\ref{sec:results} we present sample results for $e^+ e^-$ and hadron-collider soft functions. All of our numerical results were generated using \softserve, and the explicit examples we consider in Section~\ref{sec:results} illustrate both the versatility and the usage of our code (template files for 
all soft functions considered in this work are provided in the \softserve~package).  Most of these
results are in fact already available in the literature at NNLO accuracy, and they hence provide strong cross-checks of 
our code, while also allowing us to study its numerical performance. Moreover, we obtain new predictions for the C-parameter, as well as thrust-axis and broadening-axis angularities. We finally conclude in Section~\ref{sec:conclude} and provide some technical details of our analysis 
in the Appendix.

%% file: Measure.tex
\section{Measurement function}
\label{sec:measure}

\subsection{General considerations}

We are concerned with soft functions that arise in processes with two massless, coloured, hard partons. 
A typical factorisation theorem of a dijet observable takes the form
\begin{align}
d\sigma = H \cdot J_n \otimes J_{\bar n} \otimes S\,,
\label{eq:fact}
\end{align}
where the symbol $\otimes$ denotes a convolution in some kinematic variables, and $H$ is a hard function 
that contains the virtual corrections to the Born process at the large scale $Q$ of the scattering process. 
The jet functions $J_n$ and $J_{\bar n}$ encode the effects from collinear emissions into the directions 
$n^\mu$ and $\bar n^\mu$ of the hard partons, and the soft function $S$ describes the low-energetic interactions 
between the two jets (for incoming partons the collinear functions are often called beam functions). As the 
soft, long-wavelength partons cannot resolve the inner structure of the jets, the soft function only sees the 
directions of the hard partons as well as their colour charges. This is reflected in the definition 
\eqref{eq:softfun:definition} of the soft function, where the Wilson lines depend on the direction and the 
colour representation of the associated hard partons. For concreteness, we adopt the notion for $e^+ e^-$ 
dijet observables in the following, and assume that the Wilson lines are given in the fundamental colour 
representation (the notation can be generalised to other processes by means of the colour-space 
formalism~\cite{Catani:1996vz}, and we will discuss some examples for hadron-initiated process below). 
We further assume that the hard partons are in a back-to-back configuration ($n\cdot \bar n=2$, 
along with $n^2=\bar n^2=0$), which is appropriate for both $e^+ e^-$ and hadron-collider kinematics.

The soft function in the factorisation theorem \eqref{eq:fact} has a double-logarithmic evolution in the 
renormalisation scale $\mu$ and, possibly, also the rapidity scale $\nu$. In order to make the associated 
divergences explicit, we find it convenient to consider an integral transformation, which turns the convolution 
in \eqref{eq:fact} into a product. Apart from avoiding distribution-valued quantities, this considerably 
simplifies the solution of the associated RGEs. For many observables this is achieved by a Laplace 
transformation. Denoting the corresponding Laplace variable by $\tau$, we write the generic measurement 
function in the definition \eqref{eq:softfun:definition} of the soft function in the form
\begin{align}
\mathcal{M}(\tau;\lbrace k_{i} \rbrace) = 
\exp\big(-\tau\, \omega(\lbrace k_{i} \rbrace)\,\big)\,,
\label{eq:measure:general}
\end{align}
where $\omega(\lbrace k_{i} \rbrace)$ is a function of the final-state momenta $k_{i}$ that is specific to the
observable. We thus assume that the distributions can be resolved by a single Laplace transformation, which 
implies that the soft function is differential in a single kinematic variable. We further assume that the 
Laplace variable has dimension 1/mass and that the measurement cannot distinguish between the two jets,
i.e.~the function $\omega(\lbrace k_{i} \rbrace)$ is supposed to be symmetric under $n\leftrightarrow\bar n$ 
exchange. In addition, we allow for a non-trivial azimuthal dependence of the observable around the jet 
(or beam) axis. In other words, there may exist an external reference vector $v^\mu$ that singles out a direction 
in the plane transverse to the jet (or beam) direction. We finally impose two technical restrictions on the 
function $\omega(\lbrace k_{i} \rbrace)$, namely its real part must be positive and it must be 
independent of the dimensional and rapidity regulators $\epsilon$ and $\alpha$.

Before we turn to some examples, let us recapitulate the assumptions that under\-lie our approach:
\begin{enumerate}[label=\bf(A\arabic*)]
\item
{\bf Dijet factorisation theorem:}
We assume that the soft function is embedded in a factorisation theorem of the form \eqref{eq:fact}, which
refers to a process with two massless, colour-charged, hard partons. The hard partons can be in the initial or 
final state, and they are supposed to be in a back-to-back configuration ($n^2=\bar n^2=0$, $n\cdot \bar n=2$).
The soft function for such dijet observables has a double-logarithmic evolution in the renormalisation 
scale $\mu$ and, possibly, also the rapidity scale $\nu$.
\item
{\bf Measurement function:}
We assume that the measurement function can be written in the form \eqref{eq:measure:general}. 
Typically, 
this is achieved by taking a Laplace (or Fourier) transform of a momentum-space soft function, with 
$\tau$ being the associated Laplace (or Fourier) variable. In order to ensure that the phase-space 
integrals converge, we require that $\Re\big(\omega(\lbrace k_{i} \rbrace)\big)>0$. More specifically, the function $\omega(\lbrace k_{i} \rbrace)$ is allowed to vanish only for configurations with zero weight in the phase-space integrations, and it is furthermore assumed to be independent of the regulators
$\epsilon$ and $\alpha$.
\item
{\bf Mass dimension:}
We assume that the variable $\tau$ has dimension 1/mass, and the function $\omega(\lbrace k_{i} \rbrace)$, 
which only depends on the final-state momenta $k_i$, must therefore have the dimension of mass. This 
requirement could easily be relaxed to any positive mass dimension in the future, although we have not 
encountered any example that requires such a generalisation so far.
\item
{\bf $n$-$\bar{n}$ symmetry:}
We assume that the measurement cannot distinguish between the two jets, and the function
$\omega(\lbrace k_{i} \rbrace)$ is therefore symmetric under the exchange of $n^\mu$ and $
\bar n^\mu$. This requirement could again easily be relaxed in the future, at the expense of doubling 
the number of input functions that need to be provided by the \softserve~user.
\item
{\bf Single-differential observables:}
We assume that the soft function only depends on one variable $\tau$ apart from the renormalisation and 
rapidity scales $\mu$ and $\nu$. Physically, this implies that the observable is differential in one kinematic 
variable. This requirement is in fact not strictly imposed in our approach (we will discuss multi-differential 
observables in Section~\ref{sec:generalize}), but we find it instructive to develop our formalism for this 
simplified class of observables first.
\item
{\bf Azimuthal dependence:}
Although we allow for a general azimuthal dependence of the observable around the jet/beam axis, we 
point out that the function $\omega(\lbrace k_{i} \rbrace)$ is allowed to depend only on one angle
$\theta_i$ per emitted particle in the $(d-2)$-dimensional transverse plane. This implies that the 
measurement is performed with respect to an external reference vector $v^\mu$, and the angle $\theta_i$ 
is then introduced as the angle between $\vec{v}_\perp$ and $\vec{k}_{i}^\perp$ in the plane transverse 
to the jet/beam direction.\footnote{For general $N$-jet soft functions with non-back-to-back kinematics, 
it was shown that two angles per emitted particle are required in the general case~\cite{Bell:2018mkk}.} 
In addition, the function $\omega(\lbrace k_{i} \rbrace)$ may depend on relative angles $\theta_{ij}$
between two emissions, which are defined as the angles between their respective transverse momenta 
$\vec{k}_{i}^\perp$ and $\vec{k}_{j}^\perp$.
\end{enumerate}

Conditions (A1) and (A2) can be viewed as the strongest assumptions of our approach, although the generalisation 
to $N\geq 2$ jet directions with non-back-to-back kinematics is already in progress~\cite{Bell:2018mkk} 
(which also requires an extension of (A6)). Whereas the generalisation of (A5) will be discussed in 
Section~\ref{sec:generalize}, we  already mentioned that assumptions (A3) and (A4) could easily be relaxed 
in the future. We further point out that our formalism is \emph{not} limited to observables that obey the 
NAE theorem. For observables that violate NAE, the uncorrelated-emission contribution becomes 
non-trivial~\cite{BRT,Bell:2018jvf}, but our method still allows for the calculation of the correlated-emission 
contribution, and hence it yields two out of three NNLO colour structures for NAE-violating observables.

Let us now see which type of observables fall into the considered class of soft functions. First, there are 
$e^+e^-$ event-shape variables that obey a hard-jet-soft factorisation theorem of the form \eqref{eq:fact} 
in the dijet limit. As an example we consider the C-parameter distribution, which was studied within SCET 
in~\cite{Hoang:2014wka,Hoang:2015hka}. In an appropriate normalisation, its Laplace-space soft function can 
be written in the form \eqref{eq:measure:general} with
\begin{align}
\omega^C(\lbrace k_{i} \rbrace) =
\sum_i\; \frac{k_i^+ k_i^-}{k_i^++k_i^-}\,,
\label{eq:omega:Cparameter}
\end{align}
where the plus- and minus-components represent the projections onto the $n^\mu$ and $\bar n^\mu$ directions
with $k_i^+ = n \cdot k_i$ and $k_i^- = \bar n \cdot k_i$. This function indeed has the dimension of mass, 
it is symmetric under $n\leftrightarrow\bar n$ exchange, and it does not depend on the regulators 
$\epsilon$ and $\alpha$. It is furthermore strictly positive, except for the trivial configuration with 
all $k_i^\mu=0$, which has zero weight in the phase-space integrations. The C-parameter is a 
single-differential observable with a trivial azimuthal dependence since the measurement is performed with 
respect to the jet axis itself.

As a second class of observables, we consider threshold resummation at hadron colliders. The classic
example is Drell-Yan production, which was factorised in the form \eqref{eq:fact} using methods from 
SCET in~\cite{Becher:2007ty} (the collinear functions are the standard parton distribution 
functions in this case). In position space, the corresponding soft function can be written in the form 
\eqref{eq:softfun:definition} with a weight factor $\exp(-i \,x\cdot P_X)$, where $P_X^\mu$ is the total 
momentum of the soft emissions. The vector $x^\mu$ thus plays the role of the reference vector $v^\mu$ 
in this case, and in the threshold kinematics it can be expanded as $x^\mu= (x^0,\vec{0})$ in the 
centre-of-mass frame of the collision. As its spatial components vanish, the observable again has a trivial 
azimuthal dependence around the beam axis. In terms of $\tau = i x^0/2$, the position-space soft function 
can then be expressed in the form \eqref{eq:measure:general} with
\begin{align}
\omega^{DY}(\lbrace k_{i} \rbrace) =
\sum_i\; (k_i^++k_i^-)\,,
\label{eq:omega:threshold}
\end{align}
and one easily verifies that assumptions (A1)-(A6) are again satisfied for this observable.

We finally consider another class of hadron-collider soft functions, which arise in the context of 
transverse-momentum resummation. Taking again the Drell-Yan process as an example, the corresponding SCET 
analysis, which now involves beam functions that describe the effects from energetic initial-state radiation, 
can be found in~\cite{Becher:2010tm}. The respective position-space soft function can then 
be written in a similar form as the one for threshold resummation, except that the reference vector 
$x_\perp^\mu$ is now purely transverse to the beam direction. It therefore induces a non-trivial azimuthal 
dependence, which is precisely of the form we anticipated in (A6). Writing $\tau = | x_\perp |/2$, one 
finds that this soft function can again be written in the form \eqref{eq:measure:general} with 
\begin{align}
\omega^{p_T}(\lbrace k_{i} \rbrace) =
-2 i \;\sum_i\; |k_i^\perp| \cos \theta_i \,.
\label{eq:omega:pT}
\end{align}
With the usual exponential damping factor of a Fourier transform in mind, we can then argue that its 
real part is positive as required by assumption (A2). The function itself, however, now vanishes for 
non-trivial kinematic configurations, which single out a complicated hypersurface in the phase-space integrations. 
These  configurations still have zero weight in the phase-space measure -- as required by (A2) -- but we will 
see later that our numerical results are less accurate for this observable in comparison with other examples that do not suffer from this problem. As the function $\omega^{p_T}(\lbrace k_{i} \rbrace)$ is purely imaginary, we will also explain later in Section~\ref{sec:generalize} that the numerical implementation of the transverse-momentum-dependent soft function in \softserve~requires special attention. One easily verifies that the remaining assumptions (A3)-(A5) 
are fulfilled for this observable. 

The above examples should not be understood as an exhaustive list of observables that can be treated in our 
formalism; they should rather help to illustrate what kind of restrictions are imposed by assumptions (A1)-(A6). 
Other observables relevant e.g.~for jet-veto resummation and jet-grooming observables also fall in the 
considered class of dijet soft functions. We will discuss further examples in Section~\ref{sec:results}.

\subsection{Specific parametrisations}
\label{sec:parametrisations}

After these general considerations, we now present the specific form we assume for the measurement function in 
our calculation. At NNLO the measurement is performed on either zero, one, or two emitted partons.

According to (A3), the function $\omega(\lbrace k_{i} \rbrace)$ is supposed to have the dimension of mass, 
and since it only depends on the final-state momenta $k_i$, it must vanish if there is no emission. We 
can therefore write the zero-emission measurement function in the form
\begin{align}
\mathcal{M}_0(\tau)=1
\label{eq:measure:zero}
\end{align}
for all observables we consider.

For one emission, we have to find a phase-space parametrisation that allows us to control the implicit 
soft and collinear divergences in the phase-space integrations. We choose the variables 
\begin{align}
y_k = \frac{k_+}{k_-}\,, \qquad\qquad 
k_T = \sqrt{k_+k_-}\,, \qquad\qquad
t_k = \frac{1-\cos\theta_k}{2}\,,
\label{eq:parametr:nlo}
\end{align}
where $y_k$ is a measure of the rapidity, $k_T$ is the magnitude of the transverse momentum, and $t_k$ 
parametrises the azimuthal dependence around the jet axis (with 
$\theta_k=\sphericalangle(\vec{v}_\perp,\vec{k}_\perp)$ as described in (A6)). The inverse of this 
transformation is then given by $k_{-} = k_{T}/\sqrt{y_k}$, $k_{+} = \sqrt{y_k} \,k_{T}$ and 
$\cos\theta_k = 1-2t_k$.

In terms of these variables, we will see in the following section that the soft divergence arises in the limit 
$k_T\to 0$. The variable $k_T$ is in fact the only dimensionful quantity in this parametrisation, and according 
to (A3) the function $\omega(\lbrace k\rbrace)=\omega(y_k, k_T, t_k)$ must therefore be linear in $k_T$.
The collinear divergences, on the other hand, emerge in the limits $y_k\to 0$ and $y_k\to \infty$. The 
$n$-$\bar{n}$ symmetry from (A4) then allows us to focus on one of the collinear limits, of which we choose the 
former. It turns out that the function $\omega(y_k, k_T, t_k)$ may vanish or diverge as $y_k\to 0$, and that we 
have to control its scaling in this limit to properly extract the collinear divergence. Taken together, this 
motivates the following ansatz for the one-emission measurement function:
\begin{align}
\mathcal{M}_1(\tau; k) = \exp\big(-\tau\, k_{T}\, y_k^{n/2}\, f(y_k,t_k)\,\big)\,,
\label{eq:measure:one}
\end{align}
where the power $n$ is fixed by the requirement that the function $f(y_k,t_k)$ is finite and non-zero in the 
limit $y_k\to0$. For one emission, the observable is thus characterised by a parameter $n$ and a function 
$f(y_k,t_k)$ that encodes the angular and rapidity dependence.\footnote{It follows from (A2) that $\Re\big(f(y_k,t_k)\big)>0$ and that $f(y_k,t_k)$ is assumed to be independent of the 
regulators $\epsilon$ and $\alpha$.}

Finding a suitable phase-space parametrisation for the double-emission contribution is much more involved. On 
the one hand the divergence structure of the matrix elements is more complicated, and on the other hand 
the measurement function must be controlled in various singular limits (whereas we only had to consider 
the limit $y_k\to 0$ in \eqref{eq:measure:one}). Moreover, we find that different parametrisations are 
needed for correlated and uncorrelated emissions. In this work we focus on the former, for which we introduce 
the variables 
\begin{subequations}
\label{eq:parametr:nnlo}
\begin{align}
p_T =  \sqrt{(k_{+} + l_{+})(k_{-} + l_{-})} \,, \quad\quad
y &= \frac{k_{+} + l_{+}}{k_{-} + l_{-}} \,, 
\quad\quad
a = \sqrt{\frac{k_{-}l_{+}}{k_{+}l_{-}}} \,,
\quad\quad
b =\sqrt{\frac{k_{+}k_{-}}{l_{+}l_{-}}} \,,
\end{align}
along with the angular variables
\begin{align}
t_k = \frac{1-\cos\theta_k}{2}\,, \qquad\qquad 
t_l = \frac{1-\cos\theta_l}{2}\,, \qquad\qquad
t_{kl} = \frac{1-\cos\theta_{kl}}{2}\,.
\label{eq:parametr:nnlo:angles}
\end{align}
\end{subequations}
The variables $p_T$ and $y$ are thus functions of the sum of the light-cone momenta, the quantity $a$ is 
a measure of the rapidity difference of the emitted partons, and $b$ is the ratio of their transverse 
momenta.\footnote{The variable $p_T$ should not be confused with the total transverse momentum of the 
emitted partons.} In general the measurement function now depends on three angles since the emitted partons 
may not only see the reference vector $v^\mu$, but they will also see each other. The angles in 
\eqref{eq:parametr:nnlo:angles} are then introduced as
$\theta_k=\sphericalangle(\vec{v}_\perp,\vec{k}_\perp)$,
$\theta_l=\sphericalangle(\vec{v}_\perp,\vec{l}_\perp)$,
and $\theta_{kl}=\sphericalangle(\vec{k}_\perp,\vec{l}_\perp)$, and the inverse transformation is now given by
$k_- = a\, b\, p_{T}/(1+a b)/\sqrt{y}$,
$k_+ = b\, \sqrt{y}\, p_{T}/(a+b)$,
$l_- = p_{T}/(1+a b)/\sqrt{y}$,
$l_+ = a\,\sqrt{y}\,p_{T}/(a+b)$,
and
$\cos\theta_i = 1-2t_i$
for $i\in\{k,l,kl\}$.

After using the symmetries under $n\leftrightarrow\bar n$ and $k\leftrightarrow l$ exchange, we will see 
in the following section that the implicit phase-space divergences now arise in four limits:
\begin{itemize}
\item
$p_T\rightarrow 0$, which corresponds to the situation in which both partons become soft;
\item
$y \rightarrow 0$, which reflects the fact that one of the partons becomes collinear to the jet direction $n^\mu$;
\item
$b \rightarrow 0$, which implies that the parton with momentum $k^\mu$ becomes soft;
\item
$a\to 1$ and $t_{kl}\to 0$, which means that the emitted partons become collinear to each other.
\end{itemize}
The first limit can in fact be treated in analogy to the limit $k_T\to 0$ in the one-emission case;
since $p_T$ is the only dimensionful variable in the parametrisation \eqref{eq:parametr:nnlo}, we know 
that the function $\omega(\lbrace k,l\rbrace)=\omega(p_T,a,b,y, t_k,t_l,t_{kl})$ must be linear in $p_T$. 
Yet, we still have to control the measurement function in the remaining three limits to make sure that 
we can properly extract the associated divergences.

What helps us in this situation is the underlying assumption in the factorisation theorem \eqref{eq:fact} 
that the observable is infrared safe. The one- and two-emission measurement functions are therefore not 
independent from each other, and we will derive explicit relations between them in the limit where one of 
the partons becomes soft ($b\to 0$) or both partons merge into a single parton ($a\to 1$ and $t_{kl}\to 0$) 
below. The last two limits from the above list are thus, as we say, \emph{protected by infrared safety}, 
which means that we are guaranteed that the measurement function does not vanish in these limits since 
it must fall back to the one-emission case (which does not vanish for a generic 
configuration of the remaining parton). We therefore only have to consider the limit $y \rightarrow 0$ 
explicitly, which can be treated similarly to the limit $y_k\to 0$ in the one-emission case. Our ansatz 
for the correlated double-emission measurement function therefore reads
\begin{align}
\label{eq:measure:NNLO:corr}
\mathcal{M}_2^{corr}(\tau; k,l) = \exp\big(-\tau\, p_{T}\, y^{n/2}\, 
F(a,b,y,t_k,t_l,t_{kl})\,\big)\,,
\end{align}
where the function $F(a,b,y,t_k,t_l,t_{kl})$ is supposed to be finite and non-zero in the limit $y\to0$. 
Interestingly, this is achieved by factorising the same power of the variable $y$ as in the 
one-emission case -- see \eqref{eq:measure:one}. We  explain in Appendix~\ref{twoloopn} why this is so, 
and we address the physical meaning of the parameter $n$ in the next section.\footnote{We
again demand that $\Re\big(F(a,b,y,t_k,t_l,t_{kl})\big)>0$ and that $F(a,b,y,t_k,t_l,t_{kl})$ is independent 
of any regulators as required by assumption (A2).}

In order to extract the divergences of the bare soft function, we find it convenient 
to map the phase-space integrations onto a unit hypercube in the variables \eqref{eq:parametr:nlo} and 
\eqref{eq:parametr:nnlo}. While this can easily be achieved by exploiting the $n$-$\bar n$ symmetry 
for one emission, we will see in the following section that this leads to two independent regions in the two-emission 
case, which we label by the letters ``A'' and ``B''. Our formulae therefore depend on two different versions of the two-emission 
measurement function, which are defined as
\begin{align}
F_A(a,b,y,t_k,t_l,t_{kl}) &= F(a,b,y,t_k,t_l,t_{kl})\,,
\nonumber\\
F_B(a,b,y,t_k,t_l,t_{kl}) &=\begin{dcases} F(1/a,b,y,t_k,t_l,t_{kl}) &\,\text{or} \\ F(a,1/b,y,t_k,t_l,t_{kl}) &\,\text{or}\\ y^{-n}F(a,b,1/y,t_k,t_l,t_{kl}) \,.\end{dcases}
\end{align}
Further explanations about the origin of these regions and the different
representations of the measurement function in region B can be found in Section~\ref{sec:doublereal}.

\begin{table}
\center
\setlength{\extrarowheight}{10pt}
\scalebox{1}{\begin{tabular}{|c|c|c|c|}
\hline \hline
Observable & $n$ & $f(y_k,t_k)$ & $F(a,b,y,t_k,t_l,t_{kl})$\\[10pt]
\hline \hline
C-parameter & $1$ & $\displaystyle\frac{1}{1+y_k}$ & $\displaystyle\frac{ab}{a(a+b)+(1+ab)y}+\frac{a}{a+b+a(1+ab)y}$\\[10pt]
\hline
Threshold resum. & $-1$ & $1+y_k$  & $1+y$\\[10pt]
\hline
$p_{T}$ resum. & 0 & $- 2i (1-2 t_k)$ & $\displaystyle-2 i \;\sqrt{\frac{a}{(1+a b)(a+b)}}\; \Big(b (1 - 2 t_k) + 1 - 2 t_l\Big)$\\[10pt]
\hline
\end{tabular}}
\caption{One- and two-emission measurement functions for the $e^+ e^-$ event shape C-parameter, threshold and 
transverse-momentum resummation in Drell-Yan production.
\label{tab:measure}}
\end{table} 

Before we come back to the explicit examples that we discussed towards the end of the last section, we derive 
the constraints from infrared safety that we mentioned earlier. To this end, we write the variables 
$p_T$ and $y$ from \eqref{eq:parametr:nnlo} in terms of $y_k$ and $k_T$ from \eqref{eq:parametr:nlo} 
and the analogous variables $y_l$ and $l_T$ for the second emitted parton,
\begin{align}
p_T =  \sqrt{k_T^2 + \left(\sqrt{\frac{y_k}{y_l}} + \sqrt{\frac{y_l}{y_k}}\right) k_T l_T + l_T^2}
\,, \quad\quad
y = \sqrt{y_k y_l} \;\,\frac{\sqrt{y_k} \,k_T+\sqrt{y_l} \,l_T}{\sqrt{y_l} \,k_T+\sqrt{y_k} \,l_T} \,.
\end{align}
In the limit in which the parton with momentum $k^\mu$ becomes soft, i.e.~$k_T\to 0$, we thus see that 
$p_T\to l_T$ and $y\to y_l$. Infrared safety then tells us that the value of the observable should 
not change under infinitesimally soft emissions, 
\begin{align}
\label{eq:irsafety:soft}
\mathcal{M}_2^{corr}(\tau; k,l) \xlongrightarrow{k^\mu\to0\;} \mathcal{M}_1(\tau; l) \,,
\end{align}
which leads to the relation
\begin{align}
F(a,0,y_l,t_k,t_l,t_{kl}) = f(y_l,t_l)\,.
\label{eq:infrared:soft}
\end{align}
We can derive a similar relation between the one- and two-emission measurement functions in the limit
in which the two partons with momenta $k^\mu$ and $l^\mu$ become collinear to each other. In this situation, 
which implies $y_k\to y_l$ and $t_k\to t_l$, we see that $p_T\to k_T + l_T$ and $y\to y_l$. As the value of the observable 
should again not change under collinear emissions, 
\begin{align}
\mathcal{M}_2^{corr}(\tau; k,l) \xlongrightarrow{k^\mu~\!\!\parallel~\!l^\mu\;} \mathcal{M}_1(\tau; k+l) \,,
\end{align}
we obtain
\begin{align}
F(1,b,y_l,t_l,t_l,0) = f(y_l,t_l)\,.
\label{eq:infrared:collinear}
\end{align}
Relations \eqref{eq:infrared:soft} and \eqref{eq:infrared:collinear} follow from the fundamental assumption that the 
observable that we factorised in \eqref{eq:fact} must be infrared safe. These relations are thus expected to 
hold for all observables we consider, and -- as argued before -- they guarantee that the measurement function 
does not vanish in two of the critical limits that we discussed above.

Starting from the observable definitions in \eqref{eq:omega:Cparameter} -- \eqref{eq:omega:pT}, we can now 
easily derive the measurement functions for the C-parameter, threshold and transverse-momentum resummation in 
the phase-space parametrisations that we use in our calculation. The result is shown in Table~\ref{tab:measure}, 
which illustrates that some observables have a non-trivial rapidity dependence, while others are sensitive to 
the reference vector $\vec{v}_\perp$ and therefore depend on the angular variables $t_k$ and $t_l$. From these 
expressions, we can verify that the functions $f(y_k,t_k)$ and $F(a,b,y,t_k,t_l,t_{kl})$ are finite in the 
limits $y_k\to 0$ and $y\to 0$, respectively, as this was the basis for extracting the corresponding values of 
the parameter $n$. We indeed see that these values can differ among the observables, and we will learn later that 
the case $n=0$ always corresponds to a SCET-2 observable. While we have so far introduced this parameter on 
purely technical grounds, we will see in the following section that it is related to the power counting of 
the momentum modes in the effective theory. Moreover, we can also easily verify that the constraints from 
infrared safety, \eqref{eq:infrared:soft} and \eqref{eq:infrared:collinear}, are satisfied for the considered
class of observables.

\subsection{Interpretation of the parameter $n$}
\label{sec:n}

We saw in the previous section that the parameter $n$ is related to the scaling of the observable in the 
soft-collinear limit, and we will indeed see later that it controls the double logarithmic contributions 
to the renormalised soft function. We also mentioned that the parameter $n$ allows us to distinguish between 
SCET-1 and SCET-2 observables, and we would like to under\-stand why the values for the C-parameter ($n=1$) 
and threshold resummation ($n=-1$) are different, given that both observables are defined within SCET-1.

To this end, we go back to the factorisation theorem \eqref{eq:fact}, which emerges in an effective field theory 
with hard, collinear, anti-collinear, and soft momentum modes. Denoting the small expansion parameter in the 
theory by $\lambda$, we associate the following power counting to the momenta 
$p^\mu = (\bar n\cdot p, n\cdot p,p_\perp^\mu)$ with
\begin{itemize}
\item
$p_h^\mu = Q(1,1,1)$
\hspace{6.6mm} (hard)
\item
$p_c^\mu = Q(1,\lambda^{2p},\lambda^{p})$ 
\; (collinear)
\item
$p_{\bar c}^\mu = Q(\lambda^{2p},1,\lambda^{p})$  
\; (anti-collinear)
\item
$p_s^\mu = Q(\lambda,\lambda,\lambda)$ 
\hspace{5.7mm} (soft)
\end{itemize}
where $Q$ is the large scale in the process, and where we have allowed for a generic scaling of the 
collinear momenta that is controlled by a parameter $p>0$. 

The factorisation theorem \eqref{eq:fact} then tells us that collinear, anti-collinear, and soft modes contribute 
to the observable at the same power, i.e.~the observable $\omega(\lbrace k_{i} \rbrace)$ must have the same 
scaling in $\lambda$ in the three regions.\footnote{One is often left with additive observables of the 
form $\omega(\lbrace k_{i} \rbrace) = \omega_c\lbrace k_{i}^c \rbrace) + \omega_{\bar c}(\lbrace k_{i}^{\bar c} \rbrace) + \omega_s(\lbrace k_{i}^s \rbrace)$,
which leads to a multiplicative factorisation theorem in Laplace space.} We know, however, that the observable 
scales as $\omega(\lbrace k_{i} \rbrace)\sim\lambda$ in the soft region, since it has 
mass dimension one -- see (A3) -- and the power counting in the soft region is directly tied to the mass 
dimension. This can easily be verified for the examples in \eqref{eq:omega:Cparameter} -- \eqref{eq:omega:pT}.

Our goal consists in establishing a relation between the parameter $n$ and the power-counting variable $p$, and 
for this purpose it is sufficient to focus on a single emission. In this case, the parameter $n$ controls the 
scaling of the observable in the collinear limit $y_k\to 0$ in the soft region. In the collinear region, on 
the other hand, we can exploit the hierarchy between the light-cone components, $k_-\gg k_+$, to express the 
observable in the form 
\begin{equation}
\label{eq:omega:scaling:collinear}
\omega(\lbrace k \rbrace) = k_+^s \,k_-^{(1-s)}\, f_c(z_k,t_k)\,,
\end{equation}
where the function $f_c(z_k,t_k)$ encodes an arbitrary dependence on the splitting variable $z_k=k_-/Q$ 
and the angular variable $t_k = (1-\cos\theta_k)/2$, and we have furthermore 
used that the observable has mass dimension one. The parameter $s$ then varies among the 
observables, and we find $s=1$ for the C-parameter, $s=0$ for threshold and $s=1/2$ for transverse-momentum 
resummation.

We argued before that the observable must scale as $\lambda$ in the collinear region as well, and since 
$k_-\sim 1$ and $k_+\sim \lambda^{2p}$ for collinear momenta, we obtain $s=1/(2p)$. We can extract further 
information if we express \eqref{eq:omega:scaling:collinear} in terms of the variables from 
\eqref{eq:parametr:nlo} and if we consider the soft limit $z_k\to 0$,
\begin{equation}
\omega(\lbrace k \rbrace) = \big(\sqrt{y_k} \,k_{T}\big)^s \,\Big(\frac{k_{T}}{\sqrt{y_k}}\Big)^{(1-s)}
\, f_c(0,t_k) = k_T\, y_k^{s-1/2}\, f_c(0,t_k)\,,
\end{equation}
which must match the expression in \eqref{eq:measure:one} in the collinear limit $y_k\to 0$. In particular,
we see that the parameter $s$ controls the scaling of the observable with the rapidity-like 
variable $y_k$, which brings us to the desired relation,
\begin{equation}
n=2s-1=\frac{1}{p}-1\,.
\end{equation}
We thus see that the parameter $n$ is directly related to the power counting of the modes in the effective 
theory via $p_c^\mu = Q(1,\lambda^{2p},\lambda^{p})$, when the soft scaling is fixed to 
$p_s^\mu = Q(\lambda,\lambda,\lambda)$. For the C-parameter, for instance, the collinear modes scale as 
$p_c^\mu = Q(1,\lambda,\sqrt{\lambda})$~\cite{Hoang:2014wka}, which implies that $p=1/2$ and hence $n=1$, 
which is in line with what we have found in Table~\ref{tab:measure}. For transverse-momentum resummation, 
on the other hand, the relevant soft and collinear modes have the same virtuality, and so $p=1$ which 
translates into $n=0$ for a SCET-2 observable. However, the third example from our list appears to be 
peculiar, since $n=-1$ requires that $p\to\infty$, and the relevant collinear modes should therefore scale 
as $p_c^\mu = Q(1,\lambda^{2p},\lambda^{p})\to Q(1,0,0)$. We recall, though, that the collinear functions 
for threshold resummation are the standard parton distribution functions, and the power counting of the 
collinear modes is therefore not related to the threshold parameter $\lambda=1-M^2/\hat s$, but rather to the 
non-perturbative scale $\Lambda_{QCD}$ in this case. In other words, the relevant collinear modes 
scale as $p_c^\mu = Q(1,\varepsilon^2,\varepsilon)$ with $\varepsilon=\Lambda_{QCD}/Q$ for this observable, and since 
$\varepsilon\ll\lambda$ this is indeed consistent with $p_c^\mu \to Q(1,0,0)$.

%% file: BareCalculation.tex
\section{Calculation of the bare soft function}
\label{sec:baresoftfun}

The definition \eqref{eq:softfun:definition} of what we call a generic dijet soft function depends on a 
measurement function $\mathcal{M}(\tau;\lbrace k_{i} \rbrace)$, whose explicit form we discussed 
extensively in the previous section, as well as a matrix element of soft Wilson lines. The Wilson line
associated e.g.~with an incoming quark that travels in the $n^\mu$ direction is defined as
\begin{equation}
S_{n}(x) = \mathbf{P}\, \text{exp} \left( i g_s \int_{-\infty}^0 ds \;n \cdot A_{s} (x + sn) \right),
\end{equation}
where $\mathbf{P}$ is the path-ordering symbol, $A_{s}^\mu(x)= A_{s}^{\mu,A}(x)\, T^A$ is the soft gluon 
field and $T^A$ are the generators of SU(3) in the fundamental representation. The Wilson line associated 
with an outgoing quark has a similar representation, except that the integration now runs from 
$0$ to $+\infty$. This subtle difference leads to the opposite sign in the $i\varepsilon$-prescription of 
the associated eikonal propagators, which -- to the considered order in the perturbative expansion -- is only 
relevant for the NNLO real-virtual interference. However, as we will see later, it turns out that the 
corresponding squared matrix element does \emph{not} depend on the sign of this prescription, and our formulae 
therefore equally apply to soft functions with incoming and outgoing light-like directions. The Wilson lines 
associated with anti-quarks, moreover, are anti-path-ordered and their definition can be found 
e.g.~in~\cite{Chay:2004zn}.

At leading order in the perturbative expansion, the soft matrix element in the definition 
\eqref{eq:softfun:definition} is trivial. Together with the form \eqref{eq:measure:zero} of the zero-emission 
measurement function, this implies that the soft function is normalised to one at leading order. 
At higher orders, the bare soft function is subject to various divergences, which we control by a dimensional 
regulator $\eps=(4-d)/2$ and a rapidity regulator $\alpha$ that is needed only for SCET-2 observables. 
The latter is introduced on the level of the phase-space integrals as
\begin{equation}
\int d^dp \; \left(\frac{\nu}{n\cdot p + \bar n \cdot p}\right)^\alpha \;  \delta(p^2) \theta(p^0) \,,
\label{eq:analyticregulator}
\end{equation}
which is in the spirit of~\cite{Becher:2011dz}, except that our version respects the $n$-$\bar{n}$ symmetry 
that we assume on the level of the observable, see (A4). In this regularisation, the purely virtual 
corrections are scaleless and vanish at every order in perturbation theory. We are thus left with a single 
real-emission contribution at NLO, and with mixed real-virtual and double real-emission corrections at NNLO. 
The bare soft function can hence be written in the generic form
\begin{align}
S_0(\tau,\nu) = 1 &+ \left(\frac{Z_\alpha\alpha_s}{4\pi}\right) \,(\mu^2 \taubar^2)^\eps \;(\nu \taubar)^\alpha \, S_R(\eps,\alpha) 
\no\\
& + \left(\frac{Z_\alpha\alpha_s}{4\pi}\right)^2 (\mu^2 \taubar^2)^{2\eps} \; \bigg\{(\nu \taubar)^\alpha
\,S_{RV}(\eps,\alpha) + (\nu \taubar)^{2\alpha}
\,S_{RR}(\eps,\alpha)\bigg\} + \mathcal{O}(\alpha_s^3)\,,
\label{eq:baresoftfun:expansion}
\end{align}
where $\alpha_{s}$ is the renormalised strong coupling constant in the $\overline{\text{MS}}$ scheme, which is 
related to the bare coupling  $\alpha_s^0$ via 
$Z_\alpha \alpha_s\,\mu^{2\eps}=e^{-\eps\gamma_E}(4\pi)^\eps \alpha_s^0$ with 
$Z_\alpha = 1-\beta_0\alpha_s/(4\pi\eps)$ and $\beta_0 = 11/3\, C_A - 4/3\,T_F n_f$. We furthermore introduced 
the rescaled variable $\taubar = \tau e^{\gamma_E}$ for convenience. 

In the following we in turn address the computation of the single real-emission correction $S_R(\eps,\alpha)$, 
the mixed real-virtual interference $S_{RV}(\eps,\alpha)$, and the double real-emission contribution 
$S_{RR}(\eps,\alpha)$ for a generic dijet soft function.

\subsection{Single real emission}
\label{sec:NLO}

\begin{figure}[t]
\begin{center}
\begin{tabular}{ccccccc}
\includegraphics[width=0.19\textwidth]{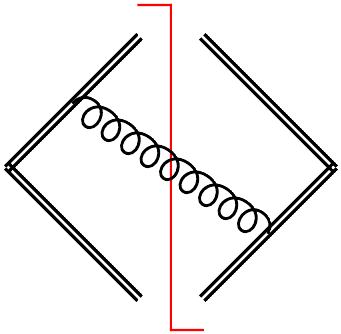} & & 
\includegraphics[width=0.19\textwidth]{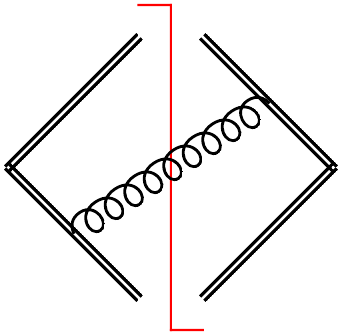} & &
\includegraphics[width=0.19\textwidth]{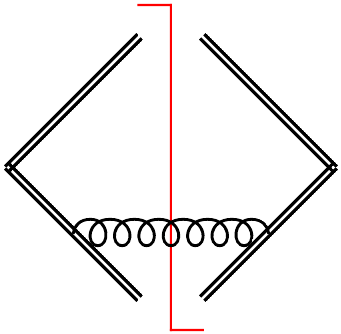} & & 
\includegraphics[width=0.19\textwidth]{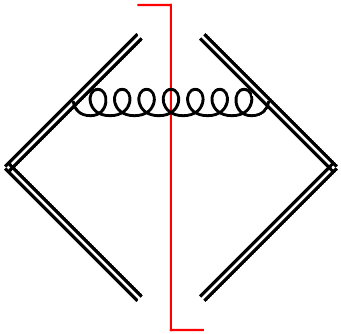}
\end{tabular}
\end{center}
\vspace{-0.5cm}
\caption{\label{fig:NLOgraphs}
Diagrams that contribute to the NLO calculation of dijet soft functions.}
\end{figure}

In the normalisation \eqref{eq:baresoftfun:expansion},
the single real-emission correction takes the form
\begin{equation}
S_R(\eps,\alpha)  = \frac{(4\pi e^{\gamma_E} \tau^2)^{-\eps}\;\taubar^{-\alpha} }{(2\pi)^{d-1}} \, 
\int d^{d}k \;\, \delta(k^{2}) \,\theta(k^{0}) 
\;\,\frac{|\mathcal{A}_R(k)|^{2}}{(n\cdot k + \bar n \cdot k)^\alpha} \;\, \mathcal{M}_1(\tau; k) \,,
\end{equation}
where $|\mathcal{A}_R(k)|^{2}$ denotes the corresponding soft matrix element and $\mathcal{M}_1(\tau; k)$ is 
the one-emission measurement function. At NLO the matrix element receives contributions from the four cut 
diagrams in Figure~\ref{fig:NLOgraphs}, where the double lines represent eikonal propagators associated 
with the $n^\mu$ and $\bar n^\mu$ Wilson lines. We find that the first two diagrams yield equal contributions, 
while the latter two vanish because they are proportional to $n^2 = 0$ or $\bar{n}^2 = 0$. The NLO 
squared matrix element is then given by
\begin{equation}
\label{eq:nlo:matrixelement}
| \mathcal{A}_R(k) |^{2} = \frac{64 \pi^2 C_{F}}{n\cdot k\;\, \bar n \cdot k}\,,
\end{equation}
where we suppressed the $i\varepsilon$-prescription of the eikonal propagators since it is irrelevant at 
this order. 

We next decompose the gluon momentum in terms of light-cone coordinates
\begin{align}
k^{\mu} &= 
k_-\;\frac{n^{\mu}}{2}  +  
k_+\; \frac{\bar{n}^{\mu}}{2} + 
k^{\mu}_{\perp}\,, 
\end{align}
with $k_- = \bar n \cdot k$, $k_+ = n \cdot k$, and $n \cdot k_\perp = \bar{n} \cdot k_\perp =0$, along 
with $k^{2}_{\perp} = - \vec{k}_{\perp}^{2}$. As the one-emission measurement function only depends on one 
angle in the $(d-2)$-dimensional transverse space, see (A6), the phase-space measure can be simplified as
\begin{align}
\label{eq:nlo:phase-space-measure}
&\int d^{d}k \;\, \delta(k^{2}) \,\theta(k^{0}) 
\\
&\quad
=\frac{\pi^{1/2-\eps}}{\Gamma(1/2-\eps)}\;
\int_0^\infty \!d|\vec{k}_\perp| \,dk_+\, dk_-\; 
|\vec{k}_\perp|^{1-2\eps}\;\,
\delta(k_+ k_- - |\vec{k}_\perp|^2)\;
\int_{-1}^1 \!d\cos\theta_k \;\sin^{-1-2\eps}\theta_k\,,
\no
\end{align}
where $\theta_k=\sphericalangle(\vec{v}_\perp,\vec{k}_\perp)$ is measured with respect to the external 
reference vector $v^\mu$. We switch to the parametrisation \eqref{eq:parametr:nlo} and use the explicit 
form \eqref{eq:measure:one} of the one-emission measurement function to obtain
\begin{align}
S_R(\eps,\alpha)  &= 
\frac{8C_F\,e^{-\gamma_E(\eps+\alpha)}}{\sqrt{\pi}}\,
\frac{\tau^{-2\eps-\alpha}}{\Gamma(1/2-\eps)}\;\int_0^\infty \! dk_T \; k_T^{-1-2\eps-\alpha}\;
\int_0^\infty \! dy_k \; \frac{y_k^{-1+\alpha/2}}{(1+y_k)^\alpha}
\no\\[0.2em]
&\quad\times
\int_0^1 \!dt_k \;(4t_k\bar{t}_k)^{-1/2-\eps} \;\;
\exp\big(-\tau\, k_{T}\, y_k^{n/2}\, f(y_k,t_k)\,\big)
\label{eq:nlo:interm}
\end{align}
with $\bar{t}_{k} = 1 - t_{k}$. As the $k_T$-dependence is universal among the considered class of observables, 
this integration can also be performed explicitly. We further use the $n$-$\bar{n}$ symmetry of the observable, 
which implies $y_{k} \to 1/ y_{k}$ in the given parametrisation, to map the $y_{k}$-integration over the 
interval $[1, \infty]$ to an integral over $[0,1]$. We then arrive at the following master formula for the 
computation of the single real-emission correction:
\begin{align}
S_R(\eps,\alpha)  &= 
\frac{16C_F\,e^{-\gamma_E(\eps+\alpha)}}{\sqrt{\pi}}\,
\frac{\Gamma(-2\eps-\alpha)}{\Gamma(1/2-\eps)}\;
\int_0^1 \! dy_k \; \frac{y_k^{-1+n\eps+(n+1)\alpha/2}}{(1+y_k)^\alpha}
\no\\[0.2em]
&\quad\times
\int_0^1 \!dt_k \;(4t_k\bar{t}_k)^{-1/2-\eps} \;\;
f(y_k,t_k)^{2\eps+\alpha}\,,
\label{eq:nlo:master}
\end{align}
which is valid for arbitrary dijet soft functions that fall into the considered class of observables and which 
are characterised by the parameter $n$ and the function $f(y_k,t_k)$. 

Upon expanding in the regulators $\alpha$ and $\eps$, the result exposes divergences, whose origin can be more 
clearly identified in \eqref{eq:nlo:interm}. First, there is a soft singularity that arises in the limit 
$k_T\to 0$ and gives rise to the factor $\Gamma(-2\eps-\alpha)$. Second, the $y_k$-integral in 
\eqref{eq:nlo:interm} diverges in the collinear limits $y_k\to 0$ and $y_k\to \infty$, i.e.~when the gluon is 
emitted into the $n^\mu$ or $\bar n^\mu$ directions. Due to the $n$-$\bar{n}$ symmetry, we can focus on one 
of these limits, and from \eqref{eq:nlo:master} we finally read off that the collinear divergence is not 
regularised in dimensional regularisation for $n=0$, which is precisely the SCET-2 case we identified in 
Section~\ref{sec:n}. 

For $n\neq0$, on the other hand, the rapidity regulator $\alpha$ can be set to zero, and the expansion of 
the SCET-1 soft function starts with a $1/\eps^2$ pole, whose coefficient is controlled by the parameter $n$. 
This can be seen explicitly if we rewrite the divergent rapidity factor in terms of distributions according to
\begin{equation}
\label{eq:yexp}
y_{k}^{-1 + n \epsilon} = \frac{\delta(y_{k})}{n\epsilon} + \left[\frac{1}{y_{k}} \right]_{+} + n \epsilon \left[\frac{\ln y_{k}}{y_{k}} \right]_{+} + \,\dots
\end{equation}
As the function $f(y_k,t_k)$ is by construction finite and non-zero in the limit $y_k\to 0$, the remaining 
integrations in the expansion of \eqref{eq:nlo:master} are well-defined and suited for a numerical integration. 
We in fact already presented the SCET-1 NLO master formula in~\cite{Bell:2015lsf}, and an earlier derivation 
along similar lines -- although less general -- was given in~\cite{Hoang:2014wka}. A similar NLO formula, 
valid also for non-back-to-back configurations, was presented in~\cite{Kasemets:2015uus}.

For SCET-2 soft functions with $n=0$, it is evident from \eqref{eq:nlo:master} that the $y_k$-integration 
produces a $1/\alpha$ pole. It is in this case important that the $\alpha$-expansion is performed before the 
$\eps$-expansion, since the $\alpha$-regulator is supposed to regularise rapidity divergences only. The 
expansion of the factor $\Gamma(-2\eps-\alpha)$ therefore generates $1/\eps$ and $\alpha/\eps^2$ terms, 
which yield $1/(\alpha\eps)$ and $1/\eps^2$ poles on the level of the bare \mbox{SCET-2} soft function, 
whose coefficients are related since they descend from the same Gamma function.

\subsection{Real-virtual interference}

\begin{figure}[t]
\begin{center}
\begin{tabular}{ccccccc}
\includegraphics[width=0.19\textwidth]{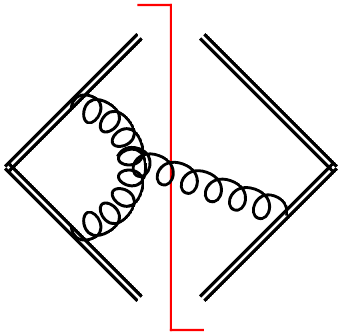} & & 
\includegraphics[width=0.19\textwidth]{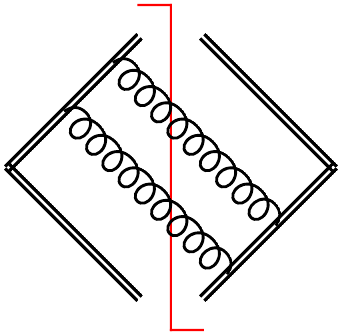} & &
\includegraphics[width=0.19\textwidth]{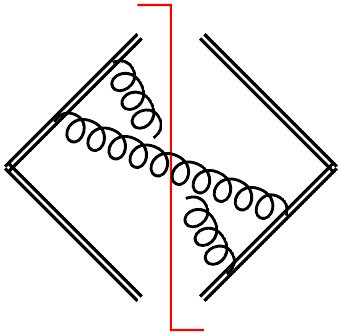} & & 
\includegraphics[width=0.19\textwidth]{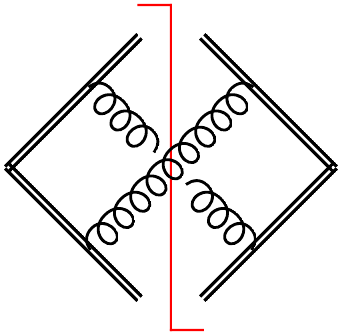}
\\[1.2em]
\includegraphics[width=0.19\textwidth]{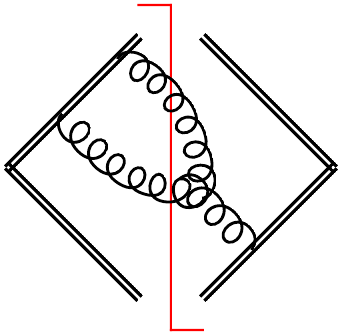} & & 
\includegraphics[width=0.19\textwidth]{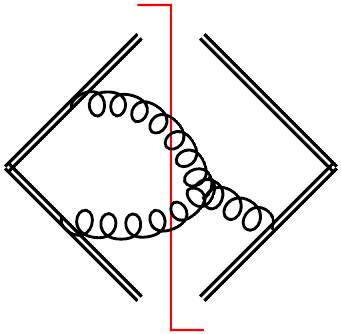} & &
\includegraphics[width=0.19\textwidth]{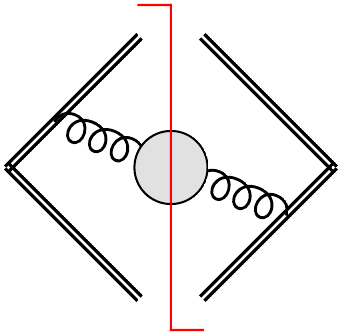} & & 
\includegraphics[width=0.19\textwidth]{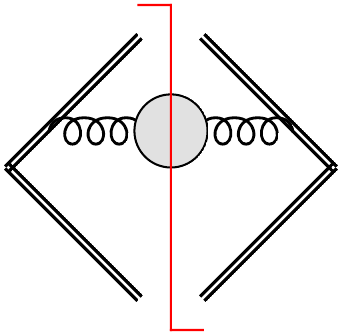}
\end{tabular}
\end{center}
\vspace{-0.5cm}
\caption{\label{fig:NNLOgraphs}
Diagrams that contribute to the NNLO calculation of dijet soft functions. The first diagram represents the 
mixed real-virtual contribution, and the other diagrams contribute to the double real-emission correction. The shaded 
circle in the last two diagrams represents the one-loop gluon self-energy corrections.}
\end{figure}

The mixed real-virtual contribution is structurally identical to the single real-emission term. We now start 
from
\begin{equation}
\label{eq:nnlo:real-virtual:start}
S_{RV}(\eps,\alpha)  = \frac{(4\pi e^{\gamma_E} \tau^2)^{-\eps}\;\taubar^{-\alpha} }{(2\pi)^{d-1}} \,
\int d^{d}k \;\, \delta(k^{2}) \,\theta(k^{0}) 
\;\,\frac{|\mathcal{A}_{RV}(k)|^{2}}{(n\cdot k + \bar n \cdot k)^\alpha} \;\, \mathcal{M}_1(\tau; k) \,,
\end{equation}
where the only difference is the soft matrix element $|\mathcal{A}_{RV}(k)|^{2}$, which can be calculated 
from the first diagram in Figure~\ref{fig:NNLOgraphs}. Interestingly, the one-loop correction now depends 
on the $i\varepsilon$-prescription of the eikonal propagators on the amplitude level, but this dependence 
drops out in the interference with the Born diagram (see also~\cite{Catani:2000pi,Kang:2015moa}). One finds
\begin{equation}
\label{eq:nnlo:rv:matrixelement}
| \mathcal{A}_{RV}(k) |^{2} = - \,\frac{64 \pi^4 \,C_A C_F\,e^{-\gamma_E \eps}\,\tau^{-2\eps}}{(n\cdot k)^{1+\eps}\;(\bar n\cdot k)^{1+\eps}} \;\,
\frac{\Gamma(-\eps)\,\cot(\pi\eps)}{\Gamma(-2\eps)\,\sin(\pi\eps)}\,,
\end{equation}
which again resembles the NLO matrix element \eqref{eq:nlo:matrixelement}, except that its expansion now 
starts with a $1/\eps^2$ pole, which is to be multiplied with the $1/(\alpha\eps)$ and $1/\eps^2$ poles 
of the subsequent phase-space integrations. The very fact that the matrix 
element \eqref{eq:nnlo:rv:matrixelement} does not depend on the rapidity regulator $\alpha$ -- which is 
implemented only on the level of the phase-space integrals in \eqref{eq:nnlo:real-virtual:start} -- is a 
key advantage of the regularisation prescription from~\cite{Becher:2011dz}.

The subsequent calculation then follows along the same lines outlined in the previous section, and 
the master formula for the computation of the real-virtual interference becomes
\begin{align}
S_{RV}(\eps,\alpha)  &= 
-16C_AC_F\,e^{-\gamma_E(2\eps+\alpha)} \;\,
\frac{\pi^{3/2}\, \Gamma(-\eps)\, \Gamma(-4 \eps-\alpha)\, \cot(\pi\eps)}{\Gamma(-2 \eps)\,\Gamma(1/2 - \eps)\,\sin(\pi\eps)}
\no\\[0.2em]
&\quad\times
\int_0^1 \! dy_k \; \frac{y_k^{-1+2n\eps+(n+1)\alpha/2}}{(1+y_k)^\alpha}\;
\int_0^1 \!dt_k \;(4t_k\bar{t}_k)^{-1/2-\eps} \;\;
f(y_k,t_k)^{4\eps+\alpha}\,,
\label{eq:rv:master}
\end{align}
which we again already presented in the SCET-1 case in~\cite{Bell:2015lsf}.

\subsection{Double real emissions}
\label{sec:doublereal}

For the double real-emission contribution, we start from
\begin{align}
S_{RR}(\eps,\alpha) &= \frac{(4\pi e^{\gamma_E} \tau^2)^{-2\eps}\;\taubar^{-2\alpha} }{(2\pi)^{2d-2}} \, 
\int d^{d}k \;\, \delta(k^{2}) \,\theta(k^{0})\, 
\int d^{d}l \;\, \delta(l^{2}) \,\theta(l^{0}) 
\no\\[0.2em]
&\quad \times
\frac{|\mathcal{A}_{RR}(k,l)|^{2}}{(n\cdot k + \bar n \cdot k)^\alpha\,(n\cdot l + \bar n \cdot l)^\alpha} \;\, \mathcal{M}_2(\tau; k,l) \,,
\label{eq:nnlo:doublereal:start}
\end{align}
where $\mathcal{M}_2(\tau; k,l)$ is the two-emission measurement function. The respective soft matrix element 
now follows from the two-particle cut diagrams in Figure~\ref{fig:NNLOgraphs}, which give rise to three colour 
structures -- $C_{F}^{2}$, $C_{F}C_{A}$ and $C_{F}T_{F}n_{f}$ -- of which the latter two are covered in this 
paper. The corresponding squared matrix elements are given by
\begin{align}
\label{eq:nnlo:matrixelement}
|\mathcal{A}_{RR}^{(n_f)}(k,l)|^{2} &= 2048 \pi^{4} \,C_{F} T_{F} n_{f} \;\, 
\frac{2k\cdot{l} \, (k_{-} + l_{-}) \, (k_{+} + l_{+}) - (k_{-}l_{+}-k_{+}l_{-})^{2}}{(k_{-} + l_{-})^{2}\,(k_{+} + l_{+})^{2}\,(2k\cdot{l})^{2}} \,,
\\[0.2em]
|\mathcal{A}_{RR}^{(C_A)}(k,l)|^{2} &= 512 \pi^4 C_F C_A \bigg\{
\frac{k_-^2 l_+ (2k_++l_+) + 2k_- l_-(k_+^2-k_+ l_+ + l_+^2) + k_+ l_-^2(k_++2l_+)}{
k_- k_+ l_- l_+ (k_-+l_-)(k_++l_+)(2k \cdot l)} 
\no\\[0.2em]
&\qquad 
-\frac{k_-(2k_+ + l_+)+ l_-(k_+ + 2l_+)}{
k_- k_+ l_- l_+ (k_-+l_-)(k_++l_+)} 
+ \frac{2(1-\eps) (k_+ l_- - l_+
k_-)^2}{(k_-+l_-)^2(k_++l_+)^2(2k \cdot l)^2}\bigg\}\,,
\no
\end{align}
where we again suppressed the $i\varepsilon$-prescription of the propagators since it is irrelevant for the 
subsequent calculation. In comparison to \eqref{eq:nlo:matrixelement} and \eqref{eq:nnlo:rv:matrixelement}, 
we observe that the singularity structure of the double real-emission contribution is much more complicated, 
and that it gives rise to overlapping divergences that are encoded e.g.~in $(k_-+l_-)$. The propagator 
\mbox{$2 k\cdot l = k_- l_+ + k_+ l_- - 2 |\vec{k}_\perp|\, |\vec{l}_\perp| \cos\theta_{kl}$}, moreover, 
depends on the angle $\theta_{kl}$ between the two emissions in the transverse plane. Nevertheless, our 
basic strategy for the evaluation of the double real-emission correction is the same as before: We switch 
to the parametrisation \eqref{eq:parametr:nnlo}, use the explicit form \eqref{eq:measure:NNLO:corr} of the 
two-emission measurement function, perform the observable-independent integrations and use symmetry arguments 
to map the integration domain onto a unit hypercube. However, the last two steps require us to find a suitable 
parametrisation for the angular integrations and to understand the implications of the $n$-$\bar{n}$ and $k$-$l$ 
symmetries, which we will address in the next two sections, before we present the master formula for the 
computation of the double real-emission contribution.

\subsubsection{Angular parametrisation}
\label{sec:angles}

\begin{figure}[t]
\centering
\includegraphics{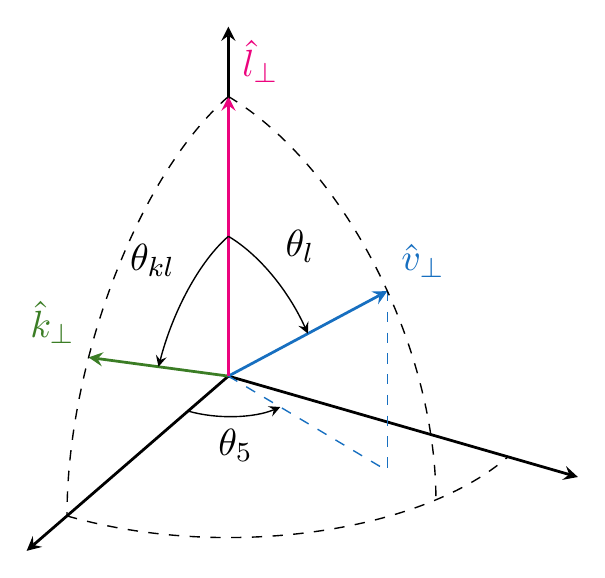}
\caption[Angular parametrisation]{The angular
parametrisation of the transverse space.
\label{fig:Angles}}
\end{figure}

According to assumption (A6), the two-emission measurement function  in general depends on three 
angles, $\theta_k=\sphericalangle(\vec{v}_\perp,\vec{k}_\perp)$, 
$\theta_l=\sphericalangle(\vec{v}_\perp,\vec{l}_\perp)$,
and $\theta_{kl}=\sphericalangle(\vec{k}_\perp,\vec{l}_\perp)$, and we would like to perform the integration 
over the remaining angles in the $(d-2)$-dimensional transverse space explicitly. This is similar in 
spirit to \eqref{eq:nlo:phase-space-measure}, where we retained the dependence on $\theta_k$, the only 
angle that arises in the one-emission measurement function.

To do so, we parametrise the vectors in the transverse space as
\begin{align}
\vec{l}_\perp&=|\vec{l}_\perp|\;
\big(1,0,0,\ldots,0\big)\,,
\no \\[0.2em]
\vec{k}_\perp&=|\vec{k}_\perp|
\;
\big(\cos\theta_{kl},\sin\theta_{kl}, 0,\ldots,0\big) \,,
\no\\[0.2em]
\vec{v}_\perp&=|\vec{v}_\perp|\;
\big(\cos \theta_{l}, \sin \theta_{l} \cos\theta_5, \sin \theta_{l} \sin\theta_5, 0, \ldots,0\big)\,,
\end{align}
which is illustrated in Figure~\ref{fig:Angles} (for convenience we show unit vectors 
$\hat{k}_\perp$, $\hat{l}_\perp$, and $\hat{v}_\perp$ that point into the 
$\vec{k}_\perp$, $\vec{l}_\perp$, and $\vec{v}_\perp$ directions, respectively). In this parametrisation, 
the angular part of the phase-space measure becomes 
\begin{align}
\int d\Omega_{d-2}^{(k)} \;d\Omega_{d-2}^{(l)}
&=\frac{4\pi^{1/2-2\eps}}{\Gamma(-\eps)\,\Gamma(1/2-\eps)}\;
\int_{-1}^1 \!d\cos\theta_{kl} \,
\int_{-1}^1 \!d\cos\theta_{l} \,
\int_{-1}^1 \!d\cos\theta_{5}
\no\\[0.2em]
&\quad\times
\sin^{-1-2\eps}\theta_{kl}\;\sin^{-1-2\eps}\theta_{l}\;\sin^{-2-2\eps}\theta_{5}\,.
\end{align}
We thus have singled out a three-dimensional subspace spanned by the angles 
$\{\theta_{kl},\theta_{l},\theta_{5}\}$, in which the angle $\theta_k$ is given by
\begin{equation}
\cos\theta_{k}= \cos\theta_{kl}\, \cos\theta_{l} + \sin\theta_{kl}\, \sin\theta_{l}\,\cos\theta_{5}\,. 
\label{eq:cosk} 
\end{equation}
This choice is of course arbitrary, but it is convenient since the matrix element depends on the 
angle $\theta_{kl}$ through the propagator $2 k\cdot l$. In order to resolve the corresponding 
divergence, we want to keep the expression of said propagator simple, and $\theta_{kl}$ should 
therefore be one of the integration variables. The angles $\theta_{k}$ and $\theta_{l}$, on the 
other hand, only enter the calculation through the measurement function, which at the end of the 
day represents a weight factor for the numerical integrations. We therefore do not mind that the 
analytic expression of the measurement function becomes complicated once we express $\theta_k$ in 
terms of the integration variables through \eqref{eq:cosk}.

We next map the integration domain onto the unit hypercube by substituting as usual  
$\cos\theta_i = 1-2t_i$ for $i\in\{k,l,kl,5\}$. 
In terms of $\bar t_i = 1- t_i$, we then have
\begin{equation}
t_{k}= t_l + t_{kl} - 2 t_{l} t_{kl} - 2\sqrt{t_{l}\bar t_{l}t_{kl}\bar t_{kl}}\,(1-2t_5) \,,  
\label{eq:tk:resolved}
\end{equation}
and we arrive at
\begin{align}
\int d\Omega_{d-2}^{(k)} \;d\Omega_{d-2}^{(l)}
&=\frac{32\pi^{1/2-2\eps}}{\Gamma(-\eps)\,\Gamma(1/2-\eps)}\;
\int_{0}^1 \!dt_{kl} \,
\int_{0}^1 \!dt_{l} \,
\int_{0}^1 \!dt_{5}
\no\\[0.2em]
&\quad\times
(4t_{kl}\bar t_{kl})^{-1/2-\eps}\;(4t_{l}\bar t_{l})^{-1/2-\eps}\;(4t_{5}\bar t_{5})^{-1-\eps}\,,
\end{align}
which is almost the final expression, except that the $t_5$-integration suffers from spurious divergences 
that arise in the limits $t_5\to 0$ and $t_5\to 1$. These divergences are clearly unphysical, and they 
indeed cancel once they are combined with the prefactor $1/\Gamma(-\eps)$. They simply arise because we 
are resolving more angles than exist in four space-time dimensions. 

Yet the $t_5$-divergences forbid a naive $\eps$-expansion on the integrand level, and we must therefore 
treat them in our formalism as if they were regular divergences. To this end, we first disentangle the 
two divergences by splitting the integration domain at $t_5=1/2$, and we subsequently rescale the two 
contributions as $t_5 \rightarrow t_5'/2$ and $t_5 \rightarrow 1- t_5'/2$, respectively. This yields 
for both cases
\begin{align}
\int d\Omega_{d-2}^{(k)} \;d\Omega_{d-2}^{(l)}
&\;\to\;\frac{16\pi^{1/2-2\eps}}{\Gamma(-\eps)\,\Gamma(1/2-\eps)}\;
\int_{0}^1 \!dt_{kl} \,
\int_{0}^1 \!dt_{l} \,
\int_{0}^1 \!dt_{5}'
\no\\[0.2em]
&\quad\;\;\times
\big(4t_{kl}\bar t_{kl}\big)^{-1/2-\eps}\;\big(4t_{l}\bar t_{l}\big)^{-1/2-\eps}\;\big(t_{5}'(2-t_{5}')\big)^{-1-\eps}\,,
\label{eq:nnlo:angularmeasure}
\end{align}
where we now have to pay attention that we integrate over two copies of the actual integrand, one with 
the substitution $t_5 \rightarrow t_5'/2$ and the second one
with $t_5 \rightarrow 1-t_5'/2$. But the integrand only depends implicitly on the variable $t_5$ 
through relation \eqref{eq:tk:resolved}, and we therefore simply have to sum over two contributions 
in which the angle $\theta_k$, and hence the variable $t_k=(1-\cos\theta_k)/2$, is resolved 
as\footnote{Notice that this definition of $t_{k}^\pm$ differs from the one we used in~\cite{Bell:2018vaa}.}
\begin{equation}
t_{k}^\pm= t_l + t_{kl} - 2 t_{l} t_{kl} \pm 2\sqrt{t_{l}\bar t_{l}t_{kl}\bar t_{kl}}\,(1-t_5') \,.  
\label{eq:tkplusminus}
\end{equation}

\subsubsection{Symmetry considerations}

We find it convenient to further map the entire integration domain onto the unit hypercube, and one can see 
in \eqref{eq:nnlo:angularmeasure} that this has already been achieved for the angular integrations. We 
therefore only have to consider remappings that involve the remaining variables $\{p_T, y, a, b\}$ in the 
parametrisation \eqref{eq:parametr:nnlo}, which are a priori all defined on the interval $[0, \infty]$.

The idea is again similar in spirit to what we have seen in the single-emission case. There we arrived at 
the representation \eqref{eq:nlo:interm}, in which the integration over the variables $k_T$ and $y_k$ both 
run from $0$ to $\infty$. After performing the integration over $k_T$ analytically, we used the $n$-$\bar{n}$ 
symmetry to map the $y_k$-integration onto the unit interval. In the present case, the $p_T$-integration can 
similarly be performed analytically since the $p_T$-dependence is universal among the considered class of 
dijet soft functions -- see \eqref{eq:measure:NNLO:corr}. We then split the integrations over $y$, $a$ and 
$b$ at the value one, and substitute $y\to 1/y$, $a\to 1/a$, and $b\to 1/b$ to map the $[1, \infty]$ 
intervals onto $[0,1]$. Explicitly, this leads to eight different contributions
\begin{align}
&\int_0^\infty \! da \,
\int_0^\infty \! db \,
\int_0^\infty \! dy \;\;
\mathcal{I}(a,b,y)
\no\\[0.2em]
&\qquad
=\int_0^1 \! da \,
\int_0^1 \! db \,
\int_0^1 \! dy \;\;
\bigg\{\mathcal{I}(a,b,y) + \mathcal{I}(1/a,b,y)
+ \mathcal{I}(a,1/b,y) + \mathcal{I}(1/a,1/b,y)
\no\\[0.2em]
& \hspace{1.7cm}
+ \mathcal{I}(a,b,1/y) + \mathcal{I}(1/a,b,1/y)
+ \mathcal{I}(a,1/b,1/y) + \mathcal{I}(1/a,1/b,1/y) \bigg\}\,,
\label{eq:nnlo:symmetry:aby}
\end{align}
where $\mathcal{I}(a,b,y)$ symbolically represents the integrand (after $p_T$-integration), which 
\mbox{implicitly} depends on the angular variables $t_{kl}$, $t_l$, and $t_5'$ that we introduced in the 
previous section. Our goal thus consists in exploiting the symmetries under $n\leftrightarrow\bar n$ 
and $k\leftrightarrow l$ exchange to reduce the number of independent integrations.

We first consider the $n$-$\bar{n}$ symmetry, which is satisfied on the level of the observable because 
of (A4), and which is also respected by the form \eqref{eq:analyticregulator} that we use for the rapidity 
regulator. It is easy to see that under $n\leftrightarrow\bar n$ exchange
\begin{equation}
a \rightarrow \frac{1}{a}\,, \qquad 
b \rightarrow b\,, \qquad 
y \rightarrow \frac{1}{y}\,, \qquad 
t_k \rightarrow t_k\,, \qquad 
t_l \rightarrow t_l\,, \qquad 
t_{kl} \rightarrow t_{kl}\,.
\end{equation}
Obviously, the measurement cannot distinguish between the two emitted partons, and the integrand is therefore 
also symmetric under $k\leftrightarrow l$ exchange, which implies
\begin{equation}
a \rightarrow \frac{1}{a}\,, \qquad 
b \rightarrow \frac{1}{b}\,, \qquad 
y \rightarrow y\,, \qquad 
t_k \rightarrow t_l\,, \qquad 
t_l \rightarrow t_k\,, \qquad 
t_{kl} \rightarrow t_{kl}\,.
\end{equation}

\begin{figure}[t]
\begin{subfigure}[t]{.32\textwidth}
\includegraphics{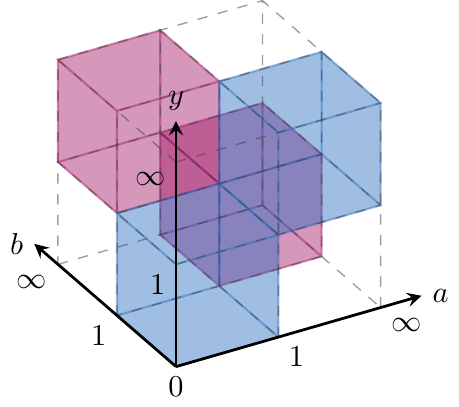}
\caption{$n \leftrightarrow \bar n$ exchange}
\end{subfigure}
\begin{subfigure}[t]{.32\textwidth}
\includegraphics{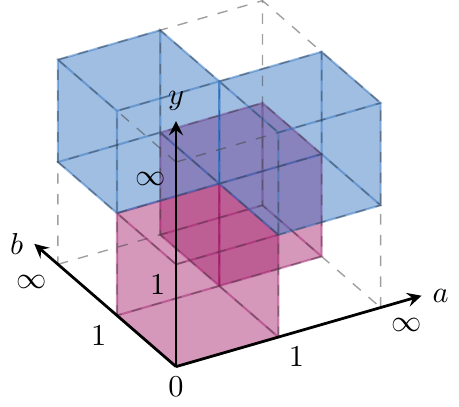}
\caption{$k \leftrightarrow l$ exchange}
\end{subfigure}
\begin{subfigure}[t]{.32\textwidth}
\includegraphics{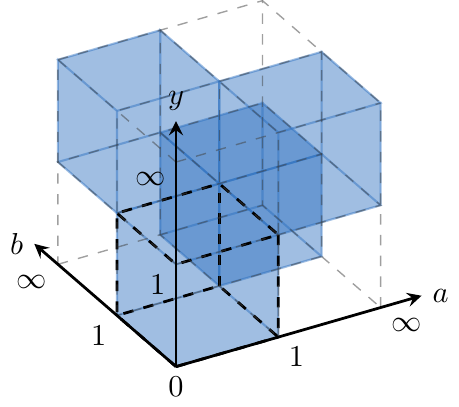}
\caption{Reduced integration region}
\end{subfigure}
\caption{Cubes of the same colour in~(a) and~(b) will yield the same result in the integration because of the
stated symmetries.  Concatenating these symmetries then allows the reduction of the integration domain to two unit cubes, one of which is highlighted in~(c), whereas
the second one emerges from it by inversion of one of the variables $a$, $b$, or $y$.
\label{fig:Symmetries}}
\end{figure}

In order to illustrate how we can make use of these symmetry considerations, let us for the moment focus on 
observables which do not depend on the angles $\theta_k$ and $\theta_l$. As the matrix 
element \eqref{eq:nnlo:matrixelement} does not depend on these angles either, the integrand 
in \eqref{eq:nnlo:symmetry:aby} is  of the form $\mathcal{I}(a,b,y,t_{kl})$.\footnote{Recall that 
$\mathcal{I}(a,b,y)$ is a short-hand notation for $\mathcal{I}(a,b,y,t_{kl},t_l,t_5')$ in 
\eqref{eq:nnlo:symmetry:aby}.} We can then exploit the $n$-$\bar{n}$ and $k$-$l$ symmetries to reduce the 
integration to two regions with
\begin{align}
&\int_0^\infty \! da \,
\int_0^\infty \! db \,
\int_0^\infty \! dy \;\;
\mathcal{I}(a,b,y,t_{kl})
\no\\[0.2em]
&\qquad
=4\, \int_0^1 \! da \,
\int_0^1 \! db \,
\int_0^1 \! dy \;\;
\bigg\{\mathcal{I}(a,b,y,t_{kl}) + \mathcal{I}(1/a,b,y,t_{kl}) \bigg\}\,,
\label{eq:nnlo:symmetry:abytkl}
\end{align}
where the form of the second term is not unique, as we show now.
This reduction is illustrated in Figure~\ref{fig:Symmetries}, where the effect of the symmetry 
transformations is shown for selected regions of the integration domain in figures (a) and (b). 
If plotted as eight stacked cubes in the three-dimensional $\{a,b,y\}$-space, the two symmetries 
ultimately enforce that the result of the integration in each of the four cubes marked in blue in 
figure (c) is the same. The eight cubes thus fall into two groups of four each, and the integration 
reduces to the form shown in \eqref{eq:nnlo:symmetry:abytkl}, where the first term corresponds to the 
blue cube that is marked with dashes in figure (c). The second term, on the other 
hand, represents one of the three white cubes adjacent to this cube, and we see that it can be 
recovered by inverting one of the variables $a$, $b$, or $y$, each corresponding to one of the 
adjacent white cubes.

Up to this point we have assumed that the measurement function does not depend on the angles $\theta_k$ 
and $\theta_l$. In the general case, the preceding discussion still caries through, except that the 
$k$-$l$ symmetry now also exchanges the angles $\theta_k$ and $\theta_l$. We may therefore expect that 
we end up with four different regions in this case, since the symmetry transformation in 
figure (b) exchanges the role of the angular integrations. We are, however, always 
free to \emph{rename} $\theta_k\leftrightarrow\theta_l$ in the transformed regions, which brings us 
back to \eqref{eq:nnlo:symmetry:abytkl}. We should also point out that the symmetry considerations 
shall be exploited on the level of the full solid angle measure 
$d\Omega_{d-2}^{(k)} \, d\Omega_{d-2}^{(l)}$, i.e.~\emph{before} we choose which angle will be expressed 
through the integration variables (see the discussion in the previous section). In our setup, we 
essentially \emph{define} the angle $\theta_k$ as the one we want to express in terms of the integration 
variables via \eqref{eq:tkplusminus}.

In conclusion we find that the integration domain in the variables $a$, $b$, and $y$ can be mapped onto 
the unit hypercube, and in doing so we obtain two contributions which we denote by the letters 
``A'' and ``B''. Region A corresponds to the blue cube in Figure~\ref{fig:Symmetries}(c) that is 
highlighted by the dashed lines, and the corresponding integrand in \eqref{eq:nnlo:symmetry:abytkl} 
is just the original integrand $\mathcal{I}(a,b,y,t_{kl},t_l,t_5')$. Region B refers to any of the 
adjacent white cubes in Figure~\ref{fig:Symmetries}(c), which can also be mapped onto the unit 
hypercube by inverting \emph{either} of the variables $a$, $b$, or $y$.

\subsubsection{Master formula}

We have now assembled all ingredients necessary to derive the master formula for the double real-emission 
correction.
Starting from the representation \eqref{eq:nnlo:doublereal:start}, we first introduce light-cone coordinates 
and switch to the parametrisation \eqref{eq:parametr:nnlo}. After inserting the explicit 
form \eqref{eq:measure:NNLO:corr} of the two-emission measurement function, we may then perform the 
integration over the dimensionful variable $p_T$ explicitly. We further use the symmetry arguments 
that we discussed in the previous section to map the integration domain onto the unit hypercube, 
and we resolve the angular phase-space measure as described in Section~\ref{sec:angles}. We then obtain
\begin{align}
&S_{RR}^{(X)}(\eps,\alpha)  
= -2C^{(X)} \;
\frac{e^{-2\gamma_E(\eps+\alpha)}\,\Gamma(-4 \eps-2\alpha)}{\pi^{3/2}\, \Gamma(- \eps)\,\Gamma(1/2 - \eps)}\;
\int_0^1 \! da \int_0^1 \! db  \int_0^1 \! dy 
\int_{0}^1 \!dt_{kl}  \int_{0}^1 \!dt_{l} 
\int_{0}^1 \!dt_{5}'
\no\\[0.2em]
&\qquad\times
\frac{a^{-2\eps}\,b^{-2\eps-\alpha}\,y^{-1+2n\eps+(n+1)\alpha}\,(a+b)^{2\eps+2\alpha}\,(1+a b)^{2\eps+2\alpha}}{\big(a+b+a(1+a b) y\big)^\alpha\;\big(a(a+b)+(1+a b)y\big)^\alpha}\;\, k^{(X)}(a,b,t_{kl})
\no\\[0.2em]
&\qquad\times
\big(4t_{kl}\bar t_{kl}\big)^{-1/2-\eps}\;\big(4t_{l}\bar t_{l}\big)^{-1/2-\eps}\;\big(t_{5}'(2-t_{5}')\big)^{-1-\eps} 
\no\\[0.2em]
&\qquad\times
\bigg\{F_A(a,b,y,t_k^+,t_l,t_{kl})^{4\eps+2\alpha}
+ F_B(a,b,y,t_k^+,t_l,t_{kl})^{4\eps+2\alpha}
+ (t_k^+\to t_k^-)
\bigg\}\,,
\label{eq:rr:master}
\end{align}
with $t_k^\pm$ from \eqref{eq:tkplusminus} and the colour factor is given by 
$C^{(n_f)}=C_{F}T_{F}n_{f}$ or $C^{(C_A)}=C_{F}C_A$. The integration kernels read
\begin{align}
k^{(n_f)}(a,b,t_{kl}) &=
\frac{128a}{(a+b)^2(1+a b)^2} \;\bigg\{
\frac{b (1-a^2)^2}{[(1-a)^2+4a t_{kl}]^2}
- \frac{(a+b)(1+ab)}{(1-a)^2+4a t_{kl}} \bigg\}\,,
\\[0.2em]
k^{(C_A)}(a,b,t_{kl}) &=
-\frac{32}{ab(a+b)^2(1+a b)^2} \;\bigg\{
\frac{2(1-\eps)a^2b^2(1-a^2)^2}{[(1-a)^2+4a t_{kl}]^2}
- (a+b)(1+ab)
\nonumber\\[0.2em]
&\qquad\times
\bigg[ b(1+a^2)+2a(1+b^2) -\frac{b(1-a^2)^2+2a(1+a^2)(1+b^2)}{(1-a)^2+4a t_{kl}}
\bigg] \bigg\}\,.
\no
\end{align}
As explained in the previous section, the master formula of the double real-emission correction 
consists of two contributions with measurement function
\begin{align}
F_A(a,b,y,t_k,t_l,t_{kl}) &= F(a,b,y,t_k,t_l,t_{kl})
\end{align}
in region A, whereas the one in region B is obtained by inverting either of the variables 
$a$, $b$, or $y$ with
\begin{align}
\label{eq:rr:fb}
F_B(a,b,y,t_k,t_l,t_{kl}) &=\begin{dcases} F(1/a,b,y,t_k,t_l,t_{kl}) &\,\text{or} \\ F(a,1/b,y,t_k,t_l,t_{kl}) &\,\text{or}\\ y^{-n}F(a,b,1/y,t_k,t_l,t_{kl}) \,.\end{dcases}
\end{align}
We stress that the three representations of this function need not be identical, since the symmetry arguments 
only guarantee that their integrals in \eqref{eq:rr:master} are equal, but not necessarily the integrands. 
One is therefore free to derive the measurement function in region B using any of the expressions on the 
right-hand side of \eqref{eq:rr:fb}.

From \eqref{eq:rr:master} we can analyse the divergence structure of the double real-emission correction. 
First, we find an explicit divergence that is encoded in the factor $\Gamma(-4\eps-2\alpha)$, which is 
associated with the limit $p_T\to0$, i.e.~the configuration where both emitted partons become soft. 
Second, we observe that the $y$-integral diverges in the limit $y\to 0$, which reflects the fact that 
one of the partons becomes collinear to the jet direction $n^\mu$. Similar to the single-emission case, 
this divergence yields a $1/\alpha$ pole for SCET-2 observables with $n=0$. Third, we identify an 
overlapping divergence in the limit in which the two emitted partons become collinear to each other, 
i.e.~$a\to 1$ and $t_{kl}\to 0$. Fourth, the $C_{F}C_A$ contribution displays an additional divergence 
in the limit $b\to0$, which implies that the parton with momentum $k^\mu$ becomes soft (due to the 
$k$-$l$ symmetry, the configuration with $l^\mu\to0$ is mapped onto the same constraint). Finally, we 
observe that the expression diverges in the limit $t_5'\to0$, which is an unphysical divergence that is 
cancelled by the prefactor $1/\Gamma(- \eps)$.

For $n\neq0$, the expansion of the $C_{F}T_{F}n_{f}$ structure thus starts with a $1/\eps^3$ divergence, 
whereas the $C_{F}C_A$ term has a $1/\eps^4$ pole because of the additional $b\to0$ singularity. Except 
for the overlapping divergence, the singularities are in fact already factorised and can easily be 
isolated using an expansion in terms of plus-distributions in analogy to \eqref{eq:yexp}. As we explained 
in detail in Section~\ref{sec:parametrisations}, the functions $F_{A/B}(a,b,y,t_k,t_l,t_{kl})$ are by 
construction finite and non-zero in the singular limits\footnote{We note that the limit $t_5'\to0$ is 
indirectly protected by infrared safety, which can be seen as follows.  The two-emission measurement function reduces to the one-emission function 
in the limit $b\to0$ -- see \eqref{eq:infrared:soft} -- and it is therefore finite and independent of 
the angle $\theta_{5}$ in this limit. If, for finite values of $b$, the limit $t_5'\to0$ were to cause 
the observable to vanish or diverge, it would mean that the combined limit $t_5'\to0$ \emph{and} $b\to 0$ 
was discontinuous. This is, however, not allowed because it would enable us to infer the presence of  
an infinitesimally soft second emission.}, and the remaining integrations are therefore well-defined upon 
an expansion in the dimensional regulator $\eps$. For SCET-2 soft functions with $n=0$, on the other hand, 
the $y$-integration generates a $1/\alpha$ pole which leads to additional $1/(\alpha\eps^2)$ divergences 
for the $C_{F}T_{F}n_{f}$ and $1/(\alpha\eps^3)$ divergences for the $C_{F}C_A$ colour structures.

In order to isolate the overlapping divergence, one could apply a standard sector decomposition 
strategy~\cite{Binoth:2000ps}, but we prefer to resolve it by means of an additional substitution,

\begin{align}
\label{eq:uvsubstitution}
a=1-u(1-v)\,, \qquad\qquad t_{kl}=\frac{u^2 v}{1-u(1-v)}\,.
\end{align}
This change of variables matches a unit hypercube in the variables $(a,t_{kl})$ onto another unit 
hypercube in the new variables $(u,v)$, and it maps the overlapping divergence that arises in the 
limit $a\to 1$ and $t_{kl}\to 0$ onto the line $u\to0$. The critical propagator then takes a particularly 
simple form
\begin{align}
\frac{1}{(1-a)^2+4a t_{kl}} \;=\; \frac{1}{u^2(1+v)^2}
\end{align}
and it turns out that the singularity from the limit $u\to0$ is completely factorised, at the expense of 
increasing the complexity of the integrand. This is, however, a minor price to pay since the integrations 
are eventually performed numerically, and the above substitution does not worsen the convergence of the 
numerical integrations (see Section~\ref{sec:numerics} for further refinements we implement to improve the 
numerical convergence).

%% file: Renorm.tex
\section{Renormalisation}
\label{sec:renormalize}

While the main objective of our work is the computation of bare dijet soft functions, we also extract the 
anomalous dimensions and matching corrections that are needed for resummations within SCET. This requires 
us to make some additional assumptions about the structure of the underlying RG equations. For many 
observables -- including all examples that we discuss  in Section~\ref{sec:results} -- the soft function 
renormalises multi\-plicatively in Laplace (or Fourier) space. We therefore focus on this particular class 
of observables in this section, leaving other soft functions that renormalise directly in momentum 
(or cumulant) space -- like certain jet-veto observables -- for a future study~\cite{BRT}.

\subsection{\si~observables}
\label{sec:ren:sceti}

For \si~observables with $n\neq 0$, we can set the additional regulator $\alpha=0$, and the expansion of the 
bare soft function takes the generic form
\begin{align}
S_0(\tau) = 1 &+ \left(\frac{Z_\alpha\alpha_s}{4\pi}\right) \,(\mu^2 \taubar^2)^\eps \; \bigg\{ \frac{x_2}{\eps^2} + \frac{x_1}{\eps}  + x_0 + x_{-1}\,\eps + x_{-2}\,\eps^2 + \mathcal{O}(\eps^3)\bigg\}
\no\\[0.2em]
& + \left(\frac{Z_\alpha\alpha_s}{4\pi}\right)^2 (\mu^2 \taubar^2)^{2\eps} \; \bigg\{ \frac{y_4}{\eps^4} + \frac{y_3}{\eps^3}  + \frac{y_2}{\eps^2} + \frac{y_1}{\eps} + y_0 + \mathcal{O}(\eps)\bigg\} + \mathcal{O}(\alpha_s^3)\,,
\end{align}
where $x_i$ and $y_i$ are the NLO and NNLO coefficients at order $1/\epsilon^{i}$, respectively. The former 
are obtained by expanding the master formula \eqref{eq:nlo:master} of the single real-emission contribution, 
while the latter are given by the sum of the real-virtual interference \eqref{eq:rv:master} and the double 
real-emission correction \eqref{eq:rr:master}. It should be understood that the coefficients carry colour 
factors;  these are given by $C_F$ in the case of the $x_i$, while the $y_i$ are sums of three different 
numbers multiplying the colour factors $C_F^2$, $C_F T_F n_f$, and $C_F C_A$. The correlated emission 
formulae provided in this paper yield the $C_F T_F n_f$ and $C_F C_A$ contributions, whereas the 
calculation of the $C_F^2$ correction is not covered in this work. For soft functions that obey the 
NAE theorem, this contribution is however proportional to the square of the one-loop correction and our 
results in Section~\ref{sec:results} are therefore complete for this particular class of observables.

We now assume that the soft function renormalises multiplicatively in Laplace space, $S = Z_S S_0$, and 
that the renormalised soft function fulfils the RGE
\begin{align}
\label{eq:RG}
\frac{\rd}{\rd \ln\mu}  \; S(\tau,\mu)
&= - \frac{1}{n} \,\bigg[ 4 \,\Gamma_{\mathrm{cusp}}(\alpha_s) \, 
\ln(\mu\taubar) 
-2 \gamma^{S}(\alpha_s) \bigg] \; S(\tau,\mu) \,,
\end{align}
where $\Gamma_{\mathrm{cusp}}(\alpha_s)$ is the cusp anomalous dimension and $\gamma^{S}(\alpha_s)$ 
denotes the (non-cusp) soft anomalous dimension. The parameter $n\neq0$ in the RGE is related to the 
power counting of the modes in the effective theory, as we explained in detail in Section~\ref{sec:n}. 
We find it convenient to define the non-cusp anomalous dimensions with a prefactor $2/n$, similar to the 
conventions we used in~\cite{Bell:2018vaa}.
Expanding the anomalous dimensions as
\begin{align}
\Gamma_{\mathrm{cusp}}(\alpha_s) = \sum_{m=0}^{\infty} \left(\frac{\alpha_{s}}{4 \pi} \right)^{m+1} \Gamma_{m}\,, \qquad \quad 
\gamma^{S}(\alpha_s)= \sum_{m=0}^{\infty} \left(\frac{\alpha_{s}}{4 \pi} \right)^{m+1} \gamma^{S}_{m}\,,
\end{align} 
and using $Z_\alpha = 1-\beta_0\alpha_s/(4\pi\eps) + \mathcal{O}(\alpha_s^2)$, one can show that the RGE 
is solved to two-loop order by
\begin{align}
S(\tau,\mu) &= 
1 + \left( \frac{\alpha_s}{4 \pi} \right) 
\left\{ -\frac{2\Gamma_0}{n} \,L^2 
+ \frac{2\gamma_0^S}{n} \,L 
+ c_1^S \right\}
+\left( \frac{\alpha_s}{4 \pi} \right)^2 
\bigg\{ \frac{2 \Gamma_0^2}{n^2} L^4 
-  4\Gamma_0\left( \frac{\gamma_0^S}{n^2} + \frac{\beta_0}{3n} \right) 
L^3 
\nonumber\\[0.2em]  
 &\quad
 - 2 \left( \frac{\Gamma_1}{n} - \frac{(\gamma_0^S)^2}{n^2} 
 - \frac{\beta_0 \gamma_0^S}{n} +\frac{\Gamma_0 c_1^S}{n} \right) L^2 
+ 2 \left (\frac{\gamma_1^S}{n} +\frac{\gamma_0^S c_1^S}{n} +\beta_0 c_1^S \right) L + c_2^S \bigg\} 
\label{eq:Sren}
\end{align}
with $L=\ln(\mu\taubar)$. The $Z$-factor $Z_{S}$ satisfies the same RGE (\ref{eq:RG}), and its explicit 
solution to two-loop order is given by
\begin{align} 
Z_{S} &= 1 + \left( \frac{\alpha_s}{4 \pi} \right) 
\left[ \frac{\Gamma_0}{n}\,\frac{1}{\eps^2} + 
\frac{2\Gamma_0 L - \gamma_0^S}{n}\, \frac{1}{\eps}
\right]
+\left( \frac{\alpha_s}{4 \pi} \right)^2 
\bigg[ \frac{\Gamma_0^2}{2n^2}\,\frac{1}{\eps^4}
+\Gamma_0 \left( \frac{2\Gamma_0}{n^2}\,L
-\frac{\gamma_0^S}{n^2}
 - \frac{3\beta_0}{4n}\right)\,\frac{1}{\eps^3}
\nonumber\\[0.2em]  
 &\quad
+\bigg( \frac{2\Gamma_0^2}{n^2}\,L^2 
- \Gamma_0 \Big(
\frac{2\gamma_0^S}{n^2} + \frac{\beta_0}{n} \Big)L
+\frac{\Gamma_1}{4n} + \frac{(\gamma_0^S)^2}{2n^2}
+ \frac{\beta_0\gamma_0^S}{2n}
\bigg) \, \frac{1}{\eps^2}
 + \frac{2\Gamma_1 L - \gamma_1^S}{2n}\, \frac{1}{\eps}
\bigg]\,.
\label{eq:ZS}
\end{align}
The universal expansion coefficients appearing in \eqref{eq:Sren} and \eqref{eq:ZS} read
\begin{equation} 
\Gamma_0 = 4 C_F\,,
\qquad
\Gamma_1 = 4 C_F
\left\{ \left(\frac{67}{9}-\frac{\pi^2}{3}\right) C_A
- \frac{20}{9} T_F n_f \right\},
\qquad
\beta_0 = \frac{11}{3} C_A - \frac43T_F n_f\,.
\label{eq:ren:univsersalcoeffs}
\end{equation}
Equipped with this knowledge, we can extract the non-cusp soft anomalous dimension directly from the 
$1/\eps$ coefficients of the bare soft function using the relations
\begin{align}
\gamma_0^S&=n x_1\,,
 \no\\[0.1em] 
\gamma_1^S &= 2n\big(y_1-x_2\, x_{-1}-x_0\left(x_1+\beta_0\right)\big)\,,
\label{eq:gamma:coeffs}
\end{align}
while the non-logarithmic coefficients of the renormalised soft function follow from the finite terms via
\begin{align}
c_1^S&=x_0\,,
\no \\[0.1em] 
c_2^S&=y_0-\left(x_1+\beta_0\right)x_{-1}-x_2\,x_{-2}\,.
\label{eq:c:coeffs}
\end{align}
A strong check of our calculation is provided by the requirement that the higher poles $1/\eps^j$ with 
$j=2,3,4$ must vanish in the product of the $Z$-factor and the bare soft function.

\subsection{\sii~observables}
\label{sec:ren:scetii}

The \sii~case is slightly more complicated, owing to the double expansion in the regulators 
$\alpha$ and $\eps$. Starting from \eqref{eq:baresoftfun:expansion}, the expansion of a bare 
SCET-2 soft function takes the generic form
\begin{align}
\label{eq:ren:scet2bare}
S_0(\tau,\nu) &= 1 + \left(\frac{Z_\alpha\alpha_s}{4\pi}\right) \,(\mu^2 \taubar^2)^\eps \;(\nu \taubar)^\alpha 
\bigg[ \frac{1}{\alpha} \bigg(
\frac{x^1_1}{\eps}  + x^1_0 + x^1_{-1}\,\eps + x^1_{-2}\,\eps^2 \bigg) 
\\[0.1em]
&\qquad + 
\frac{x^0_2}{\eps^2}  + \frac{x^0_1}{\eps} + x^0_0 + x^0_{-1}\,\eps
+\alpha \bigg(
\frac{x^{-1}_3}{\eps^3} + \frac{x^{-1}_2}{\eps^2} + \frac{x^{-1}_1}{\eps}  + x^{-1}_0  \bigg) 
+\mathcal{O}\Big(\frac{\eps^3}{\alpha},\eps^2,\alpha\eps,\alpha^2\Big)\bigg]
\no\\[0.1em]
&\quad + \left(\frac{Z_\alpha\alpha_s}{4\pi}\right)^2 (\mu^2 \taubar^2)^{2\eps} \; \Bigg[ (\nu \taubar)^{2\alpha} \,\bigg[ \frac{1}{\alpha^2} \bigg(
\frac{y^2_2}{\eps^2} + \frac{y^2_1}{\eps}  + y^2_0 \bigg) 
\no\\[0.1em]
&\qquad + 
\frac{1}{\alpha} \bigg(
\frac{y^1_3}{\eps^3}  + \frac{y^1_2}{\eps^2}  + \frac{y^1_1}{\eps}  + y^1_0 \bigg) + \frac{y^0_4}{\eps^4} + \frac{y^0_3}{\eps^3}  + \frac{y^0_2}{\eps^2} + \frac{y^0_1}{\eps} + y^0_0 
+\mathcal{O}\Big(\frac{\eps}{\alpha^2},\frac{\eps}{\alpha},\eps,\alpha\Big)
\bigg]\no\\[0.1em]
&\qquad + (\nu \taubar)^{\alpha} \,\bigg[ 
\frac{1}{\alpha} \bigg(
\frac{z^1_3}{\eps^3}  +\frac{z^1_2}{\eps^2}  +\frac{z^1_1}{\eps}  + z^1_0\bigg) + \frac{z^0_4}{\eps^4} + \frac{z^0_3}{\eps^3}  + \frac{z^0_2}{\eps^2} + \frac{z^0_1}{\eps} + z^0_0
+\mathcal{O}\Big(\frac{\eps}{\alpha},\eps,\alpha\Big)\bigg] \Bigg],
\no
\end{align}
where $x^i_j$, $y^i_j$, and $z^i_j$ label the $1/(\alpha^i\epsilon^{j})$ coefficients of the single 
real-emission, double real-emission, and real-virtual interference term, respectively. We recall that 
the rapidity regulator is implemented on the level of the phase-space integrals, which explains the 
different powers of $(\nu \taubar)$ in the NNLO correction. The coefficients again carry colour factors 
given by $C_F$ for the $x^i_j$, $C_F C_A$ for the $z^i_j$, and with $y^i_j$ containing three contributions 
proportional to $C_F^2$, $C_F T_F n_f$, and $C_F C_A$, of which the latter two are covered in this paper. 
The $C_F^2$ contribution is, on the other hand, again proportional to the square of the NLO correction for soft 
functions that obey the NAE theorem.

In the following we adopt the notation of the \emph{collinear anomaly} 
approach~\cite{Becher:2010tm,Becher:2011pf} to extract the relevant quantities for resummations in SCET-2. 
The formalism is equivalent to the \emph{rapidity renormalisation group} (RRG) advocated in~\cite{Chiu:2012ir}, 
and we briefly comment on the translation into the RRG framework -- including some subtleties about the 
choice of the rapidity regulator -- at the end of this section.

In the collinear anomaly language, the bare soft function can be written in the form
\begin{align}
\label{eq:ren:scet2factorisation}
S_0(\tau,\nu)
&= (\nu^2\taubar^2)^{-\mathcal{F}_0(\tau)} \;W_0^S(\tau)\,,
\end{align}
where we made the $\nu$-dependence explicit and we suppressed the terms divergent in $\alpha$, which cancel 
between the soft and collinear functions. The bare collinear anomaly exponent $\mathcal{F}_0(\tau)$ controls 
the logarithmic dependence on the rapidity scale $\nu$, and the bare soft remainder function $W_0^S(\tau)$ 
collects the terms that are not associated with the rapidity divergences. The latter is in fact meaningless 
in the collinear anomaly framework without knowledge about the corresponding collinear remainder function 
$W_0^C(\tau)$, since only their product obeys a well-defined RG equation in the $\overline{\text{MS}}$ 
scheme~\cite{Becher:2011pf}. As the collinear remainder function is not known in the chosen regularisation 
scheme for most of the observables we consider in Section~\ref{sec:results}, we disregard the soft remainder 
function $W_0^S(\tau)$ and focus on the collinear anomaly exponent $\mathcal{F}_0(\tau)$ in the following.

Following the procedure described in~\cite{Becher:2012qc}, we can extract the bare anomaly exponent from the 
bare SCET-2 soft function $S_0(\tau,\nu)$. This extraction is in fact subtle since the anomaly exponent is 
related  to the coefficient of the logarithm $\ln (\nu^2\taubar^2)$ -- see \eqref{eq:ren:scet2factorisation} -- 
rather than the associated $1/\alpha$ divergences~\cite{Becher:2012qc}. In terms of the expansion coefficients 
of the bare soft function from \eqref{eq:ren:scet2bare}, we find that the bare anomaly exponent takes the form
\begin{align}
&\mathcal{F}_0(\tau) = -\frac12 \left(\frac{Z_\alpha\alpha_s}{4\pi}\right) \,(\mu^2 \taubar^2)^\eps \; \bigg\{ \frac{x^1_1}{\eps}  + x^1_0 + x^1_{-1}\,\eps + x^1_{-2}\,\eps^2 + \mathcal{O}(\eps^3)\bigg\}
- \left(\frac{Z_\alpha\alpha_s}{4\pi}\right)^2 (\mu^2 \taubar^2)^{2\eps} 
\no\\[0.1em]
&\quad \times \bigg\{ \bigg( y^1_3 + \frac{z^1_3}{2} - x^0_2 x^1_1 \bigg) \frac{1}{\eps^3}
+\bigg( y^1_2 + \frac{z^1_2}{2} - x^0_2 x^1_0 - x^0_1 x^1_1 \bigg) \frac{1}{\eps^2}
+\bigg( y^1_1 + \frac{z^1_1}{2} - x^0_2 x^1_{-1} 
\no\\[0.1em]
&\quad\qquad - x^0_1 x^1_0 - x^0_0 x^1_1 \bigg) \frac{1}{\eps} 
+y^1_0 + \frac{z^1_0}{2} - x^0_2 x^1_{-2} - x^0_1 x^1_{-1} 
- x^0_0 x^1_0 - x^0_{-1} x^1_1 + \mathcal{O}(\eps)\bigg\}.
\end{align} 
Owing to its place in the exponent, the anomaly coefficient renormalises additively in Laplace space, 
$\mathcal{F}_0=\mathcal{F}+Z_\mathcal{F}$, and the renormalised anomaly exponent satisfies the RGE
\begin{align}
\label{eq:RGE:SCET-2}
\frac{\rd}{\rd \ln\mu}  \; \mathcal{F}(\tau,\mu)
&= 2 \,\Gamma_{\mathrm{cusp}}(\alpha_s)\,,
\end{align}
which to two-loop order is solved by
\begin{align}
\mathcal{F}(\tau,\mu) &= 
\left( \frac{\alpha_s}{4 \pi} \right) 
\Big\{ 2\Gamma_0 \,L 
+ d_1 \Big\}
+\left( \frac{\alpha_s}{4 \pi} \right)^2 
\Big\{ 2 \beta_0\Gamma_0\, L^2 
 + 2 \left( \Gamma_1 + \beta_0 d_1 \right) L + d_2 \Big\},
 \label{eq:d1d2}
\end{align}
where again $L=\ln(\mu\taubar)$ and the expansion coefficients of the cusp anomalous dimension and the 
beta function can be found in \eqref{eq:ren:univsersalcoeffs}. The $Z$-factor $Z_\mathcal{F}$ satisfies a 
similar RGE as the anomaly coefficient, and its explicit form to two-loop order reads
\begin{align} 
Z_\mathcal{F}&= \left( \frac{\alpha_s}{4 \pi} \right) 
\left\{ \frac{\Gamma_0}{\eps} \right\}
+\left( \frac{\alpha_s}{4 \pi} \right)^2 
\bigg\{ - \frac{\beta_0\Gamma_0}{2\eps^2}
+ \frac{\Gamma_1}{2\eps}
\bigg\}\,.
\end{align}
We can then extract the non-logarithmic terms of the renormalised anomaly coefficient \eqref{eq:d1d2} 
using the relations
\begin{align}
d_1&=-\frac{x^1_0}{2}\,,
 \no\\[0.1em] 
d_2 &= -y^1_0 - \frac{z^1_0}{2}  + 
 x^0_{-1} x^1_1 +  x^0_{0} x^1_0
+  x^0_{1} x^1_{-1} +  x^0_{2} x^1_{-2} + \frac{\beta_0 x^1_{-1}}{2}\,.
\label{eq:scet2:d1d2extraction}
\end{align}
The cancellation of $1/\eps^j$ divergences with $j=1,2,3$ in the renormalised anomaly exponent then provides 
another strong check of our calculation.

We finally translate our findings into the RRG framework from~\cite{Chiu:2012ir}. Here the renormalisation is 
implemented directly on the level of the soft function rather than the anomaly exponent, $S = Z_S S_0$, and 
the Z-factor $Z_S$ absorbs both $1/\eps$ and $1/\alpha$ divergences according to a modified 
$\overline{\text{MS}}$ prescription. Furthermore, the renormalised soft function satisfies the RRG equation 
\begin{align}
\label{eq:RRG}
\frac{\rd}{\rd \ln\nu}  \; S(\tau,\mu,\nu)
&= \bigg[ 4 \,A_\Gamma(\mu_s,\mu) 
-2 \gamma_{\nu}(\mu_s) \bigg] \; S(\tau,\mu,\nu) \,,
\end{align}
where
\begin{align}
  A_\Gamma(\mu_1, \mu_2) &= - \int_{\alpha_s (\mu_1)}^{\alpha_s (\mu_2)} d\alpha\;
  \frac{\Gamma_{\mathrm{cusp}}(\alpha)}{\beta (\alpha)}\,,
\end{align}
which is solved by
\begin{align}
S(\tau,\mu,\nu)&=
\left(\frac{\nu}{\nu_s}\right)^{4 \,A_\Gamma(\mu_s,\mu) 
-2 \gamma_{\nu}(\mu_s)}\, S(\tau,\mu,\nu_s)\,.
\label{eq:RRG:solution}
\end{align}
The solution can be compared to \eqref{eq:ren:scet2factorisation} in the collinear anomaly approach, 
bearing in mind that a similar relation holds among the renormalised quantities in this case. With the 
all-order solution to the RGE \eqref{eq:RGE:SCET-2},
\begin{align}
\mathcal{F}(\tau,\mu) &= 
-2 \,A_\Gamma(\mu_s,\mu) 
+\mathcal{F}(\tau,\mu_s),
\end{align}
we can then identify the $\nu$-anomalous dimension in the RRG approach with the collinear anomaly exponent,
\begin{align}
\gamma_{\nu}(\mu_s) &=\mathcal{F}(\tau,\mu_s)\,.
\label{eq:RRG:gammanu}
\end{align}
The comparison between \eqref{eq:ren:scet2factorisation} and \eqref{eq:RRG:solution} in addition allows 
us to express the renormalised soft remainder function as $W^S(\tau,\mu)=S(\tau,\mu,\nu_s=1/\bar\tau)$. 
Interestingly, the latter has a well-defined $\mu$-evolution in the RRG framework that is governed by the 
RGE
\begin{align}
\frac{\rd}{\rd \ln\mu}  \; S(\tau,\mu,\nu_s=1/\bar\tau)
&= \bigg[ 4 \,\Gamma_{\mathrm{cusp}}(\alpha_s) \, 
\ln(\mu\taubar) 
-2 \gamma^{S}(\alpha_s) \bigg] \; S(\tau,\mu,\nu_s=1/\bar\tau) \,,
\label{eq:RRG:muRGE}
\end{align}
whereas the soft remainder function does not obey a simple RGE in the collinear anomaly approach 
(only the product of the soft and collinear remainder functions does so). Moreover, we find that the 
RGE \eqref{eq:RRG:muRGE} is \emph{not} satisfied by our solution for SCET-2 soft functions, and the 
problem can be traced back to the way we have implemented the rapidity regulator. In other words, the 
RRG approach intrinsically makes specific assumptions about the form of the rapidity regulator, which 
in particular must be implemented on the level of connected webs~\cite{Chiu:2012ir}. We are not aware 
that this difference between the collinear anomaly and the RRG approach has been made so clearly in the 
literature before.

To summarise, for SCET-2 observables we determine the collinear anomaly exponent 
$\mathcal{F}(\tau,\mu)$ in \eqref{eq:d1d2} or, equivalently, the $\nu$-anomalous dimension 
$\gamma_{\nu}(\mu_s)$ in \eqref{eq:RRG:gammanu} using the relations in \eqref{eq:scet2:d1d2extraction}. 
As our calculation yields the full bare soft function in \eqref{eq:ren:scet2bare}, it also determines 
the bare soft remainder function $W_0^S(\tau)$, which is a useful input in the collinear anomaly 
approach if the corresponding collinear remainder function $W_0^C(\tau)$ is known in the same 
regularisation scheme. Our results for the soft remainder function are, on the other hand, not 
consistent with the RRG framework since we did not implement the rapidity regulator on the level of 
connected webs.

%% file: Generalisations.tex
\section{Revisiting our assumptions}
\label{sec:generalize}

Having established the theoretical framework for the calculation of the correlated-emission contribution to 
dijet soft functions, we now return to the list of assumptions that we outlined in Section~\ref{sec:measure}. 
In particular, we can now better understand why these assumptions were made and how some of them could possibly 
be relaxed in the future. In this section, we in fact already introduce two extensions of our formalism that are 
valid for multi-differential and Fourier-space soft functions. We now address each of the assumptions from 
Section~\ref{sec:measure} in turn.


\subsection*{(A1) Beyond dijet factorisation}

The soft functions we consider in this work are defined in terms of two light-like Wilson lines, and they are 
supposed to be embedded in a dijet factorisation theorem of the form \eqref{eq:fact}. The soft function is, 
moreover, assumed to have a double-logarithmic evolution in the scales $\mu$ and, possibly, $\nu$. Soft 
functions that are blind to the jet directions only have a single-logarithmic evolution and thus they cannot 
be computed directly in the current formalism.\footnote{In some cases it may be possible to write such soft 
functions as a difference of two dijet soft functions.} This can clearly be seen from \eqref{eq:measure:one} 
and \eqref{eq:measure:NNLO:corr}, where it is not possible to extract a value for the parameter $n$ if the 
observable is exactly zero in the collinear limit. As the parameter $n$ controls the double-logarithmic terms in 
the RGE \eqref{eq:RG}, the current formalism cannot be applied to problems with single logarithms per loop 
order.

We further assumed that the two hard, massless partons are in a back-to-back configuration, i.e.~$n\cdot \bar n=2$. 
Although the generalisation to  arbitrary kinematical configurations is relatively straight-forward, we plan 
to relax this assumption only in the context of general $N$-jet soft functions~\cite{Bell:2018mkk}. It would 
also be interesting to extend the formalism to processes with massive hard partons ($n^2\neq0$), which is 
relevant for top-quark related processes.

\subsection*{(A2) Loosening the constraints on the $\omega(\lbrace k_{i} \rbrace)$ measure}

Core to our approach is the structure of the function $\omega(\lbrace k_{i} \rbrace)$ appearing in the exponent 
of the measurement function \eqref{eq:measure:general}, and whose explicit one-emission and two-emission 
parametrisations were given in \eqref{eq:measure:one} and \eqref{eq:measure:NNLO:corr}, respectively.  
While we required that $\Re\big(\omega(\lbrace k_{i} \rbrace)\big)>0$, we already mentioned in (A2) that 
the observables are allowed to vanish for configurations with zero weight in the phase-space integrations.  
We now address this caveat more carefully, and we further elaborate on both the treatment of complex numbers
and the regulator dependence of the measurement function.

\subsubsection*{Integrable divergences}

According to the master formulae \eqref{eq:nlo:master}, \eqref{eq:rv:master}, and \eqref{eq:rr:master}, 
the functions $f(y_k,t_k)$ and $F_{A,B}(a,b,y,t_k,t_l,t_{kl})$ enter our formulae for the numerical 
integrations -- after expansion in the various regulators -- in terms of logarithms. It is therefore crucial 
that these functions do not vanish in the singular limits of the matrix elements, since otherwise the delta 
function associated with the divergence would put the argument of the logarithm to zero. The functions may, 
however, vanish for non-singular configurations with zero weight in the phase-space integrals, since the 
logarithms only constitute integrable divergences in this case. Of course, the logarithmic divergences may 
still pose a challenge for the numerical integrations, but for all examples we consider in 
Section~\ref{sec:results} the integrable divergence seem to be under control. If, on the other hand, the 
functions $f(y_k,t_k)$ and $F_{A,B}(a,b,y,t_k,t_l,t_{kl})$ vanish over wide ranges of phase space (with 
non-zero weight), the logarithmic divergence is no longer integrable and these situations are therefore 
excluded by assumption (A2).

\subsubsection*{Fourier transforms and complex numbers}

Whereas our formalism assumes that $\Re\big(\omega(\lbrace k_{i} \rbrace)\big)>0$, the 
\softserve~implementation requires that $\omega(\lbrace k_{i} \rbrace)\big)$ is strictly real and
non-negative. Whenever the soft function is defined in Fourier rather than Laplace space, the one-emission 
measurement function will appear as $[i \widetilde{f}(y_k,t_k)]^{2\eps+\alpha}$ in the NLO master formula 
\eqref{eq:nlo:master} with a real-valued function $\widetilde{f}(y_k,t_k)$ (and similarly for the NNLO master formulae). 
In some cases like the one for threshold resummation in Drell-Yan production -- see Section~\ref{sec:measure} -- 
we can absorb the imaginary unit into the definition of the Laplace variable $\tau$, which brings the soft 
function into the standard form we assume in our framework. In other cases, such as the one for 
transverse-momentum resummation, this would leave us, however, with a real-valued function $\widetilde{f}(y_k,t_k)$ 
that can take on both positive and negative values, which is not allowed for the \softserve~implementation.
We can nevertheless use our framework to calculate the real part of such soft functions by carefully treating 
the imaginary unit in the master formulae as a phase. A detailed description of this method is given in 
Appendix~\ref{afourier}, and as an example that exploits the Fourier-space extension we compute the soft 
function for transverse-momentum resummation in Section~\ref{sec:results}.

\subsubsection*{Regulator dependence}

According to (A2), the function $\omega(\lbrace k_{i} \rbrace)$ is assumed to be independent of the rapidity 
regulator $\alpha$ and the dimensional regulator $\eps$. While this is, of course, always fulfilled for a 
physical observable, this restricts the set of transformations that one may use to bring the soft function 
into the form \eqref{eq:measure:general}. In particular, we are currently aware of a single observable that 
gives rise to an $\eps$-dependent $\omega(\lbrace k_{i} \rbrace)$ measure:  the $e^+ e^-$ event shape jet 
broadening. Due to soft recoil effects, the broadening soft function requires a $(d-2)$-dimensional Fourier 
transform on top of the Laplace transformation to resolve all distributions~\cite{Becher:2011pf}, 
which brings the soft function into the form \eqref{eq:measure:general} with an $\eps$-dependent function $\omega(\lbrace k_{i} \rbrace;\eps)$.

The main effect of the regulator dependence shows up in the expansion of the Laurent series
in $\eps$, where different orders of the function $f(y_k,t_k;\eps)$ would contribute at different orders of the 
Laurent 
series. As the entire framework hinges on a precise understanding of the behaviour of the input functions in 
the singular limits, the discussion of the implications of infrared safety would have to be extended to the 
different orders in the function $f(y_k,t_k;\eps)$ itself. This could then lead to additional restrictions 
on the form of the regulator dependence. Such a discussion lies outside the scope of the present paper, and 
will only be revisited in the future if the specific need arises.

\subsection*{(A3) Mass dimension $\neq$ 1}

Relaxing our assumption about the mass dimension of the $\omega(\lbrace k_i \rbrace)$ measure is probably 
the easiest generalisation in our list. According to \eqref{eq:measure:one}, the mass dimension determines 
the power of the variable $k_T$ in the measurement function, which is later integrated out analytically 
in \eqref{eq:nlo:interm} (similar arguments hold for the two-emission measurement function and the 
variable $p_T$).  However, the latter integration is perfectly convergent and well-behaved for any
positive, non-zero mass dimension, and the only changes manifest in different numerical constants in some 
places, like the argument of the Gamma functions or certain exponents.

Switching to different mass dimensions would therefore only result in slightly modified master formulae in our
approach. While this could easily be implemented, we have not yet encountered any observable which would 
require such a tune.


\subsection*{(A4) Broken $n$-$\bar{n}$ symmetry}

In its current form we assume that  the $\omega(\lbrace k_{i} \rbrace)$ measure is symmetric under 
$n \leftrightarrow \bar n$ exchange, which is not necessarily the case for all observables. 
In Section~\ref{sec:results} we consider e.g.~the hemisphere soft function, which depends on two invariant 
masses $M_L$ and $M_R$ that are not invariant, but rather mapped onto each other, under 
$n \leftrightarrow \bar n$ exchange.

Relaxing the $n$-$\bar{n}$ symmetry is again possible at the expense of doubling the number of input functions 
that need to be provided by the \softserve~user. The required input currently includes the measurement 
functions $F_A$ and $F_B$ and the parameter $n$ (the function $f$ is internally determined using the 
infrared-safety constraints). If the $n$-$\bar{n}$ symmetry is given up, four functions $F_A$ through $F_D$ and two 
parameters $n_{(A,B)}$ and $n_{(C,D)}$ would be required instead. The latter reflects the fact that the 
rapidity scaling of the observable may differ between the two light-like directions if the $n$-$\bar n$ 
symmetry is broken. The extension from two to four input functions can also easily be seen from the symmetry 
considerations in Section~\ref{sec:doublereal}. 

In practice, a broken $n$-$\bar{n}$ symmetry can already be emulated with the current version of 
\softserve~by averaging separate results for the $\{F_A,F_B\}$ and $\{F_C,F_D\}$ sets, similar to a procedure 
we use for the hemisphere soft function in Section~\ref{sec:results} (with special attention required to get 
the angular dependence correct). The substitutions to generate the relevant input functions are then
$y\to 1/y$ to derive $F_C$ from $F_A$, and $a\to 1/a$ or $b\to 1/b$ to derive both $F_B$ from $F_A$, and $F_D$ 
from $F_C$\footnote{In Figure~\ref{fig:Symmetries} regions C and D build up the upper layer of 
four cubes.}.


\subsection*{(A5) Multi-differential observables}

Soft functions for exclusive observables typically depend on more than one kinematic 
variable, requiring multiple Laplace transformations to resolve all 
distributions. We can in such cases choose the first Laplace variable $\tau_1$ to have dimension 
1/mass, and keep the remaining variables $\tau_i$ for $i\geq2$ dimensionless. 
This generalises our ansatz \eqref{eq:measure:general} for the measurement function to
\begin{align}
\mathcal{M}(\tau_1,\tau_2,\ldots;\lbrace k_{i} \rbrace) = 
\exp\big(-\tau_1\, \omega(\lbrace k_{i} \rbrace;\tau_2,\ldots)\,\big)\,,
\end{align}
from which one can derive the one- and two-emission measurement functions
$f(y_k,t_k;\tau_2,\ldots)$ and $F(a,b,y,t_k,t_l,t_{kl};\tau_2,\ldots)$ via the usual
procedure. In essence, multi-differential soft function can thus be computed by treating
the Laplace variables $\tau_i$ for $i\geq2$ as parameters, which need to be sampled over.
We demonstrate this strategy in Section~\ref{sec:results} by calculating two double-differential
soft functions: the hemisphere soft function in $e^+ e^-$ collisions and the soft function for 
exclusive Drell-Yan production.


\subsection*{(A6) Extended angular dependence}

Finally, we note that the angular parametrisation of our integrals, which we presented in detail in 
Section~\ref{sec:angles}, is made under the assumption of a back-to-back kinematic setup for the soft Wilson 
lines.  For dijet observables where this is not the case, the angular parametrisation is not sufficient, since
more dynamic angles can be resolved by the measurement function. Our master formulae therefore need to be 
revisited for non-back-to-back dijet observables, and we in fact already implemented such a generalisation
in the $N$-jet extension of our formalism~\cite{Bell:2018mkk}.

%% file: Numerics.tex
\section{Numerical implementation}
\label{sec:numerics}

The master formulae we derived in the preceding sections are in principle complete as 
they render the sources of all singularities manifest and allow for a numerical evaluation across 
a wide field of observables with only a few required properties. In fact, the master formulae can already be 
used to derive semi-analytic expressions for the anomalous dimensions and collinear anomaly 
exponents \citep{Bell:2018vaa}, although the matching corrections seem to be a bit out of reach in such an
approach due to their complexity. Solving the equations analytically is nevertheless possible in some isolated 
cases, in particular in the absence of a non-trivial angular dependence (see the C-parameter in Section~\ref{sec:results}).

Still, the master formulae are not yet ideally suited for a numerical implementation because of the presence
of an overlapping divergence in the variables $a$ and $t_{kl}$. The overlapping divergence could in principle 
be resolved by multiple sector decomposition steps (it produces three sectors due to the square in $(1-a)^2$) 
before the subtraction and the numerical evaluation. This is precisely what the programs 
{\tt{SecDec}} \citep{Carter:2010hi,Borowka:2012yc,Borowka:2015mxa} and its successor {\tt{pySecDec}} 
\citep{Borowka:2017idc} were designed to do, and we can indeed evaluate the master formulae with these 
programs\footnote{We in fact extensively used {\tt{(py)SecDec}} to cross-check our \softserve~numbers.}. 
 However, there are a few ways to improve on {\tt{(py)SecDec}} as regards our purposes, chiefly because there 
 are simplifications possible that {\tt{(py)SecDec}} --- as a program designed with a larger scope of 
 applications in mind --- cannot easily exploit.

The core insight that motivates our tailored numerical approach is the substitution \eqref{eq:uvsubstitution},
which removes the overlapping divergence in favour of a monomial divergence
$u^{-1-2\epsilon}$, at the cost of increasing the complexity of the integrand. Disentangling the divergence 
between $a$ and $t_{kl}$ means that all divergences are now present in monomial form, which makes a
subtraction and expansion procedure trivial. It should be noted that using this substitution can also
speed up {\tt{(py)SecDec}} runs, as the sector decomposition steps are of course no longer needed in this case
either anymore.

The program we subsequently wrote is called \texttt{SoftSERVE}, and it is publicly available at 
\url{https://softserve.hepforge.org}. \texttt{SoftSERVE} mainly implements the master formulae from this paper 
in C++ syntax, and integrates them using the Cuba library~\cite{Hahn:2004fe}. It is therefore subject to the 
same assumptions and capabilities of the formalism we developed in this paper, with one additional constraint: 
It is limited to strictly real measurement functions\footnote{The case of Fourier-space soft functions is 
special, and will be revisited at the end of this section and in Appendix~\ref{afourier}.}. Below, we will lay 
out the main reasons for forgoing \texttt{(py)SecDec} and writing a dedicated program, while delegating the 
technical details of the C++ implementation and the ultimate structure of the program to the 
\texttt{SoftSERVE} manual, which is provided alongside the program. 

The single most important motivation for writing a dedicated program is the problem of rounding errors in 
conjunction with plus-distributions. The rather technical details of this problem can be found in the 
\texttt{SoftSERVE} manual although, put succinctly, problems can arise if plus-distributions are integrated 
against functions that involve large cancellations in the limit originating from the plus-distribution. In such 
cases the plus-distribution can artificially inflate rounding errors. As this problem arises from rounding 
errors due to cancellations large enough to exhaust a typical \texttt{double} type variable's width (i.e. 
the number of stored digits), the solution is of course to use data types storing more digits. For our 
approach the problem can only appear in the measurement function, as all other factors appearing are 
free of large cancellations. The best and most efficient solution is therefore to write a program that can 
use (slow) multi-precision arithmetic for the calculations related to the measurement functions (and only if 
told to do so), and that evaluates everything else using (fast) \texttt{double} precision floating-point 
arithmetic.

Having written a dedicated wrapper for the master formulae, a vanilla run of our program 
(or a vanilla run of {\tt{(py)SecDec}} using the master formulae) reveals problems with numerical 
convergence that can be traced back to the appearance of square-root and logarithmic divergences. While these 
are analytically integrable divergences, a numerical approach that relies on sampling the integrand function 
runs into problems, particularly if the numerical integration involves adaptive variance-reduction techniques. 
Creative substitutions are again the solution to this problem. In particular, we can remove integrable 
divergences at both ends of a given integration over a variable $x\in[0,1]$ by substituting
\begin{equation}
x=1-\left(1-\xi^i\right)^j
\end{equation}
with suitably chosen parameters $i,j\geq 1$. Values of $i=2,4,\ldots$ then remove logarithmic, square-root,
\ldots divergences at $x=0$ while $j$ does so at $x=1$.  As an example, consider
\begin{align}
\int_0^1 \text{d}x \, \frac{\ln (1-x)+\ln x}{\sqrt{1-x}} 
= \int_0^1 \text{d}\xi\;
8\xi(1-\xi^2)\,\Big(4\ln(1-\xi^2)+\ln\big(1-(1-\xi^2)^4\big)\Big),
\end{align}
where we used $i=2$ and $j=4$. The expression on the right-hand side is thus no longer divergent at either 
endpoint of the integration domain after the substitution.

This performance improvement comes at a minor cost: In addition to the functions $F_{A,B}$ and the
parameter $n$, a second parameter $m$ is needed that can easily be derived from the functions $F_{A,B}$. 
As its significance is purely performance-related, we explain this in more detail in the \softserve~manual.

Applying these substitutions obscures the original definition of the observable and what ultimately serves 
as input into the program. We therefore inject a wrapping layer of substitutions between the user input and 
the numerics, to allow a user to input the relevant data using the physical parametrisation of the measurement
function in terms of the variables $a$, $b$, $y$, $t_k$, $t_l$ and $t_{kl}$, which have physical 
meaning and allow more thorough checking of the input for consistency. In addition, it is easy to derive the 
one-loop input from the two-loop input in this form using the infrared-safety constraints \eqref{eq:infrared:soft} and
\eqref{eq:infrared:collinear}, which we also exploit and implement. 

An important boost in performance furthermore arises due to the fact that all divergences giving rise to a 
pole are sourced from monomials like $y^{-1+n \epsilon}$, 
\begin{equation}
y^{-1+n \epsilon}= \frac{\delta(y)}{n \epsilon} + \left[\frac{1}{y}\right]_+ 
+ n \epsilon \left[\frac{\ln y}{y}\right]_+ + \,\ldots
\end{equation}
This feature means that the leading pole coefficients need only to be integrated over domains with reduced 
dimensionality, as the 
delta distributions render some variables spurious. The full integrands appearing at each order are thus sums 
over expressions that depend on a reduced set of variables. That sum then again depends on all 
variables\footnote{As a pedagogical example, consider a function $f(x,y)=g(x)+h(y)$ that depends on two 
variables, although its summands $g(x)$ and $h(y)$ depend only on one.}, and in \softserve~we simply relabel 
those variables to reduce the dimensionality, which can speed up the integrations significantly.

Finally, since we know exactly what \texttt{SoftSERVE} will be used for, we can supply it with scripts to facilitate the computations:
\begin{itemize}
\item \texttt{execsftsrv} runs the set of all colour structures for a given observable,
\item \texttt{sftsrvres} sums different contributions to the same colour structure,
\item \texttt{laprenorm} automates the renormalisation procedure outlined in Section~\ref{sec:renormalize},
\item \texttt{fourierconvert} allows us to treat Fourier-space soft functions.
\end{itemize}

All scripts are supplied in the initial \texttt{SoftSERVE \!0.9} release in a version without support for 
explicit calculation of the $C_F^2$ contribution, i.e.~a version designed for observables obeying NAE. 
These scripts are therefore indexed by a suffix \texttt{NAE}. Following an upcoming publication dealing 
with uncorrelated emissions~\cite{BRT} and the subsequent release of \texttt{SoftSERVE \!1.0}, 
unsuffixed scripts that run the calculations for all colour structures will be supplied as well.

It is crucial to note, however, that we are not limited in scope to observables obeying NAE. 
The $C_F T_F n_f$ and $C_F C_A$ colour structures for NAE-breaking observables can still be 
computed using the initial release of \texttt{SoftSERVE}, and the scripts above can still be used for this 
purpose. The $C_F^2$ piece must then, however, be calculated by some other means, as the results for $C_F^2$ 
presented by the \texttt{SoftSERVE \!0.9} release (or by using the \texttt{NAE} scripts in any released version) will 
simply not be correct -- they are merely proportional to the square of the one-loop result. 

The \texttt{SoftSERVE} package also comes supplied with a range of template observables serving as examples 
for an inexperienced user. Their salient features are described in the manual, and they include the observables
presented in Section~\ref{sec:results}. Additionally, two sets of integrator settings are supplied, 
the \emph{standard} and \emph{precision} settings. For almost all cases the standard settings will provide 
sufficient accuracy; they represent the recommended choice of settings. The precision settings are designed 
to increase the achieved accuracy by roughly one order of magnitude, but operate at the edge of 
\texttt{SoftSERVE}'s capabilities. They are therefore more prone to unexpected errors and more sensitive to 
suboptimal coding. The user is of course free to choose his/her own custom settings, as the input files 
provide access to the relevant flags and options of the Cuba integrators.

As a final comment, the \texttt{fourierconvert} script allows us to calculate real parts of Fourier-space soft 
functions. As referenced in Section~\ref{sec:measure}, we can treat observables with positive real part, which 
includes purely imaginary Fourier-space soft functions, where we assume an infinitesimal positive real part to 
make the analytic integration convergent. The numerical implementation is, on the other hand,
limited to strictly real measurement 
functions, which would exclude the purely imaginary Fourier-space soft functions. The precise nature of these problems 
and the analytic workaround are delegated to Appendix~\ref{afourier}, where we find that the real part of a 
Fourier-space soft function can be calculated from the absolute value of the Fourier-space measurement function, 
after a reshuffling of the Laurent series. The \texttt{fourierconvert} script performs this reshuffle, taking the correct propagation of uncertainties into account.

%% file: Results.tex
\section{Results}
\label{sec:results}

In this section we collect our results for multiple dijet soft functions that have been calculated using our 
novel algorithm and its implementation in {\texttt{SoftSERVE}}.  The relevant resummation ingredients for 
\si~soft functions were defined in Section~\ref{sec:ren:sceti}, and we present our results in the form
\begin{align} 
\gamma_0^S &= \gamma_0^{C_F} \,C_F\,,
\no\\[0.2em]
\gamma_1^S &= \gamma_1^{C_A} \,C_F C_A 
+ \gamma_1^{n_f} \,C_F T_F n_f
+ \gamma_1^{C_F} \,C_F^2\,,
\no\\[0.2em]
c_1^S &= c_1^{C_F} \,C_F\,,
\no\\[0.2em]
c_2^S &= c_2^{C_A} \,C_F C_A 
+ c_2^{n_f} \,C_F T_F n_f
+ c_2^{C_F} \,C_F^2\,.
\label{eq:scet-1:results}
\end{align}
As all soft functions we consider in this section obey the NAE theorem, we  have 
\mbox{$\gamma_1^{C_F}=0$}, $c_2^{C_F}=1/2(c_1^{C_F})^2$, and the missing NNLO coefficients can thus entirely 
be determined from the correlated-emission contribution.

For SCET-2 soft functions we similarly decompose the collinear anomaly exponent
\begin{align} 
d_1 &= d_1^{C_F} \,C_F\,,
\no\\[0.2em]
d_2 &= d_2^{C_A} \,C_F C_A 
+ d_2^{n_f} \,C_F T_F n_f
+ d_2^{C_F} \,C_F^2\,,
\label{eq:scet-2:results}
\end{align}
and we again have $d_2^{C_F}=0$ for soft functions that obey the NAE theorem. As explained in 
Section~\ref{sec:ren:scetii}, we also obtain results for the bare soft remainder function $W_0^S(\tau)$, 
which is however meaningless without its collinear counterpart $W_0^C(\tau)$ in the collinear anomaly 
framework. We therefore restrict our attention to the collinear anomaly exponent for \mbox{\sii}~observables 
in the following.

The numbers we present in this section can be reproduced using the template files that are provided in the 
\softserve~distribution (bearing in mind that Monte Carlo integrations yield statistical predictions). Whenever 
we quote numbers, we used the precision setting for the numerical integrations unless mentioned otherwise. 
For the plots we used the standard setting, since the uncertainties are anyway not visible on the scale of the
plots. The standard setting yields runtimes of a few minutes on a single 8-core machine, and it usually produces 
results with percent accuracy. Calculations using the precision setting, on the other hand, typically run for a
few hours and they yield numbers with per mille accuracy or better. The Cuba library supports parallelisation, 
and so these run-time estimates are highly dependent on the number of available cores. As the 
{\texttt{SoftSERVE}} manual contains instructions about how to set up the corresponding input files, we only 
present the relevant input functions along with the numerical results in this section.

For many of the observables we consider in this section, the soft function has been determined to NNLO before, 
either by an explicit calculation or via a fit to a fixed-order code. The available results provide useful cross 
checks for our code, and they allow us to assess its numerical performance. For those results that were not 
available in the literature, we checked our \softserve~numbers with independent {\tt{(py)SecDec}}~runs. In 
some cases the uncertainty estimate provided by the integrator is smaller than the rounding error introduced 
by the truncation to six digits, and in these cases we simply give an uncertainty for the last digit 
(i.e.~an error estimate of $10^{-6}$ for a result of $9.8696$ will show up as $9.8696(1)$).

\subsection{$e^+ e^-$ event shapes}

The formalism we have developed in this paper applies to soft functions that are defined in terms of two 
light-like directions, and $e^+ e^-$ event-shape variables that obey a hard-jet-soft factorisation theorem 
in the dijet limit are primary examples that fall into this class. In the following we present results for 
popular event shapes like C-parameter, thrust, and total jet broadening, but we also study the less known 
event shape angularities and a particular example of a double-differential observable.


\subsubsection*{C-parameter}

The C-parameter was one of the observables that we already introduced in Section~\ref{sec:measure}. Starting 
from the definition
\begin{align}
\omega^C(\lbrace k_{i} \rbrace) =
\sum_i\; \frac{k_i^+ k_i^-}{k_i^++k_i^-}\,,
\end{align}
the relevant input functions were given in Table~\ref{tab:measure},
 with $n=1$, $f(y_k,t_k)= 1/(1+y_k)$, and\footnote{Throughout this section we suppress the angular variables 
 in the arguments of the two-emission measure\-ment function if the observable does not depend on any of 
 these angles. We also remind the reader that the expression for the measurement function in region B is in 
 general not unique, due to the freedom in the definition \eqref{eq:rr:fb}.}
\begin{align}
F_A(a,b,y)&=
\frac{ab}{a(a+b)+(1+ab)y}+\frac{a}{a+b+a(1+ab)y}\,,
\no\\[0.2em]
F_B(a,b,y)&=
\frac{ab}{1+ab+a(a+b)y}+\frac{a}{a(1+ab)+(a+b)y}\,.
\end{align}
We then obtain for the C-parameter soft function
\begin{alignat}{2}
\gamma_0^{C_F}&=2\cdot 10^{-10}\pm 10^{-6}  \, &&\quad[0]\,, 
\no\\
\gamma_1^{C_A}&=15.7939(10) \, &&\quad[15.7945]\,, 
\no\\
\gamma_1^{n_f}&=3.90983(14) \, &&\quad[3.90981]\,, 
\no\\
c_1^{C_F}&=-3.28987(1) \, &&\quad[-3.28987]\,, 
\no\\
c_2^{C_A}&=-57.9814(35)  &&\quad[-58.16(26)]\,, 
\no\\
c_2^{n_f}&=43.8181(4)  \, &&\quad[43.74(6)]\,,
\end{alignat}
where the available results from~\cite{Hoang:2014wka} are listed in the square brackets. While the first four 
numbers are known analytically, the last two numbers were obtained in~\cite{Hoang:2014wka} via a fit to the 
\texttt{EVENT2} generator. Our numbers confirm these fit values, but they are significantly more accurate. 
For the C-parameter, we could actually derive analytic results following the strategy that we used for the 
derivation of anomalous dimensions in~\cite{Bell:2018vaa}, yielding
\begin{align}
c_2^{C_A}&=-\frac{2212}{81} - \frac{67\pi^2}{54} +\frac{13\pi^4}{15}
-\frac{770\zeta_3}{9} = -57.9757\,,
\no\\
c_2^{n_f}&=\frac{224}{81} + \frac{10\pi^2}{27} +\frac{280\zeta_3}{9} = 43.8182\,,
\end{align}
which confirms our \softserve~numbers at the $2\sigma$ level.


\subsubsection*{Thrust}

Thrust is the canonical event shape, which on the level of the soft function is defined as
\begin{align}
\omega^T(\lbrace k_{i} \rbrace) =
\sum_i\; \min(k_i^+,k_i^-)\,.
\end{align}
Thrust is again a \si~observable with $n=1$, $f(y_k,t_k)= 1$, and
\begin{align}
F_A(a,b,y)&=
\theta\left(\frac{a(a+b)}{1+ab}-y \right) + 
\theta\left(y-\frac{a(a+b)}{1+ab} \right) \left(\frac{a}{a+b}+\frac{ab}{(1+ab)y} \right),
\no\\[0.2em]
F_B(a,b,y)&=
\theta\left(\frac{a(1+ab)}{a+b}-y \right)+
\theta\left(y-\frac{a(1+ab)}{a+b} \right) \left(\frac{ab}{1+ab} + \frac{a}{(a+b)y} \right).
\label{eq:FaFb:thrust}
\end{align}
For the thrust soft function we find 
\begin{alignat}{2}
\gamma_0^{C_F}&=  2\cdot 10^{-12}\pm 10^{-10}  \, &&\quad[0]\,, 
\no\\
\gamma_1^{C_A}&=  15.7939(11) \, &&\quad[15.7945]\,, 
\no\\
\gamma_1^{n_f}&=  3.90987(15)  \, &&\quad[3.90981]\,, 
\no\\
c_1^{C_F}&=  -9.8696(1)  \, &&\quad[-9.8696]\,, 
\no\\
c_2^{C_A}&=   -56.5049(31) \, &&\quad[-56.4990]\,, 
\no\\
c_2^{n_f}&=  43.3906(4)  \, &&\quad[43.3905]\,,
\end{alignat}
which is in agreement with the analytic results from~\cite{Kelley:2011ng,Monni:2011gb}.

From the numerical perspective, observables like thrust are more problematic than the C-parameter because of 
the $1/y$ divergence in the second term of \eqref{eq:FaFb:thrust}, which is formally cut off by the step 
function. As the limit $y\rightarrow 0$ is a singular limit of the matrix element, the contribution from this 
phase-space region is enhanced and it therefore sources the numerical instability. As a consequence we find 
that observables like thrust are more prone to producing ill-defined results, and their uncertainty estimates 
are often less trustworthy. The presence of a plus-distribution in the subtraction also means that this problem 
cannot be solved, but merely mitigated, with some ideas laid out in the \softserve~manual.


\subsubsection*{Jet broadening}

The first \sii~observable in our list is total jet broadening, which has been studied within SCET 
in~\cite{Becher:2011pf,Chiu:2012ir,Becher:2012qc}. When the broadening is measured with respect to the 
thrust axis, it is well known that soft recoil effects complicate the resummation, which could be 
circumvented by choosing a different reference axis~\cite{Larkoski:2014uqa}. We will come back to a recoil-free 
definition of jet broadening later, but for the moment we simply switch off the recoil effects in the original 
thrust-axis definition since this allows us to compare our results to an existing NNLO calculation. We thus 
start from the definition
\begin{align}
\omega^B(\lbrace k_{i} \rbrace) =
\frac12 \,\sum_i\; \sqrt{k_i^+ k_i^-}\,,
\end{align}
which yields $n=0$ as required for a \sii~observable, along with $f(y_k,t_k)= 1/2$ and
\begin{align}
F_A(a,b,y)&= F_B(a,b,y) =
\sqrt{\frac{a}{(1+ab)(a+b)}}\,\frac{(1+b)}{2}\,.
\end{align}
Using \softserve~we then find for the collinear anomaly exponent 
\begin{alignat}{2}
d_1^{C_F}&=  -5.5452(1)  \, &&\quad[-5.5452]\,, 
\no\\
d_2^{C_A}&=  7.03648(85)  \, &&\quad[7.03605]\,, 
\no\\
d_2^{n_f}&=  -11.5393(2)  \, &&\quad[-11.5393]\,,
\end{alignat}
which is in excellent agreement with the analytic results from~\cite{Becher:2012qc}.


\subsubsection*{Angularities}

\begin{figure}[t]
\centering
\hspace{1mm}
\includegraphics[width=0.435\textwidth]{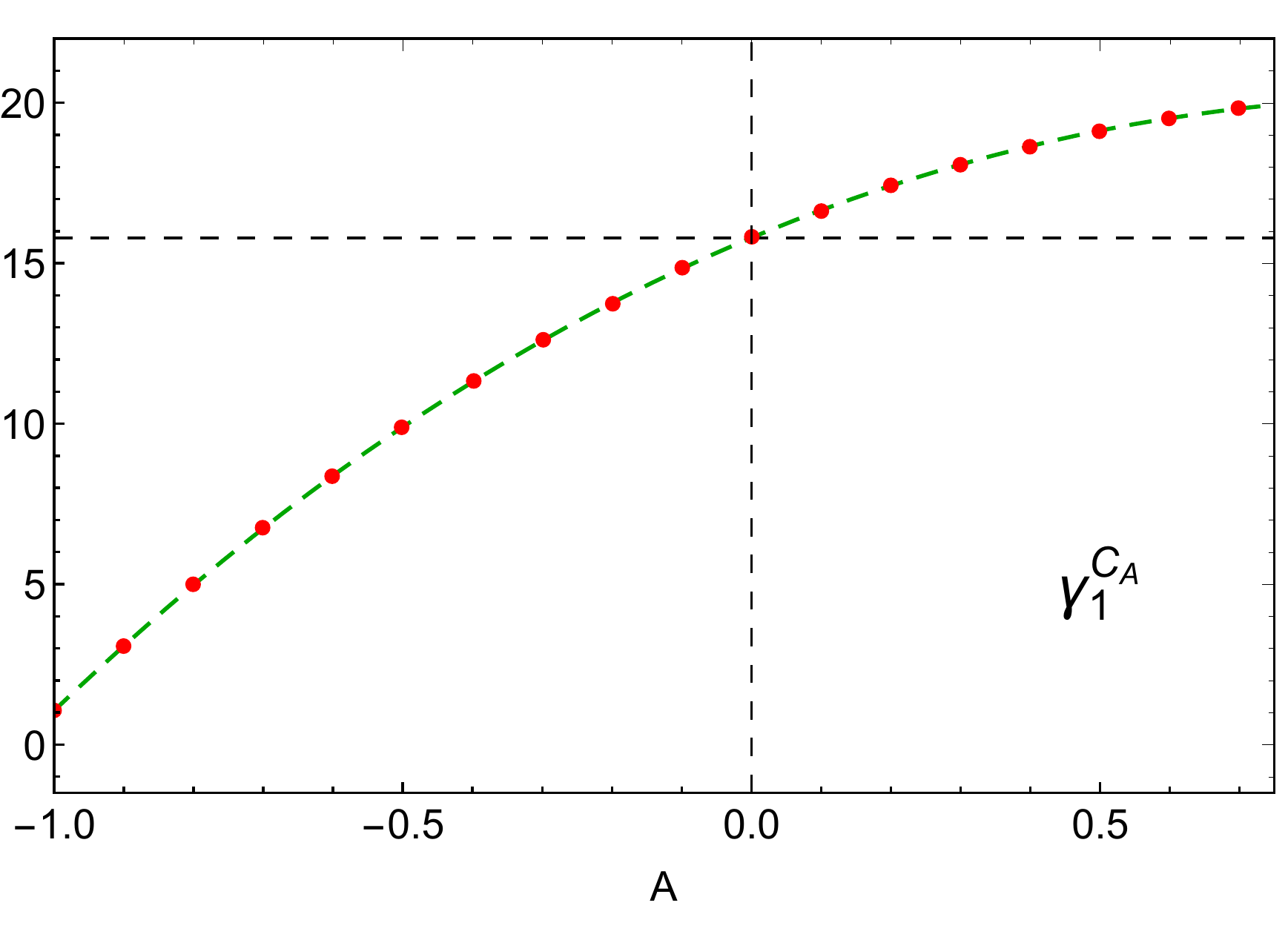}
\hspace{3mm}
\includegraphics[width=0.435\textwidth]{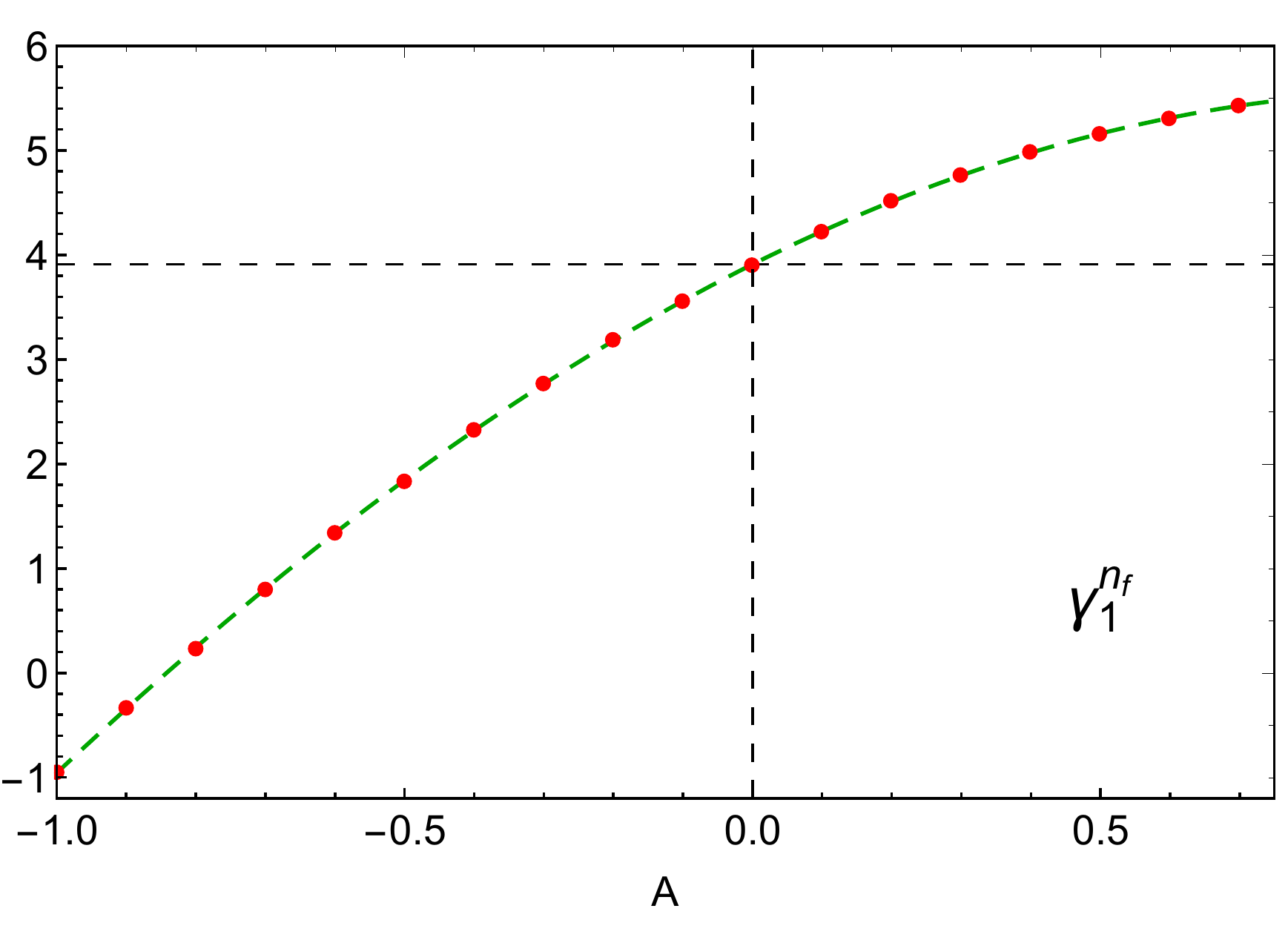}
\includegraphics[width=0.45\textwidth]{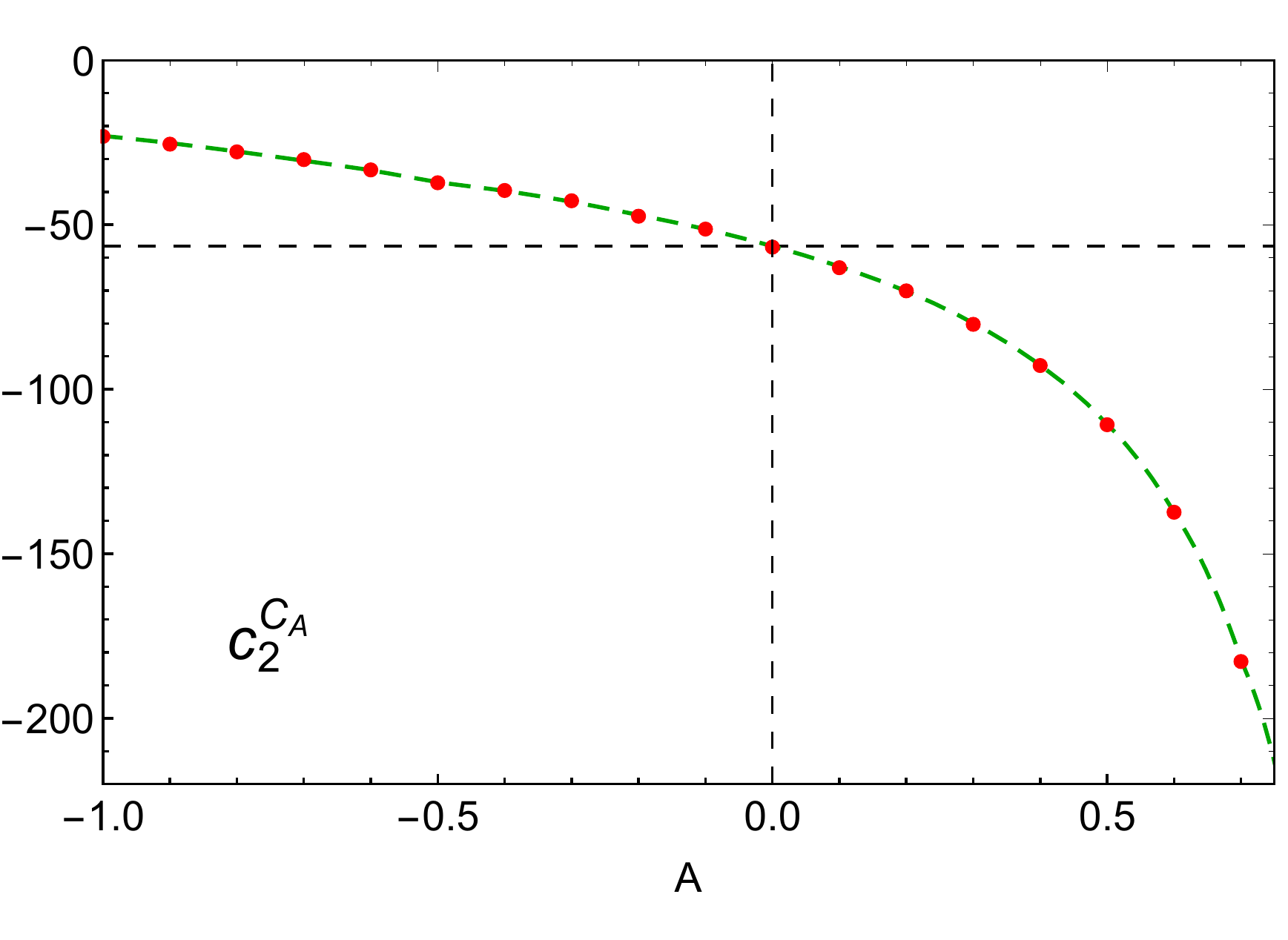}
\hspace{2mm}
\includegraphics[width=0.44\textwidth]{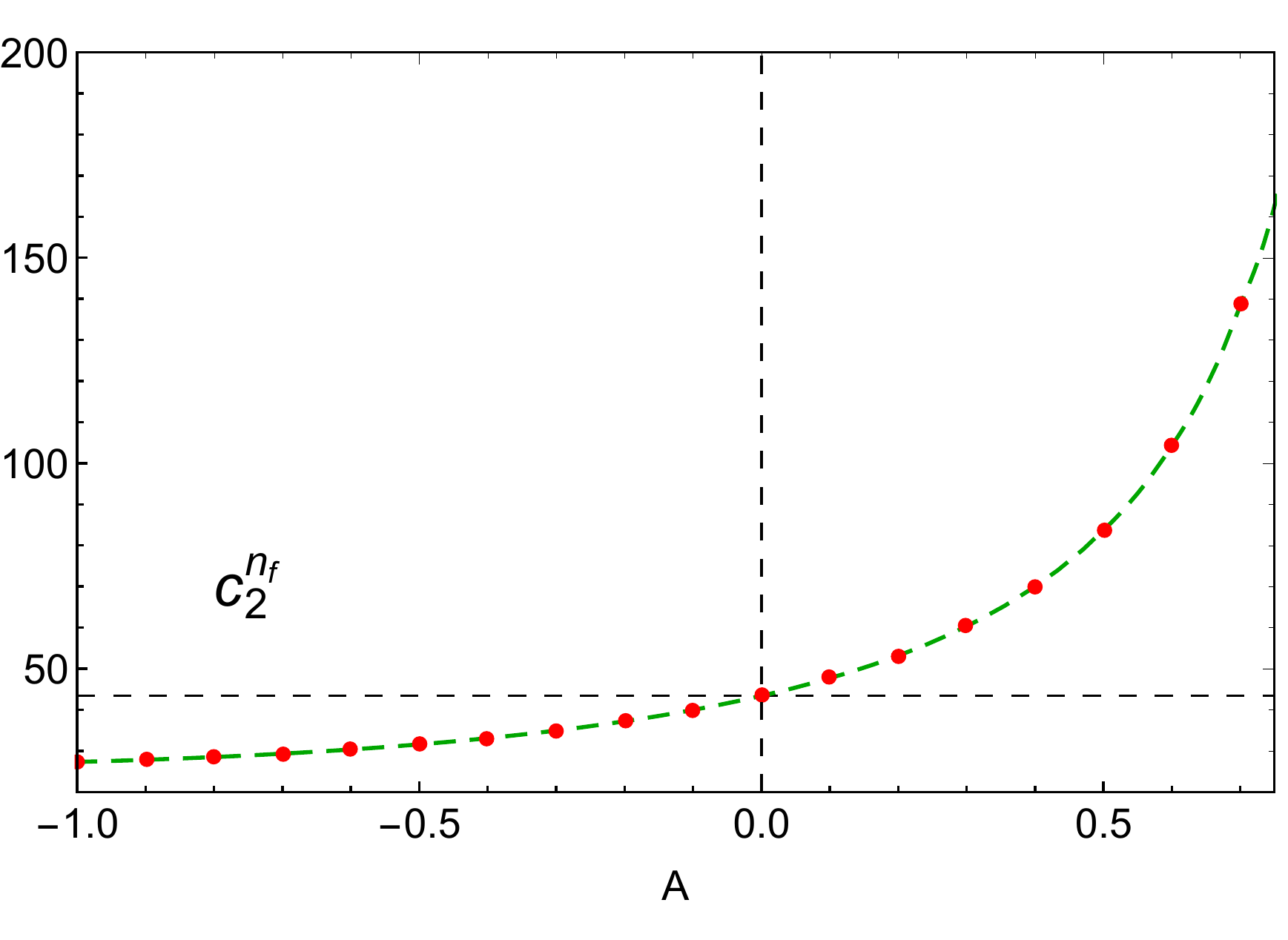}
\caption{Two-loop anomalous dimension and finite term of the renormalised angularity soft function. Red dots 
indicate values calculated with {\texttt{SoftSERVE}} and green dashed lines represent an interpolation through 
these numbers. We also highlight the thrust numbers for $A=0$, which are known analytically 
from~\cite{Kelley:2011ng,Monni:2011gb}.}
\label{fig:ANGgraphs}
\end{figure}

Angularities represent a generalisation of the thrust and broadening event shapes that depend on a continuous 
parameter $A$. According to their standard definition, the angularities are measured with respect to the thrust 
axis, and they are defined as
\begin{align}
\omega^A(\lbrace k_{i} \rbrace) =
\sum_i\; \bigg(\theta(k_i^--k_i^+){(k_i^+)}^{1-A/2} {(k_i^-)}^{A/2} + \theta(k_i^+-k_i^-) {(k_i^+)}^{A/2}{(k_i^-)}^{1-A/2} \bigg)\,,
\end{align}
which reduces to thrust for $A=0$, and is proportional to total broadening for $A=1$. For values of $A<1$ 
considered here, the angularities fall into the \si~class with $n=1-A$, $f(y_k,t_k)= 1$, and
\begin{align}
F_A(a,b,y)&=
\theta\left(\frac{a(a+b)}{1+ab}-y \right) \frac{a + a^A b}{a + b} \;
\bigg(\frac{a + b}{a(1 + a b)} \bigg)^{A/2} 
\\
&\quad+ 
\theta\left(y-\frac{a(a+b)}{1+ab} \right) 
\bigg[\frac{a}{a+b}\bigg(\frac{a + b}{a(1 + a b)} \bigg)^{A/2}
+\frac{ab\, y^{A-1}}{(1+ab)} \bigg(\frac{1 + a b}{a(a+b)} \bigg)^{A/2} \bigg]\,,
\no\\[0.2em]
F_B(a,b,y)&=
\theta\left(\frac{a(1+ab)}{a+b}-y \right) \frac{a^A + a b}{1 + a b} \;
\bigg(\frac{1 +a b}{a(a + b)} \bigg)^{A/2}
\no\\
&\quad+
\theta\left(y-\frac{a(1+ab)}{a+b} \right) 
\bigg[\frac{ab}{1+ab} \bigg(\frac{1 + a b}{a(a+b)} \bigg)^{A/2}
+ \frac{a\, y^{A-1}}{(a+b)} \bigg(\frac{a + b}{a(1 + a b)} \bigg)^{A/2}\bigg]\,.
\no
\end{align}
One easily verifies that these expressions reduce to those of thrust in the corresponding limit, 
$A \rightarrow 0$. The angularity soft function has been computed to NLO in~\cite{Hornig:2009vb}, where it was 
found that $\gamma_0^{S}(A) = 0$ and $c_1^{C_F}(A) = - \pi^2/(1-A)$. Our \softserve~numbers indeed confirm these 
expressions and our results for the two-loop soft anomalous dimension and the finite term of the renormalised 
soft function are displayed as a function of the parameter $A$ in Figure~\ref{fig:ANGgraphs}. While our results 
for the constants $c_2^{C_A}$ and $c_2^{n_f}$ are new (preliminary results were reported in~\cite{Bell:2015lsf}), 
we already derived the two-loop soft anomalous dimension in~\cite{Bell:2018vaa}. Our results have in fact 
already been utilised to resum the angularities distribution to NNLL accuracy in~\cite{Procura:2018zpn} and to 
NNLL$'$ accuracy in \citep{Bell:2018gce}, and they thus represent the first phenomenological application of 
\softserve.

As mentioned in the previous paragraph, the angularities share a numerical instability with thrust. We therefore 
find that some \softserve~runs using the precision setting produce ill-defined results, and the error bars are 
in general also less robust than for other observables.

\subsubsection*{Broadening-axis angularities}

\begin{figure}[t]
\centering
\hspace{0mm}
\includegraphics[width=0.435\textwidth]{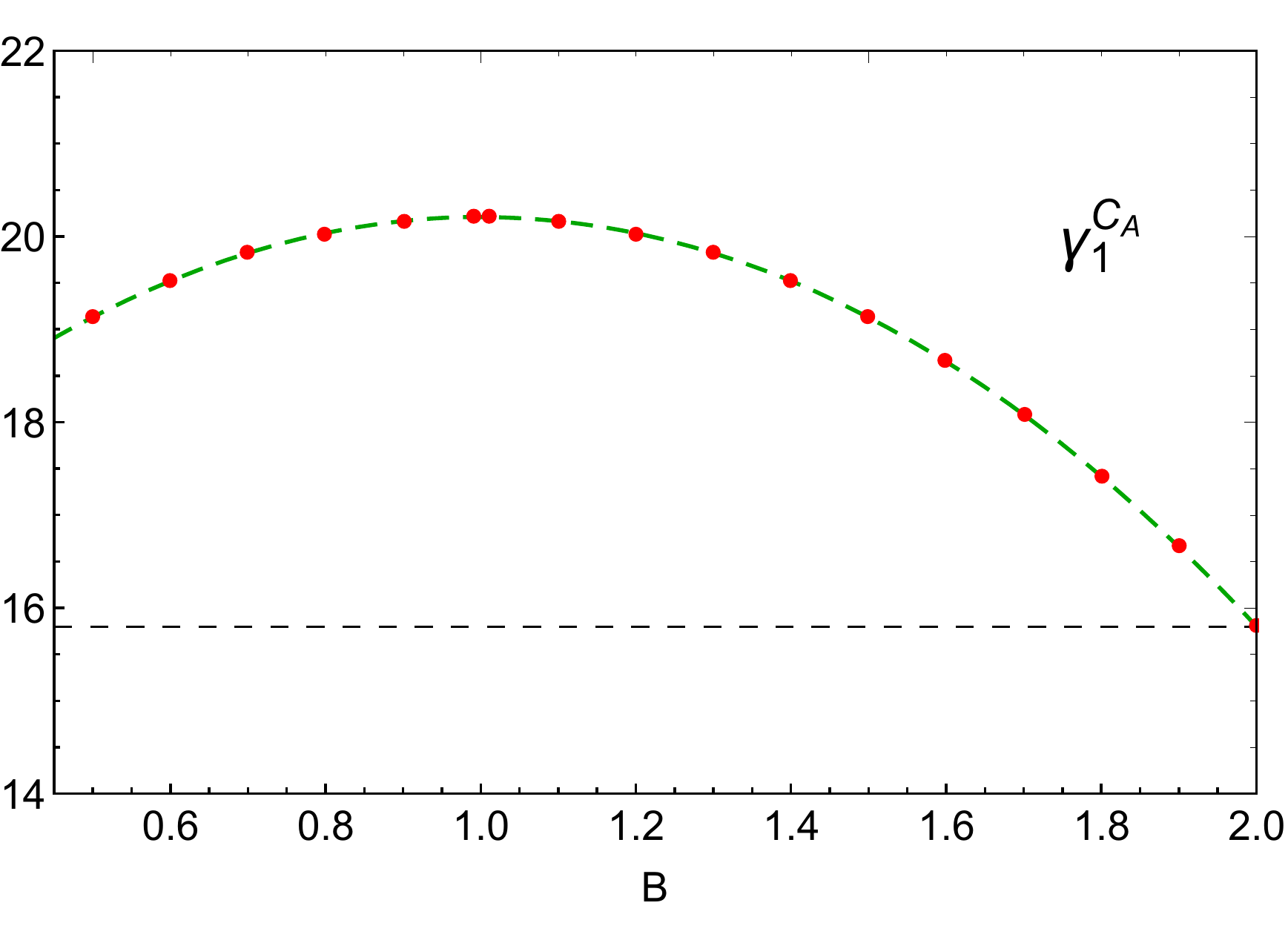}
\hspace{3mm}
\includegraphics[width=0.435\textwidth]{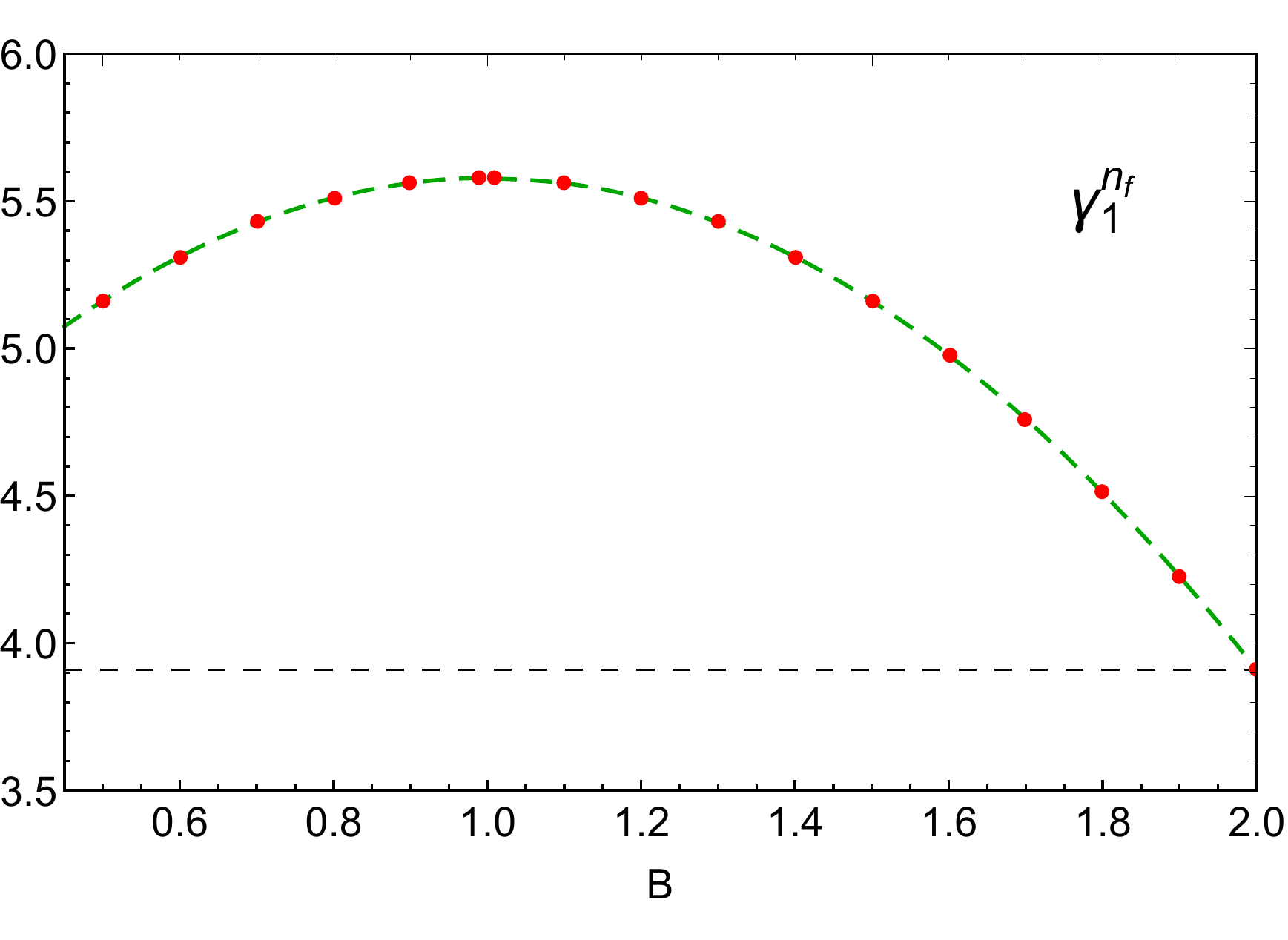}
\includegraphics[width=0.45\textwidth]{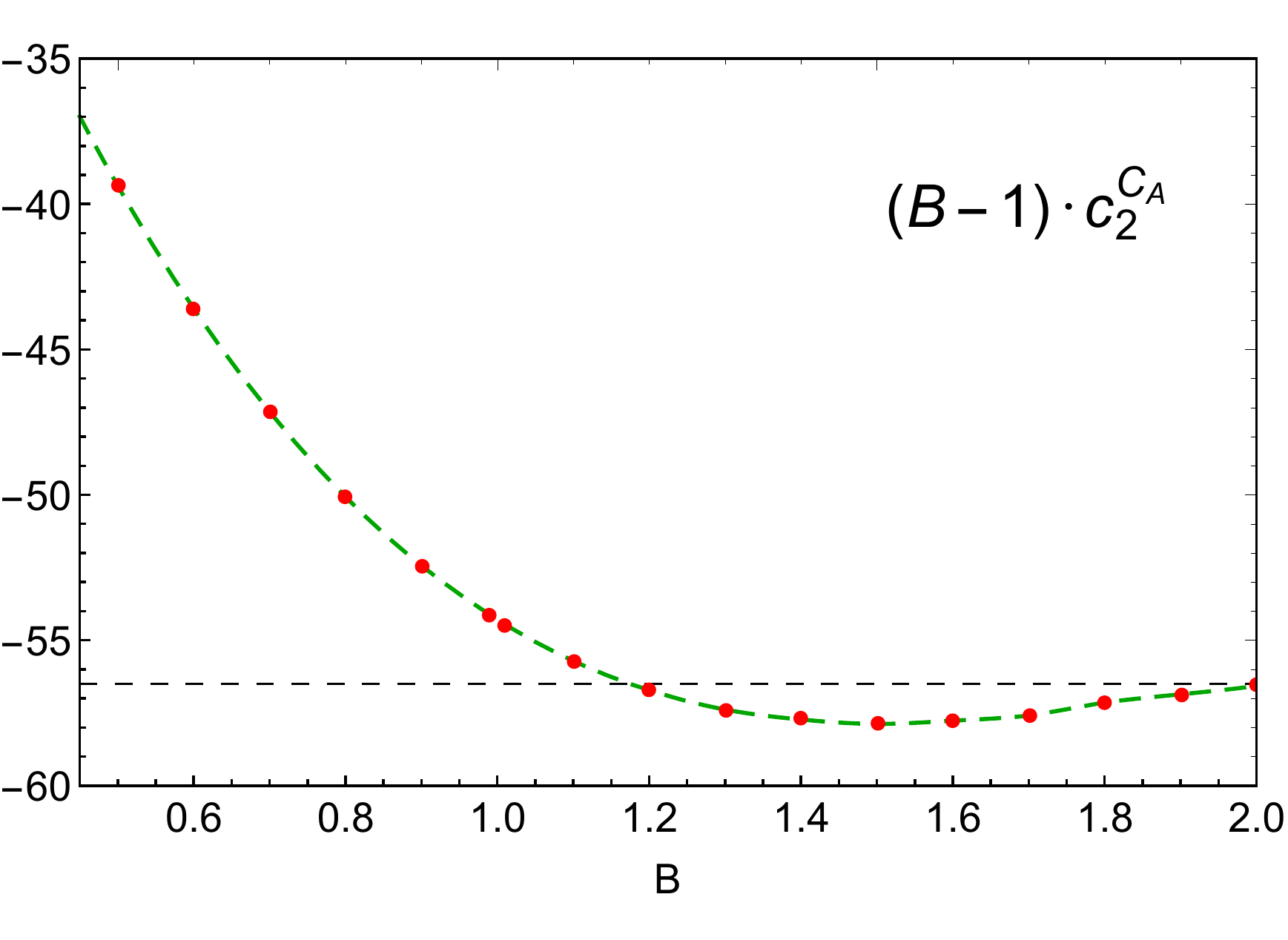}
\hspace{2mm}
\includegraphics[width=0.45\textwidth]{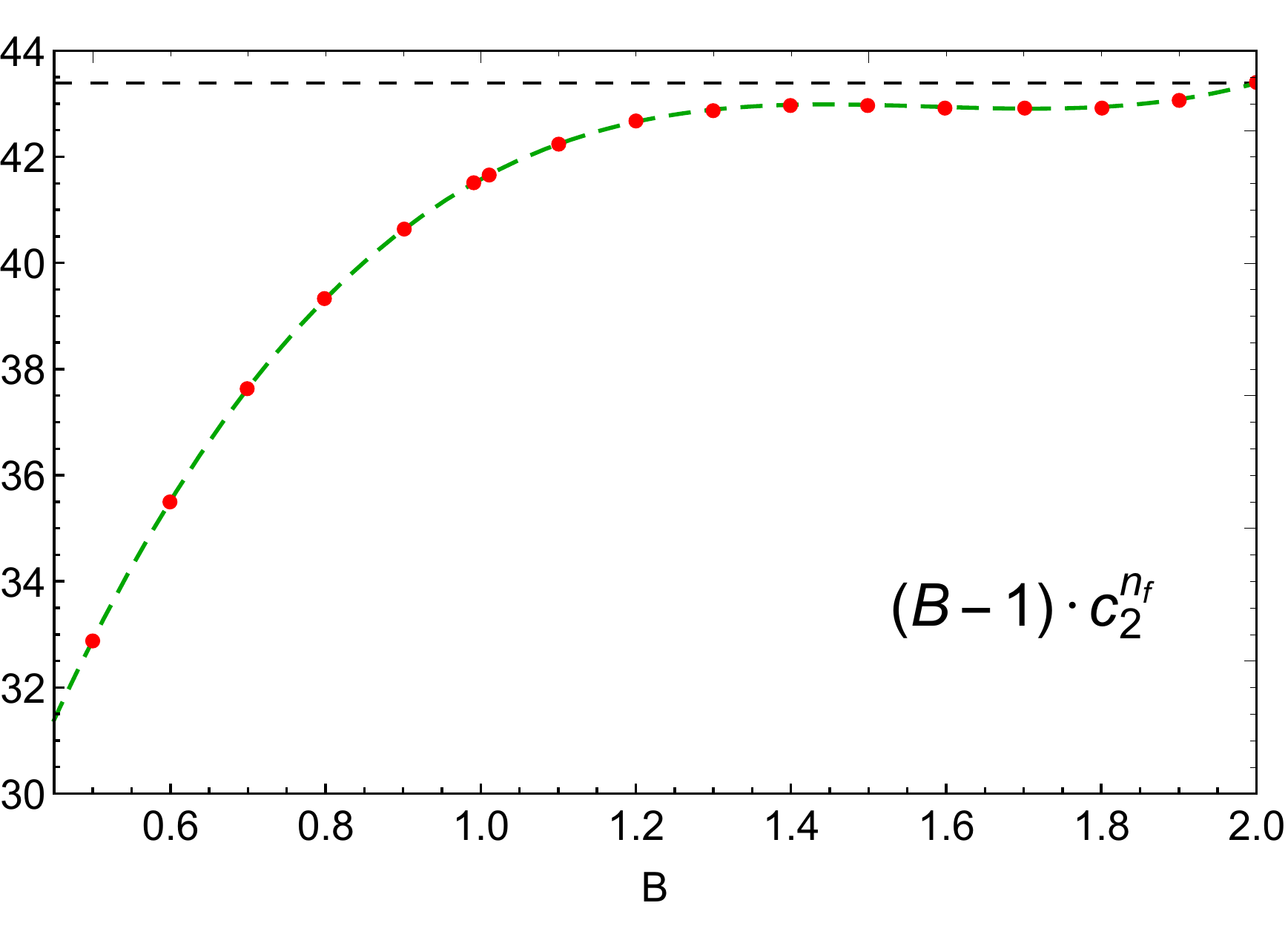}
\caption{The same as in Figure \ref{fig:ANGgraphs}, but for angularities defined with respect to the broadening 
axis (as opposed to the thrust axis).}
\label{fig:BANGgraphs}
\end{figure}

Similar to the total broadening event shape, the thrust-axis angularities are subject to soft recoil effects 
which become increasingly more important in the limit $A\to1$. One way of avoiding these complications consists 
in defining the angularities with respect to the broadening axis rather than the thrust axis. 
Following~\cite{Larkoski:2014uqa}, we then start from
\begin{align}
\omega^{BA}(\lbrace k_{i} \rbrace) =
\sum_i\; \bigg\{\Big(\theta(k_i^--k_i^+){(k_i^+)}^{B/2}  + \theta(k_i^+-k_i^-) {(k_i^-)}^{B/2}\Big) {(k_i^+ + k_i^-)}^{1-B/2}\bigg\}\,,
\end{align}
where we omitted a factor $2^{B-1}$, which makes the expression for the one-loop soft anomalous dimension 
easier.\footnote{\label{foot:normalisation}According to \eqref{eq:measure:general}, a change in the 
normalisation of $\omega(\lbrace k_{i} \rbrace)$ always results in a rescaling of the Laplace parameter 
$\tau$. This in turn  modifies the numerical value of the non-cusp anomalous dimension in the RGE \eqref{eq:RG} 
and the $d_i$ coefficients in \eqref{eq:d1d2}. We are, however, free to turn the argument around and choose 
the normalisation of $\omega(\lbrace k_{i} \rbrace)$ such that the expression of the one-loop anomalous 
dimension becomes simple.}  Thrust corresponds to $B=2$ in this notation, and we are interested 
in values $B\leq2$ that extend beyond the broadening limit $B=1$. The broadening-axis angularities are thus 
characterised by $n=B-1$, $f(y_k,t_k)= (1+y_k)^{1-B/2}$, and
\begin{align}
&F_A(a,b,y)=
\theta_1\;
\frac{b\big(a(a+b)+(1+ab)y\big)^{1-B/2} + a^{B/2}\big(a+b+a(1+ab)y\big)^{1-B/2}}{\left(a+b\right)\left(1+ab\right)^{1-B/2}}
\\[0.2em]
&\qquad+ (1-\theta_1)\,
\bigg\{\frac{a^{B/2}\big(a+b+a(1+ab)y\big)^{1-B/2}}{(a+b)(1+ab)^{1-B/2}} +
\frac{a^{B/2}\,b\big(a(a+b)+(1+ab)y\big)^{1-B/2}}{y^{B/2}(1+ab)(a+b)^{1-B/2}}  \bigg\}\,,
\no\\[0.4em]
&F_B(a,b,y)=
\theta_2\;
\frac{\big(a(1+ab)+(a+b)y\big)^{1-B/2}+ a^{B/2}\,b\big(1+ab+a(a+b)y\big)^{1-B/2}}{(1+ab)(a+b)^{1-B/2}}
\no\\[0.2em]
&\qquad+
(1-\theta_2)\,
\bigg\{\frac{a^{B/2}\,b\big(1+ab+a(a+b)y\big)^{1-B/2}}{(1+ab)(a+b)^{1-B/2}}
+\frac{a^{B/2} \big(a(1+ab)+(a+b)y\big)^{1-B/2}}{y^{B/2}(1+ab)^{1-B/2}(a+b)}\bigg\}\,,
\no
\end{align}
where we defined
\begin{align}
\theta_1  = \theta\left(\frac{a(a+b)}{1+ab}-y \right) , 
\qquad\qquad
\theta_2  = \theta\left(\frac{a(1+ab)}{a+b}-y \right) . 
\end{align}
The considered soft function is known to NLO from~\cite{Larkoski:2014uqa}, and our results confirm these 
findings with $\gamma_0^{S}(B) = 0$ and $c_1^{C_F}(B) = (B^2-3B-1)/(B-1)\, \pi^2/3$. Our NNLO results, on the 
other hand, are new and they are shown in Figure~\ref{fig:BANGgraphs} as a function of the parameter $B$. While 
we cannot run \softserve~in the SCET-1 mode for $B=1$, the plots illustrate that we can compute the anomalous 
dimension and the matching correction sufficiently close to the SCET-2 limit -- done here using $B=0.99$ and 
$B=1.01$ -- and we can even interpolate between those points to convert the soft anomalous dimension 
into the SCET-2 anomaly exponent using the formulae provided in~\cite{Bell:2018vaa}.


\subsubsection*{Hemisphere masses}

\begin{figure}[t]
\centering
\includegraphics[width=0.45\textwidth]{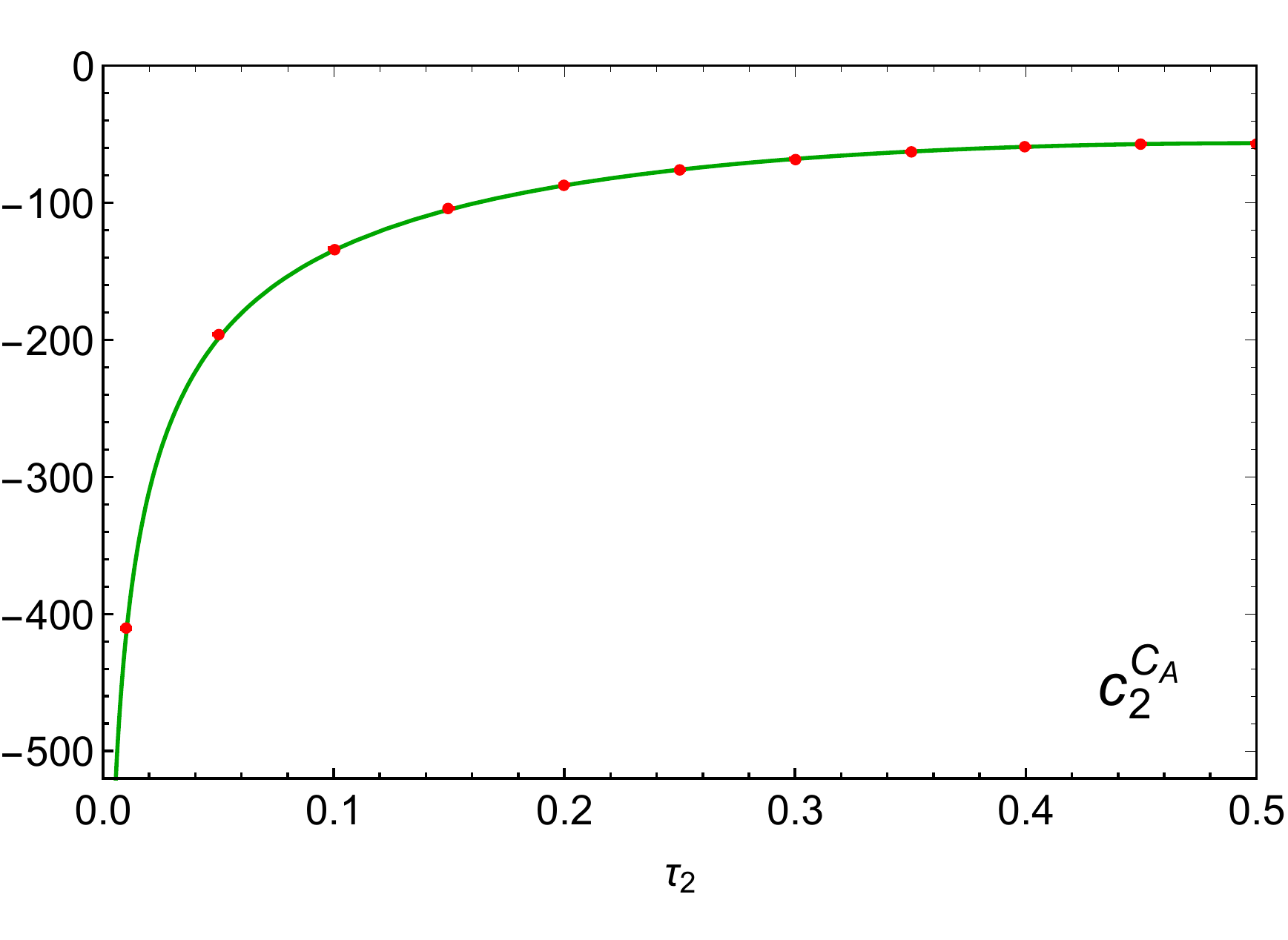}
\hspace{2mm}
\includegraphics[width=0.44\textwidth]{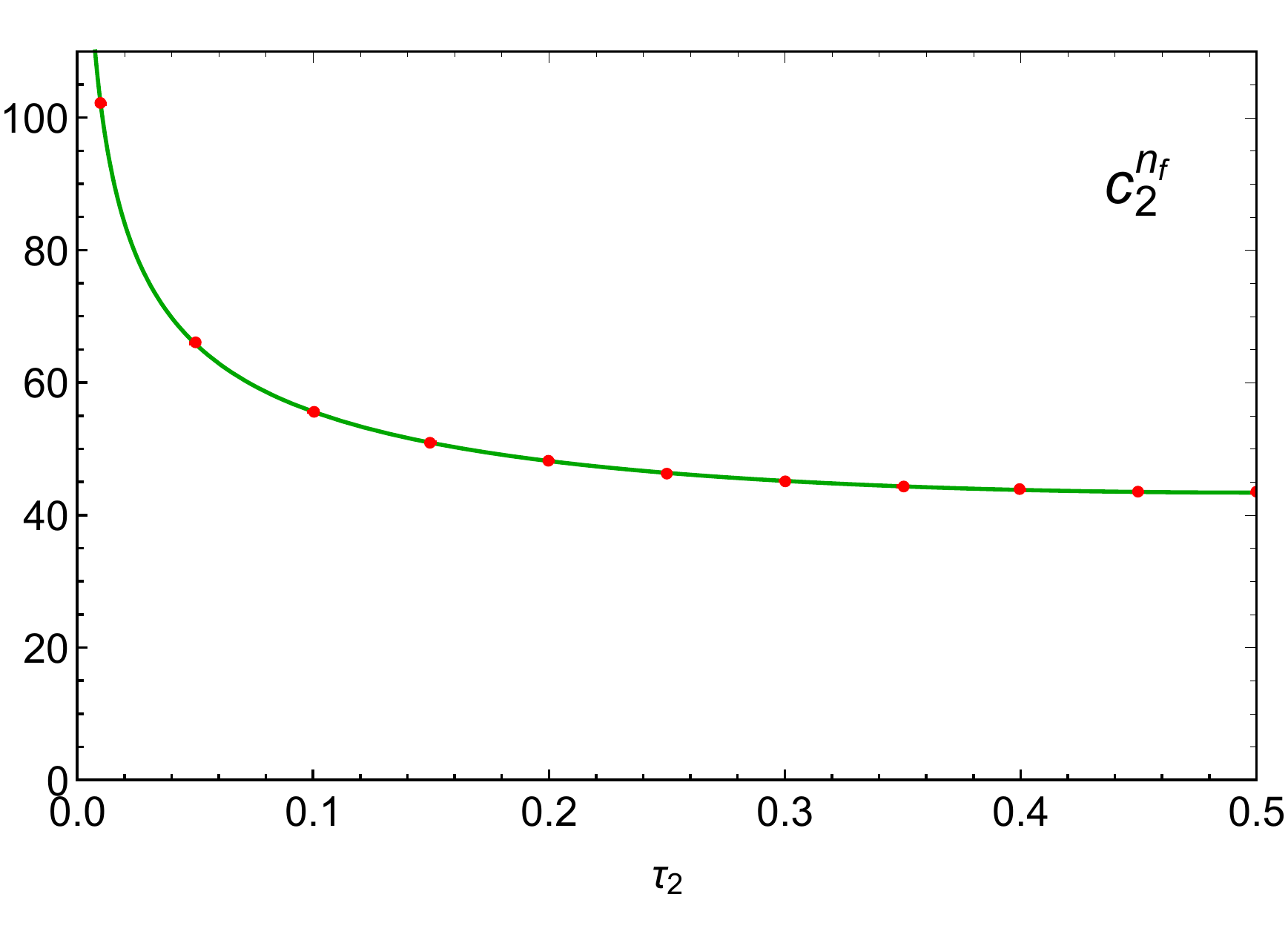}
\caption{Two-loop finite term of the renormalised hemisphere soft function. Red dots indicate values 
calculated with {\texttt{SoftSERVE}} and the green solid line represents the analytic result of 
\citep{Kelley:2011ng}.}
\label{fig:HEMMASSgraphs}
\end{figure}

We finally consider the hemisphere soft function~\cite{Hornig:2011iu,Kelley:2011ng}, which represents 
our first example of a double-differential observable.  In this case, we take two Laplace transformations 
with respect to the hemisphere masses $M_L$ and $M_R$, and we denote the respective Laplace variables by 
$\tau_L$ and $\tau_R$. We furthermore introduce the variables 
\begin{align}
\tau_1 = \tau_{L} + \tau_{R}\,, \qquad\qquad 
\tau_2 = \frac{\tau_{L}}{\tau_{L}+\tau_{R}}\,,
\end{align}
such that the dependence on the dimensionful variable $\tau_1$ factorises by construction. The observable, on 
the other hand, now depends on the second Laplace variable $\tau_2$ via 
\begin{align}
 \label{eq:MLMR:def}
\omega^{M_{L}M_R}(\lbrace k_{i} \rbrace;\tau_2) =\frac{1}{\sqrt{\tau_2\bar\tau_2}}\;
\sum_i\; \bigg(\theta(k_i^--k_i^+) \, \tau_2\, k_i^+ 
+ \theta(k_i^+-k_i^-) \,\bar \tau_2\, k_i^- \bigg)
\end{align}
with $\bar \tau_2=1-\tau_2$, and we have furthermore adjusted the normalisation for later convenience -- 
see footnote~\ref{foot:normalisation}. The observable is thus described by 
$n=1$, $f(y_k,t_k;\tau_2)= \sqrt{\tau_2/\bar \tau_2}$, and
\begin{align}
F_A(a,b,y;\tau_2)&=
\theta\left(\frac{a(a+b)}{1+ab}-y \right) \!\sqrt{\frac{\tau_2}{\bar \tau_2}}+ 
\theta\left(y-\frac{a(a+b)}{1+ab} \right) \!
\left(\frac{a\sqrt{\tau_2}}{(a+b)\sqrt{\bar \tau_2}} 
+\frac{ab\sqrt{\bar \tau_2}}{(1+ab)y\sqrt{\tau_2}} \right),
\no\\[0.2em]
F_B(a,b,y;\tau_2)&=
\theta\left(\frac{a(1+ab)}{a+b}-y \right) \!\sqrt{\frac{\bar \tau_2}{\tau_2}}+
\theta\left(y-\frac{a(1+ab)}{a+b} \right) \!
\left(\frac{ab\sqrt{\bar \tau_2}}{(1+ab)\sqrt{\tau_2}} + \frac{a \sqrt{\tau_2}}{(a+b)y\sqrt{\bar \tau_2}} \right),
\end{align}
from which we recover the thrust expressions for $\tau_2=1/2$. The analysis of the hemisphere soft function 
is, moreover, complicated by the fact that the definition in  \eqref{eq:MLMR:def} is not symmetric under 
$n$-$\bar n$ exchange, since the roles of $M_L$ and $M_R$ -- and hence $\tau_2$ and $\bar \tau_2$ -- 
are interchanged under this symmetry. This seems to be in conflict with assumption (A4), but we can simply 
restore this symmetry by  substituting  
\begin{align}
\label{eq:hemi:average}
S_0(\tau_1,\tau_2) \to  \frac{1}{2} \Big( S_0(\tau_1,\tau_2) + S_0(\tau_1,\bar \tau_2)\Big)
\end{align}
on the level of the bare soft function\footnote{This procedure illustrates the comment made in 
Section~\ref{sec:generalize} concerning the emulation of a broken $n$-$\bar n$-symmetry, 
with $S_0(\tau_1,\bar \tau_2)$ effectively representing the contribution from regions C and D.}.
As the soft anomalous dimension in \eqref{eq:gamma:coeffs} and the matching correction in \eqref{eq:c:coeffs}
depend only linearly on the $C_F C_A$ and $C_F T_F n_f$ pole coefficients, the average in 
\eqref{eq:hemi:average} can actually be directly implemented on the level of these quantities as well.

Due to the particular normalisation in \eqref{eq:MLMR:def}, the anomalous dimension of the hemisphere 
soft function matches the one for thrust,  and we display the finite terms of the two-loop soft function 
as a function of the Laplace variable $\tau_2$ in Figure~\ref{fig:HEMMASSgraphs} (the interval 
$\tau_2 \in [0.5,1]$ is just a mirror image of $\tau_2 \in [0,0.5]$ due to the symmetrisation procedure). 
In addition, we have Laplace-transformed the analytic expressions from~\cite{Kelley:2011ng}, and the result 
is shown in Figure~\ref{fig:HEMMASSgraphs} by the solid lines. As can be seen from the plots, we observe 
perfect agreement for this double-differential observable, and we of course also confirm the corresponding 
one-loop expression with $c_1^{C_F}(\tau_2) =  -8\,  \mathrm{artanh}^2(1-2\tau_2)-\pi^2$.

As a generalisation of thrust, the hemisphere soft function is prone to numerical instabilities, and for 
two out of twenty runs that were needed to produce the plots in Figure~\ref{fig:HEMMASSgraphs}, two resulted 
in ill-defined expressions for the default setting of the numerical integrator. This problem could, however, 
be remedied by increasing the value of the Divonne \texttt{border} parameter from $10^{-8}$ to 
$10^{-6}$.  While doing so, we verified that the variation of this parameter does not introduce
systematic uncertainties that are relevant at the quoted accuracy.


\subsection{Hadron-collider observables}

Another important class of soft functions that fall into the considered `dijet' category are hadron-collider 
observables for which central jets are vetoed, either via a kinematical restriction or an explicit jet 
algorithm. As the latter are usually not consistent with NAE, we postpone their discussion to a future 
study~\cite{BRT}, and we focus on soft functions that are relevant for soft-gluon resummation, 
transverse-momentum resummation, and hadronic event shapes in the following. For hadron-collider soft 
functions, only $q\bar q$--initiated processes are of the type \eqref{eq:softfun:definition}, whereas 
the Wilson lines for other channels refer to different colour representations. The soft function does, 
however, obey Casimir scaling to the considered order, and the results can therefore easily be translated 
to other channels by rescaling the expressions in \eqref{eq:scet-1:results} and \eqref{eq:scet-2:results} 
with $C_i/C_F$, where $C_i$ is the Casimir operator of the said channel.


\subsubsection*{Threshold Drell-Yan production}

The production of a lepton pair $pp \rightarrow l_{1}l_{2}X$ at threshold represents the simplest 
hadron-collider soft function we can consider. We already encountered its definition in 
Section~\ref{sec:measure}, where we found that
\begin{align}
\omega^{DY}(\lbrace k_{i} \rbrace) =
\sum_i\; (k_i^++k_i^-)\,,
\end{align}  
which yields $n=-1$, $f(y_k,t_k)= 1+y_k$, and
\begin{align}
F_A(a,b,y)&= F_B(a,b,y) = 1+y\,.
\end{align}
Using \softserve, we then find
\begin{alignat}{2}
\gamma_0^{C_F}&=  8\cdot 10^{-10}\pm 10^{-6}  \, &&\quad[0]\,, 
\no\\ 
\gamma_1^{C_A}&=  15.7941(10)  \, &&\quad[15.7945]\,, 
\no\\
\gamma_1^{n_f}&=  3.90983(14)  \, &&\quad[3.90981]\,, 
\no\\
c_1^{C_F}&=  3.28987(1)  \, &&\quad[3.28987]\,, 
\no\\
c_2^{C_A}&=  6.81309(280)  \, &&\quad[6.81287]\,, 
\no\\
c_2^{n_f}&=  -10.6857(5)  \, &&\quad[-10.6857]\,,
\end{alignat}
which nicely agrees with the analytic NNLO results from~\citep{Belitsky:1998tc}.


\subsubsection*{W-production at large transverse momentum}

We next consider the soft function for $W$-production at large transverse momentum from~\cite{Becher:2012za}. 
Although this is a soft function that involves more than two Wilson lines, it represents \emph{de facto} a 
dijet soft function, since the gluon attachments to the Wilson line $S_{n_J}$ vanish up 
to NNLO and we are furthermore free to choose $n_1\cdot n_2=2$ along with 
$n_1\cdot n_J=n_2\cdot n_J=2$ due to rescaling invariance of the Wilson lines~\cite{Becher:2012za}. 
The vector $n_J^\mu$ now introduces a non-trivial angular dependence with
\begin{align}
\omega^{W}(\lbrace k_{i} \rbrace) =
\sum_i\; n_J \cdot k_i =
\sum_i\; \Big(k_i^++k_i^- - 2\sqrt{k_i^+ k_i^-}\cos\theta_i\Big)\,,
\end{align}  
where $\theta_i=\sphericalangle(\vec{n}_{J}^{\perp},\vec{k}_{i}^{\perp})$. Up to NNLO the soft function is 
thus characterised by $n=-1$, $f(y_k,t_k) = 1 + y_k - 2\sqrt{y_k}(1-2t_k)$, and
\begin{align}
F_A(a,b,y,t_k,t_l,t_{kl}) &=  F_B(a,b,y,t_k,t_l,t_{kl}) 
\no\\[0.2em]
&= 
1 + y - 2 \;\sqrt{\frac{a y}{(1+a b)(a+b)}} \;
\Big(b (1-2t_k) + 1-2t_l\Big),
\end{align}
and we obtain
\begin{alignat}{2}
\gamma_0^{C_F}&=  7\cdot 10^{-9}\pm 2\cdot 10^{-6}  \, &&\quad[0]\,, 
\no\\
\gamma_1^{C_A}&=  15.7943(24)  \, &&\quad[15.7945]\,, 
\no\\
\gamma_1^{n_f}&=  3.90987(21)  \, &&\quad[3.90981]\,, 
\no\\
c_1^{C_F}&=  9.8696(1)  \, &&\quad[9.8696]\,, 
\no\\
c_2^{C_A}&=  -2.64371(893)  \, &&\quad[-2.65010]\,, 
\no\\
c_2^{n_f}&=  -25.3069(10)  \, &&\quad[-25.3073]\,,
\end{alignat}
which is again in agreement with the analytic results from~\cite{Becher:2012za}.


\subsubsection*{Exclusive Drell-Yan production}

\begin{figure}[t]
\centering
\includegraphics[width=0.45\textwidth]{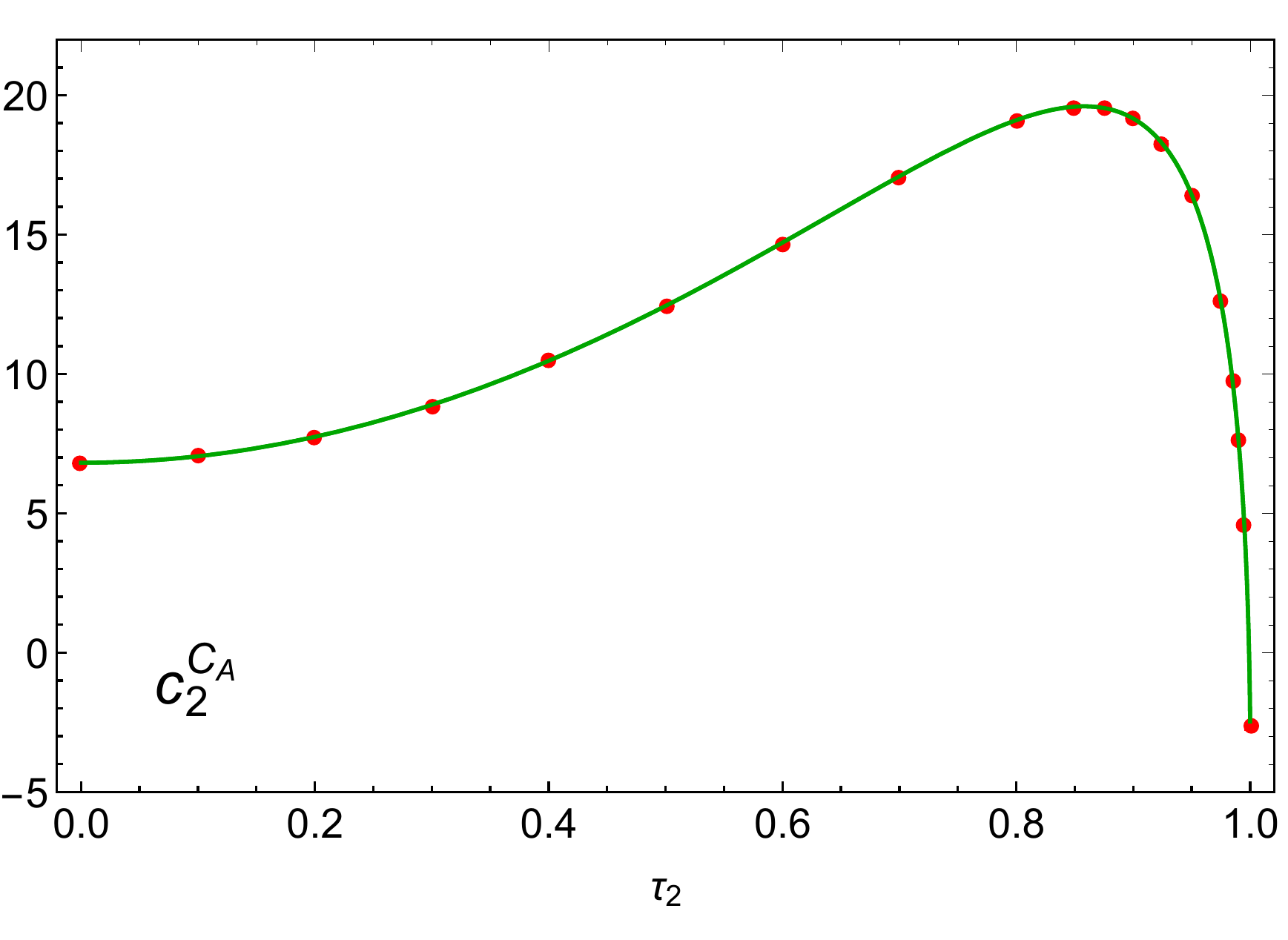}
\hspace{2mm}
\includegraphics[width=0.45\textwidth]{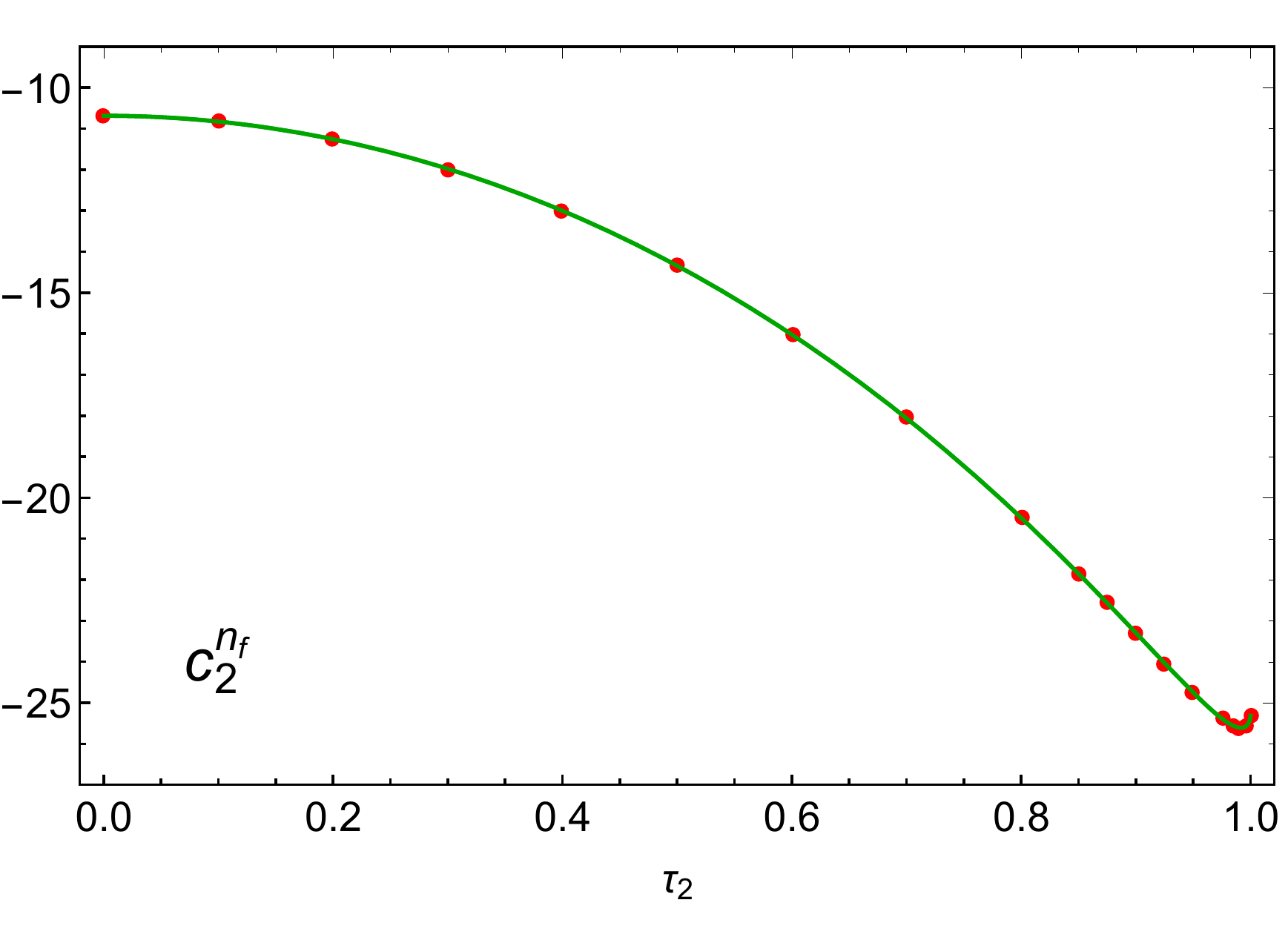}
\caption{Two-loop finite term of the exclusive Drell-Yan soft function. Red dots indicate values calculated 
with {\texttt{SoftSERVE}} and the green line represents the analytic result of~\cite{Li:2011zp}.}
\label{fig:EDY}
\end{figure}

The soft function for exclusive Drell-Yan production~\cite{Li:2011zp} is another example of a 
double-differential observable. Due to rescaling invariance of the Wilson lines, the position-space soft 
function can in this case only depend on $\tau_1 = i/2 \,\sqrt{x_+ x_-}$ and $\tau_2 = \sqrt{x_T^2/x_+x_-}$. 
Whereas the dependence on the dimensionful variable $\tau_1$ factorises, the observable then becomes a 
non-trivial function of the variable $\tau_2$ with
\begin{align}
\omega^{exDY}(\lbrace k_{i} \rbrace;\tau_2) =
\sum_i\; \Big(k_i^++k_i^- - 2\tau_2\sqrt{k_i^+ k_i^-}\cos\theta_i\Big)\,.
\end{align} 
The soft function thus generalises the two preceding examples, and we recover the threshold Drell-Yan soft 
function for $\tau_2=0$, while the one for $W$-production at large transverse momentum corresponds to 
$\tau_2=1$. In terms of our parametrisations, we find $n=-1$, 
$f(y_k,t_k;\tau_2) = 1 + y_k - 2\tau_2\sqrt{y_k}(1-2t_k)$, and
\begin{align}
F_A(a,b,y,t_k,t_l,t_{kl};\tau_2) &=  F_B(a,b,y,t_k,t_l,t_{kl};\tau_2) 
\no\\[0.2em]
&= 
1 + y - 2 \tau_2\;\sqrt{\frac{a y}{(1+a b)(a+b)}} \;
\Big(b (1-2t_k) + 1-2t_l\Big).
\end{align}
It turns out that the respective anomalous dimension is independent of $\tau_2$, and it can therefore be read 
off from the two previous examples. The finite term of the renormalised NNLO soft function is, on the other 
hand, shown in Figure~\ref{fig:EDY} together with the numbers from~\cite{Li:2011zp}. We again find perfect 
agreement with the known analytic results, which is of course also true for the one-loop constant 
$c_1^{C_F}(\tau_2)=4\,\mathrm{Li}_2(\tau_2^2) + \pi^2/3$.

\subsubsection*{Transverse-momentum resummation}

The soft function for transverse-momentum resummation in Drell-Yan production was the third example we 
considered in Section~\ref{sec:measure}. There we saw that
\begin{align}
\omega^{p_T}(\lbrace k_{i} \rbrace) =
-2 i \;\sum_i\; |k_i^\perp| \cos \theta_i \,,
\end{align}
where the factor of $i$ reflects a Fourier transformation. The measurement function is thus purely imaginary 
in this case and, unlike in the preceding example, we cannot simply factor out an imaginary unit along with 
the Laplace variable $\tau$, since $\omega^{p_T}(\lbrace k_{i} \rbrace) $ would in this case not be bounded 
to be positive, which would contradict assumption (A2). Nevertheless we argued in Section~\ref{sec:generalize} 
that we can compute Fourier-space soft functions by using the absolute value of the naive measurement function. 
The required input functions for the transverse-momentum soft function are therefore given by $n=0$, 
$f(y_k,t_k) = 2|1-2t_k|$, and
\begin{align}
F_A(a,b,y,t_k,t_l,t_{kl}) &=  F_B(a,b,y,t_k,t_l,t_{kl}) 
= 2 \;\sqrt{\frac{a}{(1+a b)(a+b)}} \;\,
\big|b (1-2t_k) + 1-2t_l\big|\,.
\end{align}
Further instructions for the computation of Fourier-space soft functions can be found in 
Appendix~\ref{afourier}. Applying the \texttt{fourierconvert} script before renormalisation then leads to
\begin{alignat}{2}
d_1^{C_F}&=  -5 \cdot 10^{-8} \pm 4\cdot 10^{-7}   \, &&\quad[0]\,, 
\no\\
d_2^{C_A}&=  -3.7572(216)  \, &&\quad[-3.7317]\,, 
\no\\
d_2^{n_f}&=  -8.2972(54)  \, && \quad[-8.2963]\,.
\end{alignat}
The slightly reduced accuracy is due to the appearance of integrable logarithmic divergences in the bulk of 
the integration region, since the absolute value can vanish for non-trivial angular configurations. 
Nevertheless, the agreement with the known results from~\cite{Becher:2010tm,Gehrmann:2014yya} is satisfactory.


\subsubsection*{Transverse thrust}

We finally consider the soft function for the hadronic event shape transverse thrust~\cite{Becher:2015gsa}. 
As this example involves four light-like Wilson lines, the computation of the full NNLO soft function clearly 
falls outside the scope of the present paper. It was shown, however, in~\cite{Becher:2015lmy} that the 
underlying anomalous dimensions can be reconstructed from the information on dijet soft functions, and we 
therefore concentrate here on the computation of the anomalous dimensions for transverse thrust.

To this end, we split the $2\rightarrow 2$ process into two toy processes for $e^+ e^-\to q\bar q$ and 
$q\bar q \to e^+ e^-$ scattering. We first consider the latter, which according to~\cite{Becher:2015gsa} 
gives rise to
\begin{align}
\omega^{TT}_{q\bar q}(\lbrace k_{i} \rbrace) =
2 c_0 \;\sum_i\; \Big(|\vec{k}_i^\perp| - |\vec{n}_\perp \cdot \vec{k}_i^\perp|\Big) \,,
\end{align}
where $c_0=e^{4G/\pi}$ depends on Catalan's constant $G\simeq 0.915966$, and the vector $\vec{n}_\perp$ 
singles out a direction in the plane transverse to the beam direction. We thus find that the toy observable 
is of SCET-2 type with $n=0$, $f(y_k,t_k) =2 c_0 (1-|1-2t_k|)$, and
\begin{align}
F_A(a,b,y,t_k,t_l,t_{kl}) &= F_B(a,b,y,t_k,t_l,t_{kl})
\no\\[0.2em]
&=2 c_0\;
\sqrt{\frac{a}{(1+a b)(a+b)}}\;
\big(b (1 - |1 - 2 t_k|) + 1 - |1 - 2 t_l|\big)\,.
\end{align}
Using \softserve~we then determine the corresponding collinear anomaly exponent, which yields
\begin{alignat}{2}
d_1^{C_F}&=3\cdot 10^{-6} \pm 3\cdot 10^{-7}  \, &&\quad[0]\,, 
\no\\
d_2^{C_A}&= 208.105(5)  \, &&\quad[208.0(1)]\,, 
\no\\
d_2^{n_f}&= -37.174(1)  \, &&\quad[-37.191(6)]\,, 
\end{alignat}
where the numerical two-loop results in the square brackets are taken from~\cite{Becher:2015lmy}. Our results 
indeed confirm these numbers, but they are again significantly more precise.

The soft function for the second toy process starts from a similar definition~\cite{Becher:2015gsa},
\begin{align}
\label{eq:omega:TT1}
\omega^{TT}_{e^+e^-}(\lbrace k_{i} \rbrace) =
4s \;\sum_i\; \Big(|\vec{k}_i^\perp| - |\vec{n}_\perp \cdot \vec{k}_i^\perp|\Big) \,,
\end{align}
where $s=\sin\theta_B$ depends on the angle between the beam and the jet axis. Its decomposition into 
light-cone coordinates is, however, significantly more involved, since the components $k_\top^\mu$ transverse 
to the thrust axis differ from the components $k_\perp^\mu$ in \eqref{eq:omega:TT1} that are transverse to the 
beam axis -- see~\cite{Becher:2015gsa}. We then find that this soft function is described by $n=1$ and
\begin{align}
f(y_k,t_k) =\frac{4s}{\sqrt{y_k}} \Bigg\{ \sqrt{  1-c_k^2 +
\left(\frac{s}{2} \Big( \frac{1}{\sqrt{y_k}}-\sqrt{y_k} \Big)+c \, c_k \right)^2 }
-\abs{ \frac{s}{2} \Big( \frac{1}{\sqrt{y_k}}-\sqrt{y_k} \Big)+c \, c_k } \Bigg\},
\end{align}
where $c=\cos\theta_B$ and $c_k=\cos\theta_k=1-2t_k$. The expressions for the two-emission measurement 
functions are rather lengthy with
\begin{align}
&F_A(a,b,y,t_k,t_l,t_{kl}) 
\no\\[0.2em]
&\quad  =4s\,\Bigg\{ b \sqrt{\left(\frac{as}{2(1+a
b)y}+\frac{c\, c_k\sqrt{a}}{\sqrt{(a+b)(1+ab)y}}-\frac{s}{2(a+b)}\right)^2+\frac{a
(1-c_k^2)}{(a+b)(1+ab)y}}
\no\\[0.2em]
&\qquad\qquad - b \abs{\frac{as}{2(1+a b)y}+\frac{c\, c_k\sqrt{a}}{\sqrt{(a+b)(1+ab)y}}-\frac{s}{2(a+b)}}
\no\\[0.2em]
&\qquad\qquad+ \sqrt{\left(\frac{s}{2(1+a b)y}+\frac{c\,c_l\sqrt{a
}}{\sqrt{(a+b)(1+ab)y}}-\frac{as}{2(a+b)}\right)^2+\frac{a(1-c_l^2)}{(a+b)(1+ab)y}} 
\no\\[0.2em]
&\qquad\qquad - \abs{\frac{s}{2(1+a
b)y}+\frac{c\, c_l\sqrt{a}}{\sqrt{(a+b)(1+ab)y}}-\frac{as}{2(a+b)}}\Bigg\}, 
\end{align}
and
\begin{align}
&F_B(a,b,y,t_k,t_l,t_{kl}) 
\no\\[0.2em]
&\quad  =4s\,\Bigg\{ b \sqrt{\left(\frac{s}{2(a+b)y}+\frac{c\, c_k \sqrt{a}}{\sqrt{(a+b)(1+ab)y}}-\frac{as}{2(1+a
b)}\right)^2+\frac{a
(1-c_k^2)}{(a+b)(1+ab)y}}
\no\\[0.2em]
&\qquad\qquad - b \abs{\frac{s}{2(a+b)y}+\frac{c\, c_k\sqrt{a}}{\sqrt{(a+b)(1+ab)y}}-\frac{as}{2(1+a b)}}
\no\\[0.2em]
&\qquad\qquad+ \sqrt{\left(\frac{as}{2(a+b)y}+\frac{c\,c_l\sqrt{a
}}{\sqrt{(a+b)(1+ab)y}}-\frac{s}{2(1+a b)}\right)^2+\frac{a(1-c_l^2)}{(a+b)(1+ab)y}} 
\no\\[0.2em]
&\qquad\qquad - \abs{\frac{as}{2(a+b)y}+\frac{c\, c_l\sqrt{a}}{\sqrt{(a+b)(1+ab)y}}-\frac{s}{2(1+ab)}}\Bigg\},
\end{align}
where we remind the reader that the last expression is not unique, due to the freedom in the definition 
\eqref{eq:rr:fb}.

The complicated structure of the measurement functions clearly exhibits that an analytic calculation is very 
challenging -- if not out of reach -- for this class of soft functions. Yet our numerical approach also suffers 
from large cancellations between the square roots and the absolute values for small values of the parameter $y$. 
This causes major problems for the numerical evaluation due to rounding errors in floating point computations. 
As this is mostly an implementation issue, we refer the user to the manual of \softserve for a detailed 
discussion. The consequence is that \softserve~must be run using multi-precision variables provided by the 
{\tt boost}~\cite{boost} or {\tt GMP/MPFR}~\cite{GMP/MPFR} libraries for this observable. This significantly 
slows down the program, and hence our transverse thrust runs used reduced accuracy settings, as specified in 
more detail in the \softserve~manual.

Although the measurement functions for this observable depend on the angle $\theta_B$ between the beam and 
jet axes, it turns out that the anomalous dimension is independent of this angle, and we obtain
\begin{alignat}{2}
\gamma_0^{C_F}&=  -2\cdot 10^{-6}\pm 4\cdot 10^{-5}  \, &&\quad[0]\,, 
\no\\
\gamma_1^{C_A}&=  -158.27(6)  \, &&\quad[-148^{+20}_{-30}]\,, 
\no\\
\gamma_1^{n_f}&=  19.3942(49)  \, &&\quad[18^{+2}_{-3}],
\end{alignat}
where the results of~\cite{Becher:2015gsa} were obtained via a fit to the \texttt{EVENT2} generator.

As explained at the beginning of this section, the finite terms of the corresponding soft functions are 
not useful, as the split into $e^+ e^-\to q\bar q$ and $q\bar q \to e^+ e^-$ observables is only 
justified on the level of the anomalous dimension and the collinear anomaly exponent. So while the 
\softserve~runs of course produce numbers for the finite terms as well, they cannot be used for any meaningful 
interpretation in this case.

%% file: Conclude.tex
\section{Conclusions}
\label{sec:conclude}

We have presented a novel formalism that allows for an automated calculation of soft functions that are defined 
in terms of two light-like Wilson lines. Our method is based on a universal phase-space parametrisation which we 
use to isolate the implicit divergences of the phase-space integrals, and which allows us to perform the 
expansion in the various regulators directly on the integrand level. The remaining integrations can then be 
performed numerically, and we developed a {\tt{C++}} package called {\tt{SoftSERVE}} which precisely performs 
these integrations using the Divonne integrator of the Cuba library.

Our method is currently restricted to the correlated-emission contribution, which is sufficient for the 
computation of NNLO soft functions that obey the NAE theorem. As exemplified by the large number of results in 
Section~\ref{sec:results}, our method is fairly general and it relies on a handful of assumptions that 
we discussed in detail in Sections~\ref{sec:measure} and \ref{sec:generalize}. Further extensions of our 
formalism that deal with the uncorrelated-emission contribution~\cite{Bell:2018jvf} and with more than two 
light-like directions including non-back-to-back kinematics~\cite{Bell:2018mkk} are currently in progress.

With the publication of the present paper, we release {\tt SoftSERVE \!0.9}, which is publicly available
at \url{https://softserve.hepforge.org/}. The current version of \texttt{SoftSERVE} not only allows for the calculation 
of bare dijet soft functions, but it also 
provides scripts for their renormalisation according to the conventions we introduced in 
Section~\ref{sec:renormalize}. In this paper we refrained from presenting the technical aspects of the numerical
implementation, which are explained in detail in the \softserve~manual.

Among the plethora of results we obtained in Section~\ref{sec:results}, our NNLO numbers for the thrust-axis 
and broadening-axis angularities are new, and we for the first time obtained the C-parameter soft function in 
an analytic form. More generally, our results for the two-loop anomalous dimensions and the
two-loop finite terms are required for NNLL and NNLL$'$ resummations within SCET, respectively. We believe
that \softserve~can open the path for high-precision SCET resummations in the future, as has recently been 
illustrated for the $e^+ e^-$ event shape angularities in~\cite{Bell:2018gce}.

Yet the process of automating resummation within SCET requires many further ingredients beyond the automated 
calculation of soft functions. At present there exists an automated NLL resummation code that is based on the
coherent-branching formalism, {\tt CAESAR}~\cite{Banfi:2004yd}, which has been extended to NNLL accuracy 
for $e^+ e^-$ event shapes in~\cite{Banfi:2014sua}. While a combination of the {\tt CAESAR} approach with 
methods from SCET is currently under investigation~\cite{Bauer:2018svx}, we believe that an alternative 
approach that is purely based on effective-field-theory techniques could be a valuable addition. We 
think that \softserve~could provide one essential pillar for such an automated resummation code in SCET.

%% file: App_Twoloopn.tex
\section{Equality of one-emission and two-emission values of $n$}
\label{twoloopn}

As mentioned in Section~\ref{sec:measure},  the leading scaling in the variables $y_k$ and $y$ is the same 
between the one-emission and two-emission measurement functions, see \eqref{eq:measure:one} and 
\eqref{eq:measure:NNLO:corr}. As we will show in this appendix, this is a non-trivial finding for generic 
observables, in particular for those that violate the NAE theorem. 

To this end, we first rewrite the two-emission measurement function in the form
\begin{equation}
\label{eq:measure:nnlo:NAEpluscluster}
\omega(\{k,l\})=\omega(\{k\})+\omega(\{l\})+\omega_c(\{k,l\}),
\end{equation}
where $\omega_c(\{k,l\})$ encodes the effects that break NAE. Whenever the emission with momentum $k^\mu$ 
becomes soft, infrared safety implies that the measurement function reduces to $\omega(\{l\})$, see
\eqref{eq:irsafety:soft}, and hence the NAE-violating term must satisfy $\omega_c(\{0,l\})=0$. Similarly, we obtain $\omega_c(\{k,0\})=0$ when 
$l^\mu$ becomes soft, and $\omega_c(\{k,\alpha k\})=0$ with $\alpha>0$, which arises when the two emissions 
become collinear to each other.

Since both $y_k=k_+/k_- \sim y$ and $y_l =l_+/l_-\sim y$ in the parametrisation \eqref{eq:parametr:nnlo},
the leading scaling in the variable $y$ in the first two terms of \eqref{eq:measure:nnlo:NAEpluscluster} 
is equal to the one-emission case. The question therefore becomes whether 
the NAE-breaking term $\omega_c(\{k,l\})$ can enforce a different scaling of the observable.

The answer lies in the factorisation theorem \eqref{eq:fact}, and the fact that the poles must cancel between 
the bare hard, jet and soft functions. As the hard function is the same for all dijet observables, the crucial 
cancellation is the one between the collinear and soft functions. The role of the parameter $n$ is two fold
in this context: First, it determines the coefficient of the leading pole in the soft function, as the monomial 
$y^{-1+n\epsilon}$ gives rise to a term $\delta(y)/n\epsilon$. And second, the analysis in Section~\ref{sec:n} 
showed that the parameter $n$ is related to the power counting of the momentum modes in the effective theory. 
Following this analysis, we thus know what the scaling of the three sectors --- soft, collinear and 
anti-collinear --- is for two emissions, as the form of the factorisation theorem implies that no new modes
appear at higher orders. As the power counting of the momentum components is uniform in the soft sector, 
the form of the NAE-breaking term $\omega_c(\{k,l\})$ is unconstrained, since consistency demands 
that its mass dimension is the same as that of the one-emission terms, which immediately puts the three terms 
in~\eqref{eq:measure:nnlo:NAEpluscluster} at the same order in the power counting. This does not hold, however, 
in the collinear sectors.

We therefore consider the collinear sector with scaling 
$p_c^\mu = Q(1,\lambda^{2/(n+1)},\lambda^{1/(n+1)})$ more closely. As we have seen in
\eqref{eq:omega:scaling:collinear}, the measurement function for one collinear emission scales in 
the soft limit $z_k\to 0$ as 
\begin{align}
\omega(\{k\}) \,\sim\, k_+^{\frac{1+n}{2}} k_-^{\frac{1-n}{2}}\,. 
\end{align}
Based on power-counting and dimensional arguments, the collinear analog of the NAE-breaking term then similarly 
scales in the soft limit as
\begin{align}
\omega_c(\{k,l\}) \,\sim\, q_+^{\frac{1+n'}{2}} q_-^{\frac{1-n'}{2}}\,, 
\end{align}
where $q$ can stand in for both $k$ or $l$ (i.e. $\sqrt{k_+ l_-}$ and $\sqrt{k_+ k_-}$ both count as 
$\sqrt{q_+ q_-}$ here).  The question then is whether $n'$ can differ from $n$. There are three possibilities:

\begin{enumerate}
\item[$n=n'$:] 
The three terms in~\eqref{eq:measure:nnlo:NAEpluscluster} contribute in the soft limit to the jet function,
and its leading poles are determined by the one-emission value of $n$. To ensure the cancellation of the leading 
poles, the soft function must follow suit and its two-emission measurement function must therefore scale as
$y^{n/2}$.
\item[$n < n'$:] 
In this case the NAE-breaking term is suppressed in the jet function in the soft limit. As 
the jet function does not see this term in the soft region, its leading poles match those of an auxiliary observable 
(the NAE version of the observable), for which the leading poles are again determined by the one-emission value 
of $n$ as described above. 
\item[$n > n'$:] 
In this case the NAE-breaking term dominates over the one-emission terms in the soft limit. But the 
observable would then not be infrared safe, which violates one of our underlying assumptions.
\end{enumerate}
We therefore conclude that the factorisation theorem \eqref{eq:fact}, in particular the absence of additional 
momentum modes, enforces the equality of the one-emission and two-emission values of $n$.

%% file: App_Fourier.tex
\section{Fourier-space soft functions}
\label{afourier}

In Section \ref{sec:generalize} we explained that the numerical evaluation of Fourier-space soft functions 
using \softserve~may require a workaround, since \softserve~assumes that the measurement function is strictly 
real and non-negative. In this appendix we describe the details of this workaround.

To begin with, we consider a one-emission measurement function of the form
\begin{align}
\mathcal{M}_1(\tau; k) = \exp\big(- i \,\tau\, k_{T}\, y_k^{n/2}\, \widetilde{f}(y_k,t_k)\,\big)\,,
\end{align}
where the factor of $i$ arises because of the Fourier transformation, $\tau$ is the respective Fourier
variable and the function $\widetilde{f}(y_k,t_k)$ is assumed to be real-valued. If the latter is non-negative, 
the factor of $i$ can simply be factorised alongside $\tau$ outside the integral, and the
\softserve~implementation proceeds in the standard form (threshold resummation for Drell-Yan production with 
$\tau=ix^0/2$ is of this type). In general, however, the function $\widetilde{f}(y_k,t_k)$ may well attain also 
negative values, as is the case e.g.~for transverse-momentum resummation with 
$\widetilde{f}(y_k,t_k)=- 2(1-2 t_k)$. As \softserve~only accepts non-negative measure\-ment functions, it 
seems as if we cannot evaluate such functions with the existing code. There exists, however, a 
workaround as long as one is interested only in the real part of the soft function (which is usually the case).

To this end, we review the analytic steps of our derivation, and we split the calculation into two parts according
to the two possible signs of the measurement function $\widetilde{f}(y_k,t_k)$. In each of these regions, we then factor 
out either $(+i)$ or $(-i)$ such that the remnant measurement function is non-negative. As the measurement 
function enters the master formula \eqref{eq:nlo:master} as $f(y_k,t_k)^{2\eps+\alpha}$, we can write the 
soft function in the form
\begin{align}
S_R\big[i\widetilde{f}\big]&=(+i)^{(2\epsilon+\alpha)}\,S_R^+\big[|\widetilde{f}|\big]
+(-i)^{(2\epsilon+\alpha)}\,S_R^-\big[|\widetilde{f}|\big]\no \\[0.2em]
&=e^{+i \frac{\pi}{2}(2\epsilon+\alpha)}\,S_R^+\big[|\widetilde{f}|\big]
+e^{-i \frac{\pi}{2}(2\epsilon+\alpha)}\,S_R^-\big[|\widetilde{f}|\big],
\end{align}
where we indicated that $S_R\big[f\big]$ in \eqref{eq:nlo:master} depends on the function 
$f(y_k,t_k) = i \widetilde{f}(y_k,t_k)$, while $S_R^\pm\big[f\big]$ depend on the magnitude of $\widetilde{f}(y_k,t_k)$. The latter are, moreover, 
restricted to an integration domain in which $\widetilde{f}(y_k,t_k)$ is either positive ($S_R^+\big[|\widetilde{f}|\big]$) 
or negative ($S_R^-\big[|\widetilde{f}|\big]$), and they are not accessible in our formalism, since we always
assume that the integrations are performed over the full domain. But if we decompose the soft function into 
its real and imaginary parts, we obtain
\begin{align}
S_R[i\widetilde{f}]&=\cos\left(\frac{\pi}{2}(2\epsilon + \alpha)\right) S_R\big[|\widetilde{f}|\big]
+i \sin\left(\frac{\pi}{2}(2\epsilon + \alpha)\right) \Big(S_R^+\big[|\widetilde{f}|\big]-S_R^-\big[|\widetilde{f}|\big]\Big),
\end{align}
where we used the fact that $S_R^\pm\big[|\widetilde{f}|\big]$ are real-valued, and that 
$S_R^+\big[|\widetilde{f}|\big]+S_R^-\big[|\widetilde{f}|\big]=S_R\big[|\widetilde{f}|\big]$, i.e.~the sum over two complementary integration
domains gives back the full integral. We thus see that the observable-dependent phase-space restrictions drop 
out in the real part of the soft function. In other words we simply have to evaluate the soft function with
$\widetilde{f}(y_k,t_k)$ replaced by its absolute value, which is yet to be multiplied by a 
regulator-depend factor.

The same arguments apply to the mixed real-virtual interference $S_{RV}\big[f\big]$ and 
the double real-emission contribution $S_{RR}\big[F\big]$, which depend on different powers of the 
measurement functions $f(y_k,t_k)^{4\eps+\alpha}$ and $F(a,b,y,t_k^+,t_l,t_{kl})^{4\eps+2\alpha}$, 
respectively, see \eqref{eq:rv:master} and \eqref{eq:rr:master}. The real part of a bare Fourier-space soft 
function then becomes
\begin{align}
\Re \left(S\big[i\widetilde{f},i\widetilde{F}\big]\right) = 1 &+ \left(\frac{Z_\alpha \alpha_{s}}{4\pi}\right) 
\left(\mu^{2} \bar{\tau}^{2} \right)^{\epsilon} (\nu \bar{\tau})^\alpha \, 
\cos\left(\frac{\pi}{2}(2\epsilon + \alpha)\right) S_R\big[|\widetilde{f}|\big]
\\ 
&+ \left( \frac{Z_\alpha \alpha_{s}}{4\pi}\right)^{2} \left( \mu^{2} \bar{\tau}^{2} \right)^{2\epsilon} 
\bigg\{ \left( \nu \bar{\tau} \right)^\alpha\,
\cos\left(\frac{\pi}{2}(4\epsilon + \alpha)\right) S_{RV}\big[|\widetilde{f}|\big]
 \no\\ &
\hspace{4cm}
+ \left(\nu \bar{\tau}\right)^{2\alpha}\,
\cos\left(\frac{\pi}{2}(4\epsilon + 2\alpha)\right) S_{RR}\big[|\widetilde{F}|\big] \bigg\}+\mathcal{O}(\alpha_{s}^{3}).
\no
\end{align}
If one is interested in calculating the real part of a Fourier-space soft function, one should thus 
run \softserve~using the absolute value of the measurement function (which is by construction real
and non-negative). The individual pieces of this calculation then need to be multiplied with different
regulator-dependent factors, which reshuffle the coefficients in the Laurent expansion. The 
\softserve~package contains a script to perform this Laurent series reshuffle keeping track of 
the error bars --- its usage is explained in the manual. As an application of 
this technique we consider the soft function for transverse-momentum resummation in 
Section \ref{sec:results}.

%% file: paper.bbl
\begin{thebibliography}{999}

\bibitem{Belitsky:1998tc}
  A.~V.~Belitsky,
  ``Two loop renormalization of Wilson loop for Drell-Yan production,''
  Phys.\ Lett.\ B {\bf 442} (1998) 307
  [hep-ph/9808389].
  
\bibitem{Becher:2005pd}
  T.~Becher and M.~Neubert,
  ``Toward a NNLO calculation of the $\bar B\to X_s\gamma$ gamma decay rate with a cut on photon energy: I. Two-loop result for the soft function,''
  Phys.\ Lett.\ B {\bf 633} (2006) 739
  [hep-ph/0512208].
  
\bibitem{Kelley:2011ng}
  R.~Kelley, M.~D.~Schwartz, R.~M.~Schabinger and H.~X.~Zhu,
  ``The two-loop hemisphere soft function,''
  Phys.\ Rev.\ D {\bf 84} (2011) 045022
  [arXiv:1105.3676 [hep-ph]].
  
\bibitem{Monni:2011gb}
  P.~F.~Monni, T.~Gehrmann and G.~Luisoni,
  ``Two-Loop Soft Corrections and Resummation of the Thrust Distribution in the Dijet Region,''
  JHEP {\bf 1108} (2011) 010
  [arXiv:1105.4560 [hep-ph]].
  
\bibitem{Hornig:2011iu}
  A.~Hornig, C.~Lee, I.~W.~Stewart, J.~R.~Walsh and S.~Zuberi,
  ``Non-global Structure of the $\mathcal{O}({\alpha}_s^2)$ Dijet Soft Function,''
  JHEP {\bf 1108} (2011) 054
   Erratum: [JHEP {\bf 1710} (2017) 101]
  [arXiv:1105.4628 [hep-ph]].
  
\bibitem{Li:2011zp}
  Y.~Li, S.~Mantry and F.~Petriello,
  ``An Exclusive Soft Function for Drell-Yan at Next-to-Next-to-Leading Order,''
  Phys.\ Rev.\ D {\bf 84} (2011) 094014
  [arXiv:1105.5171 [hep-ph]].
  
\bibitem{Kelley:2011aa}
  R.~Kelley, M.~D.~Schwartz, R.~M.~Schabinger and H.~X.~Zhu,
  ``Jet Mass with a Jet Veto at Two Loops and the Universality of Non-Global Structure,''
  Phys.\ Rev.\ D {\bf 86} (2012) 054017
  [arXiv:1112.3343 [hep-ph]].
  
\bibitem{Becher:2012za}
  T.~Becher, G.~Bell and S.~Marti,
  ``NNLO soft function for electroweak boson production at large transverse momentum,''
  JHEP {\bf 1204} (2012) 034
  [arXiv:1201.5572 [hep-ph]].
  
\bibitem{Ferroglia:2012uy}
  A.~Ferroglia, B.~D.~Pecjak, L.~L.~Yang, B.~D.~Pecjak and L.~L.~Yang,
  ``The NNLO soft function for the pair invariant mass distribution of boosted top quarks,''
  JHEP {\bf 1210} (2012) 180
  [arXiv:1207.4798 [hep-ph]].
  
\bibitem{Becher:2012qc}
  T.~Becher and G.~Bell,
  ``NNLL Resummation for Jet Broadening,''
  JHEP {\bf 1211} (2012) 126
  [arXiv:1210.0580 [hep-ph]].
  
\bibitem{vonManteuffel:2013vja}
  A.~von Manteuffel, R.~M.~Schabinger and H.~X.~Zhu,
  ``The Complete Two-Loop Integrated Jet Thrust Distribution In Soft-Collinear Effective Theory,''
  JHEP {\bf 1403} (2014) 139
  [arXiv:1309.3560 [hep-ph]].
  
\bibitem{Ferroglia:2013awa}
  A.~Ferroglia, S.~Marzani, B.~D.~Pecjak and L.~L.~Yang,
  ``Boosted top production: factorization and resummation for single-particle inclusive distributions,''
  JHEP {\bf 1401} (2014) 028
  [arXiv:1310.3836 [hep-ph]].
  
\bibitem{Czakon:2013hxa}
  M.~Czakon and P.~Fiedler,
  ``The soft function for color octet production at threshold,''
  Nucl.\ Phys.\ B {\bf 879} (2014) 236
  [arXiv:1311.2541 [hep-ph]].
  
\bibitem{vonManteuffel:2014mva}
  A.~von Manteuffel, R.~M.~Schabinger and H.~X.~Zhu,
  ``The two-loop soft function for heavy quark pair production at future linear colliders,''
  Phys.\ Rev.\ D {\bf 92} (2015) no.4,  045034
  [arXiv:1408.5134 [hep-ph]].
  
\bibitem{Boughezal:2015eha}
  R.~Boughezal, X.~Liu and F.~Petriello,
  ``$N$-jettiness soft function at next-to-next-to-leading order,''
  Phys.\ Rev.\ D {\bf 91} (2015) no.9,  094035
  [arXiv:1504.02540 [hep-ph]].
  
\bibitem{Echevarria:2015byo}
  M.~G.~Echevarria, I.~Scimemi and A.~Vladimirov,
  ``Universal transverse momentum dependent soft function at NNLO,''
  Phys.\ Rev.\ D {\bf 93} (2016) no.5,  054004
  [arXiv:1511.05590 [hep-ph]].
  
\bibitem{Luebbert:2016itl}
  T.~L\"ubbert, J.~Oredsson and M.~Stahlhofen,
  ``Rapidity renormalized TMD soft and beam functions at two loops,''
  JHEP {\bf 1603} (2016) 168
  [arXiv:1602.01829 [hep-ph]].
  
\bibitem{Gangal:2016kuo}
  S.~Gangal, J.~R.~Gaunt, M.~Stahlhofen and F.~J.~Tackmann,
  ``Two-Loop Beam and Soft Functions for Rapidity-Dependent Jet Vetoes,''
  JHEP {\bf 1702} (2017) 026
  [arXiv:1608.01999 [hep-ph]].
  
\bibitem{Li:2016tvb}
  H.~T.~Li and J.~Wang,
  ``Next-to-Next-to-Leading Order $N$-Jettiness Soft Function for One Massive Colored Particle Production at Hadron Colliders,''
  JHEP {\bf 1702} (2017) 002
  [arXiv:1611.02749 [hep-ph]].
  
\bibitem{Campbell:2017hsw}
  J.~M.~Campbell, R.~K.~Ellis, R.~Mondini and C.~Williams,
  ``The NNLO QCD soft function for 1-jettiness,''
  Eur.\ Phys.\ J.\ C {\bf 78} (2018) no.3,  234
  [arXiv:1711.09984 [hep-ph]].

\bibitem{Wang:2018vgu}
  G.~Wang, X.~Xu, L.~L.~Yang and H.~X.~Zhu,
  ``The next-to-next-to-leading order soft function for top quark pair production,''
  JHEP {\bf 1806} (2018) 013
  [arXiv:1804.05218 [hep-ph]].
  
\bibitem{Li:2018tsq}
  H.~T.~Li and J.~Wang,
  ``Next-to-next-to-leading order $N$-jettiness soft function for $tW$ production,''
  Phys.\ Lett.\ B {\bf 784} (2018) 397
  [arXiv:1804.06358 [hep-ph]].
  
\bibitem{Dulat:2018vuy}
  F.~Dulat, S.~H\"oche and S.~Prestel,
  ``Leading-Color Fully Differential Two-Loop Soft Corrections to QCD Dipole Showers,''
  Phys.\ Rev.\ D {\bf 98} (2018) no.7,  074013
  [arXiv:1805.03757 [hep-ph]].
  
\bibitem{Angeles-Martinez:2018mqh}
  R.~Angeles-Martinez, M.~Czakon and S.~Sapeta,
  ``NNLO soft function for top quark pair production at small transverse momentum,''
  JHEP {\bf 1810} (2018) 201
  [arXiv:1809.01459 [hep-ph]].

\bibitem{Li:2016ctv}
  Y.~Li and H.~X.~Zhu,
  ``Bootstrapping Rapidity Anomalous Dimensions for Transverse-Momentum Resummation,''
  Phys.\ Rev.\ Lett.\  {\bf 118} (2017) no.2,  022004
  [arXiv:1604.01404 [hep-ph]].
  
\bibitem{Moult:2018jzp}
  I.~Moult and H.~X.~Zhu,
  ``Simplicity from Recoil: The Three-Loop Soft Function and Factorization for the Energy-Energy Correlation,''
  JHEP {\bf 1808} (2018) 160
  [arXiv:1801.02627 [hep-ph]].

\bibitem{BRT}
  G.~Bell, R.~Rahn and J.~Talbert,
  in preparation.
  
\bibitem{Gatheral:1983cz}
  J.~G.~M.~Gatheral,
  ``Exponentiation of Eikonal Cross-sections in Nonabelian Gauge Theories,''
  Phys.\ Lett.\  {\bf 133B} (1983) 90.
  
\bibitem{Frenkel:1984pz}
  J.~Frenkel and J.~C.~Taylor,
  ``Nonabelian Eikonal Exponentiation,''
  Nucl.\ Phys.\ B {\bf 246} (1984) 231.

\bibitem{Bell:2018jvf}
  G.~Bell, R.~Rahn and J.~Talbert,
  ``Automated Calculation of Dijet Soft Functions in the Presence of Jet Clustering Effects,''
  PoS RADCOR {\bf 2017} (2018) 047
  [arXiv:1801.04877 [hep-ph]].
  
\bibitem{Chiu:2009yx}
  J.~y.~Chiu, A.~Fuhrer, A.~H.~Hoang, R.~Kelley and A.~V.~Manohar,
  ``Soft-Collinear Factorization and Zero-Bin Subtractions,''
  Phys.\ Rev.\ D {\bf 79} (2009) 053007
  [arXiv:0901.1332 [hep-ph]].

\bibitem{Becher:2011dz}
  T.~Becher and G.~Bell,
  ``Analytic Regularization in Soft-Collinear Effective Theory,''
  Phys.\ Lett.\ B {\bf 713} (2012) 41
  [arXiv:1112.3907 [hep-ph]].
  
\bibitem{Chiu:2012ir}
  J.~Y.~Chiu, A.~Jain, D.~Neill and I.~Z.~Rothstein,
  ``A Formalism for the Systematic Treatment of Rapidity Logarithms in Quantum Field Theory,''
  JHEP {\bf 1205} (2012) 084
  [arXiv:1202.0814 [hep-ph]].

\bibitem{Li:2016axz}
  Y.~Li, D.~Neill and H.~X.~Zhu,
  ``An Exponential Regulator for Rapidity Divergences,''
  [arXiv:1604.00392 [hep-ph]].
	
\bibitem{Bauer:2000ew}
  C.~W.~Bauer, S.~Fleming and M.~E.~Luke,
  ``Summing Sudakov logarithms in B ---> X(s gamma) in effective field theory,''
  Phys.\ Rev.\ D {\bf 63} (2000) 014006
  [hep-ph/0005275].

\bibitem{Bauer:2000yr}
  C.~W.~Bauer, S.~Fleming, D.~Pirjol and I.~W.~Stewart,
  ``An Effective field theory for collinear and soft gluons: Heavy to light decays,''
  Phys.\ Rev.\ D {\bf 63} (2001) 114020
  [hep-ph/0011336].
  
\bibitem{Bauer:2001yt}
  C.~W.~Bauer, D.~Pirjol and I.~W.~Stewart,
  ``Soft collinear factorization in effective field theory,''
  Phys.\ Rev.\ D {\bf 65} (2002) 054022
  [hep-ph/0109045].
  
\bibitem{Beneke:2002ph}
  M.~Beneke, A.~P.~Chapovsky, M.~Diehl and T.~Feldmann,
  ``Soft collinear effective theory and heavy to light currents beyond leading power,''
  Nucl.\ Phys.\ B {\bf 643} (2002) 431
  [hep-ph/0206152].
  
\bibitem{Hahn:2004fe}
  T.~Hahn,
  ``CUBA: A Library for multidimensional numerical integration,''
  Comput.\ Phys.\ Commun.\  {\bf 168} (2005) 78
  [hep-ph/0404043].
  
\bibitem{Bell:2018vaa}
  G.~Bell, R.~Rahn and J.~Talbert,
  ``Two-loop anomalous dimensions of generic dijet soft functions,''
  Nucl.\ Phys.\ B {\bf 936} (2018) 520
  [arXiv:1805.12414 [hep-ph]].

\bibitem{Bell:2018gce}
  G.~Bell, A.~Hornig, C.~Lee and J.~Talbert,
  ``$e^+ e^-$ angularity distributions at NNLL$^\prime$ accuracy,''
  JHEP {\bf 1901} (2019) 147
  [arXiv:1808.07867 [hep-ph]].
  
\bibitem{Bell:2015lsf}
  G.~Bell, R.~Rahn and J.~Talbert,
  ``Automated Calculation of Dijet Soft Functions in Soft-Collinear Effective Theory,''
  PoS RADCOR {\bf 2015} (2016) 052
  [arXiv:1512.06100 [hep-ph]].
  
\bibitem{Catani:1996vz}
  S.~Catani and M.~H.~Seymour,
  ``A General algorithm for calculating jet cross-sections in NLO QCD,''
  Nucl.\ Phys.\ B {\bf 485} (1997) 291
   Erratum: [Nucl.\ Phys.\ B {\bf 510} (1998) 503]
  [hep-ph/9605323].
  
\bibitem{Bell:2018mkk}
  G.~Bell, B.~Dehnadi, T.~Mohrmann and R.~Rahn,
  ``Automated Calculation of ${\pmb N}$-jet Soft Functions,''
  PoS LL {\bf 2018} (2018) 044
  [arXiv:1808.07427 [hep-ph]].
  
\bibitem{Hoang:2014wka}
  A.~H.~Hoang, D.~W.~Kolodrubetz, V.~Mateu and I.~W.~Stewart,
  ``$C$-parameter distribution at N$^3$LL including power corrections,''
  Phys.\ Rev.\ D {\bf 91} (2015) no.9,  094017
  [arXiv:1411.6633 [hep-ph]].
  
\bibitem{Hoang:2015hka}
  A.~H.~Hoang, D.~W.~Kolodrubetz, V.~Mateu and I.~W.~Stewart,
  ``Precise determination of $\alpha_s$ from the $C$-parameter distribution,''
  Phys.\ Rev.\ D {\bf 91} (2015) no.9,  094018
  [arXiv:1501.04111 [hep-ph]].
  
\bibitem{Becher:2007ty}
  T.~Becher, M.~Neubert and G.~Xu,
  ``Dynamical Threshold Enhancement and Resummation in Drell-Yan Production,''
  JHEP {\bf 0807} (2008) 030
  [arXiv:0710.0680 [hep-ph]].
  
\bibitem{Becher:2010tm}
  T.~Becher and M.~Neubert,
  ``{Drell-Yan} Production at Small $q_T$, Transverse Parton Distributions and the Collinear Anomaly,''
  Eur.\ Phys.\ J.\ C {\bf 71} (2011) 1665
  [arXiv:1007.4005 [hep-ph]].
  
\bibitem{Chay:2004zn}
  J.~Chay, C.~Kim, Y.~G.~Kim and J.~P.~Lee,
  ``Soft Wilson lines in soft-collinear effective theory,''
  Phys.\ Rev.\ D {\bf 71} (2005) 056001
  [hep-ph/0412110].
  
\bibitem{Kasemets:2015uus}
  T.~Kasemets, W.~J.~Waalewijn and L.~Zeune,
  ``Calculating Soft Radiation at One Loop,''
  JHEP {\bf 1603} (2016) 153
  [arXiv:1512.00857 [hep-ph]].
  
\bibitem{Catani:2000pi}
  S.~Catani and M.~Grazzini,
  ``The soft gluon current at one loop order,''
  Nucl.\ Phys.\ B {\bf 591} (2000) 435
  [hep-ph/0007142].
	
\bibitem{Kang:2015moa}
  D.~Kang, O.~Z.~Labun and C.~Lee,
  ``Equality of hemisphere soft functions for $e^+e^-$, DIS and $pp$ collisions at $\mathcal{O}(\alpha_s^2)$,''
  Phys.\ Lett.\ B {\bf 748} (2015) 45
  [arXiv:1504.04006 [hep-ph]].
  
\bibitem{Binoth:2000ps}
  T.~Binoth and G.~Heinrich,
  ``An automatized algorithm to compute infrared divergent multiloop integrals,''
  Nucl.\ Phys.\ B {\bf 585} (2000) 741
  [hep-ph/0004013].

\bibitem{Becher:2011pf}
  T.~Becher, G.~Bell and M.~Neubert,
  ``Factorization and Resummation for Jet Broadening,''
  Phys.\ Lett.\ B {\bf 704} (2011) 276
  [arXiv:1104.4108 [hep-ph]].

\bibitem{Carter:2010hi}
  J.~Carter and G.~Heinrich,
  ``SecDec: A general program for sector decomposition,''
  Comput.\ Phys.\ Commun.\  {\bf 182} (2011) 1566
  [arXiv:1011.5493 [hep-ph]].
  
\bibitem{Borowka:2012yc}
  S.~Borowka, J.~Carter and G.~Heinrich,
  ``Numerical Evaluation of Multi-Loop Integrals for Arbitrary Kinematics with SecDec 2.0,''
  Comput.\ Phys.\ Commun.\  {\bf 184} (2013) 396
  [arXiv:1204.4152 [hep-ph]].
  
\bibitem{Borowka:2015mxa}
  S.~Borowka, G.~Heinrich, S.~P.~Jones, M.~Kerner, J.~Schlenk and T.~Zirke,
  ``SecDec-3.0: numerical evaluation of multi-scale integrals beyond one loop,''
  Comput.\ Phys.\ Commun.\  {\bf 196} (2015) 470
  [arXiv:1502.06595 [hep-ph]].

\bibitem{Borowka:2017idc}
  S.~Borowka, G.~Heinrich, S.~Jahn, S.~P.~Jones, M.~Kerner, J.~Schlenk and T.~Zirke,
  ``pySecDec: a toolbox for the numerical evaluation of multi-scale integrals,''
  Comput.\ Phys.\ Commun.\  {\bf 222} (2018) 313
  [arXiv:1703.09692 [hep-ph]].
  
\bibitem{Larkoski:2014uqa}
  A.~J.~Larkoski, D.~Neill and J.~Thaler,
  ``Jet Shapes with the Broadening Axis,''
  JHEP {\bf 1404} (2014) 017
  [arXiv:1401.2158 [hep-ph]].

\bibitem{Hornig:2009vb}
  A.~Hornig, C.~Lee and G.~Ovanesyan,
  ``Effective Predictions of Event Shapes: Factorized, Resummed, and Gapped Angularity Distributions,''
  JHEP {\bf 0905} (2009) 122
  [arXiv:0901.3780 [hep-ph]].

\bibitem{Procura:2018zpn}
  M.~Procura, W.~J.~Waalewijn and L.~Zeune,
  ``Joint resummation of two angularities at next-to-next-to-leading logarithmic order,''
  JHEP {\bf 1810} (2018) 098
  [arXiv:1806.10622 [hep-ph]].

\bibitem{Gehrmann:2014yya}
  T.~Gehrmann, T.~Luebbert and L.~L.~Yang,
  ``Calculation of the transverse parton distribution functions at next-to-next-to-leading order,''
  JHEP {\bf 1406} (2014) 155
  [arXiv:1403.6451 [hep-ph]].

\bibitem{Becher:2015gsa}
  T.~Becher and X.~Garcia i Tormo,
  ``Factorization and resummation for transverse thrust,''
  JHEP {\bf 1506} (2015) 071
  [arXiv:1502.04136 [hep-ph]].

\bibitem{Becher:2015lmy}
  T.~Becher, X.~Garcia i Tormo and J.~Piclum,
  ``Next-to-next-to-leading logarithmic resummation for transverse thrust,''
  Phys.\ Rev.\ D {\bf 93} (2016) no.5,  054038
   Erratum: [Phys.\ Rev.\ D {\bf 93} (2016) no.7,  079905]
  [arXiv:1512.00022 [hep-ph]].

\bibitem{boost}
  The boost C++ libraries,
  \href{https://www.boost.org/}{https://www.boost.org/}.

\bibitem{GMP/MPFR}
  The GNU Multiple Precision Arithmetic Library,
  \href{http://gmplib.org/}{http://gmplib.org/};\\
  The GNU Multiple Precision Floating-Point Reliable Library,
  \href{https://www.mpfr.org/}{https://www.mpfr.org/}.
  
\bibitem{Banfi:2004yd}
  A.~Banfi, G.~P.~Salam and G.~Zanderighi,
  ``Principles of general final-state resummation and automated implementation,''
  JHEP {\bf 0503} (2005) 073
  [hep-ph/0407286].

\bibitem{Banfi:2014sua}
  A.~Banfi, H.~McAslan, P.~F.~Monni and G.~Zanderighi,
  ``A general method for the resummation of event-shape distributions in $e^{+} e^{-}$ annihilation,''
  JHEP {\bf 1505} (2015) 102
  [arXiv:1412.2126 [hep-ph]].

\bibitem{Bauer:2018svx}
  C.~W.~Bauer and P.~F.~Monni,
  ``A numerical formulation of resummation in effective field theory,''
  JHEP {\bf 1902} (2019) 185
  [arXiv:1803.07079 [hep-ph]].

\end{thebibliography}
